\documentclass[useAMS,usenatbib]{mn2e}
\usepackage{cancel}	
\usepackage{color}
\usepackage{graphicx, subfigure}
\usepackage[authoryear]{natbib}
\usepackage{placeins}
\usepackage{amsmath}
\usepackage{tabulary}

\def\kms{km\,s$^{-1}$}
\def\M{M$_{\odot}$}

\def\Ha{H$\alpha$}

 \def\Ni{$^{56}$Ni}
 \def\Co{$^{56}$Co}
 
 \def\Mej{$M_{\rm ej}$}
\def\Mcsm{$M_{\rm CSM}$}

\def\vph{$v_{\rm phot}$}

\def\tr{$\tau_{\rm rise}$}
\def\td{$\tau_{\rm dec}$}
\def\tm{$\tau_{m}$}
\def\ti{$\tau_{\rm in}$}

\def\Lgriz{$L_{griz}$}
\def\Mgriz{$M_{griz}$}

\def\Lp{$L_{\rm max}$}

\def\Mg{$M_{g}$}

\def\Tcol{$T_{\rm col}$}

\title[Diversity and ejected mass in SLSNe Ic]{On the diversity of super-luminous supernovae: ejected mass as the dominant factor}

\author[Nicholl et al.]{M.~Nicholl$^{1}$\thanks{Email : mnicholl03@qub.ac.uk}, S.~J.~Smartt$^1$, A.~Jerkstrand$^1$, C.~Inserra$^1$, S.~A.~Sim$^1$, T.-W.~Chen$^1$,
\newauthor{S.~Benetti$^2$, M.~Fraser$^3$, A.~Gal-Yam$^4$, E.~Kankare$^1$, K.~Maguire$^5$, K.~Smith$^1$, }
\newauthor{M.~Sullivan$^6$, S.~Valenti$^{7,8}$, D.~R.~Young$^1$, C.~Baltay$^9$, F.~E.~Bauer$^{10,11,12}$, }
\newauthor{S.~Baumont$^{13,14}$, D.~Bersier$^{15}$, M.-T.~Botticella$^{16}$, M.~Childress$^{18,19}$, M.~Dennefeld$^{20}$, }
\newauthor{M.~Della Valle$^{16}$, N.~Elias-Rosa$^2$, U.~Feindt$^{21,22}$, L.~Galbany$^{11,23}$, E.~Hadjiyska$^9$, }
\newauthor{L.~Le Guillou$^{13,14}$, G.~Leloudas$^{4,24}$, P.~Mazzali$^{15}$, R.~McKinnon$^9$, J.~Polshaw$^1$, }
\newauthor{D.~Rabinowitz$^9$, S.~Rostami$^9$, R.~Scalzo$^{17}$, B.~P.~Schmidt$^{17}$, }
\newauthor{S.~Schulze$^{10,11}$, J.~Sollerman$^{25}$, F.~Taddia$^{25}$, F.~Yuan$^{17}$}
\\
$^1$Astrophysics Research Centre, School of Mathematics and Physics, Queens University Belfast, Belfast BT7 1NN, UK \\
$^2$INAF - Osservatorio Astronomico di Padova, vicolo dell'Osservatorio 5, I-35122 Padova, Italy \\
$^3$Institute of Astronomy, University of Cambridge, Madingley Road, Cambridge, CB3 0HA, UK\\
$^4$Benoziyo Center for Astrophysics, Weizmann Institute of Science, Rehovot 76100, Israel\\
$^5$European Southern Observatory, Karl-Schwarzschild-Str. 2, 85748 Garching b. M\"unchen, Germany\\
$^6$School of Physics and Astronomy, University of Southampton, Southampton, SO17 1BJ, UK\\
$^7$Department of Physics, University of California, Santa Barbara, Broida Hall, Mail Code 9530, Santa Barbara, CA 93106-9530, USA\\
$^8$Las Cumbres Observatory, Global Telescope Network, 6740 Cortona Drive Suite 102, Goleta, CA 93117, USA\\
$^9$Department of Physics, Yale University, New Haven, CT 06520-8121, USA\\
$^{10}$Instituto de Astrof\'{\i}sica, Facultad de F\'{i}sica, Pontificia Universidad Cat\'{o}lica de Chile, 306, Santiago 22, Chile\\
$^{11}$Millennium Institute of Astrophysics, Vicu\~{n}a Mackenna 4860, 7820436 Macul, Santiago, Chile\\
$^{12}$Space Science Institute, 4750 Walnut Street, Suite 205, Boulder, Colorado 80301\\
$^{13}$Sorbonne Universites, UPMC Univ. Paris 06, UMR 7585, LPNHE, F-75005 Paris, France\\
$^{14}$CNRS, UMR 7585, Laboratoire de Physique Nucleaire et des Hautes Energies, 4 place Jussieu, 75005 Paris, France\\
$^{15}$Astrophysics Research Institute, Liverpool John Moores University, 146 Brownlow Hill, Liverpool L3 5RF, UK\\
$^{16}$INAFÐOsservatorio Astronomico di Capodimonte, Salita Moiariello 16, I-80131 Napoli, Italy\\
$^{17}$Research School of Astronomy and Astrophysics, Australian National University, Canberra, ACT 2611, Australia.\\
$^{18}$ARC Centre of Excellence for All-sky Astrophysics (CAASTRO), Australian National University, Canberra, ACT 2611, Australia.\\
$^{20}$Institut d'Astrophysique de Paris, CNRS, and Universite Pierre et Marie Curie, 98 bis Boulevard Arago, 75014, Paris, France\\
$^{21}$Institut f\"ur Physik, Humboldt-Universit\"at zu Berlin, Newtonstr. 15, 12489 Berlin, Germany\\
$^{22}$Physikalisches Institut, Universit\"at Bonn, Nu§allee 12, 53115 Bonn, Germany\\
$^{23}$Departamento de Astronom\'ia, Universidad de Chile, Casilla 36-D, Santiago, Chile\\
$^{24}$Dark Cosmology Centre, Niels Bohr Institute, University of Copenhagen, Juliane Maries vej 30, 2100 Copenhagen, Denmark\\
$^{25}$Department of Astronomy and the Oskar Klein Centre, Stockholm University, AlbaNova, SE-106 91 Stockholm, Sweden\\
}% 

\begin{document}

\maketitle

\clearpage

\begin{abstract}
We assemble a sample of 24 hydrogen-poor super-luminous supernovae (SLSNe). Parameterizing the light curve shape through rise and decline timescales shows that the two are highly correlated. Magnetar-powered models can reproduce the correlation, with the diversity in rise and decline rates driven by the diffusion timescale. Circumstellar interaction models can exhibit a similar rise-decline relation, but only for a narrow range of densities, which may be problematic for these models. We find that SLSNe are approximately 3.5 magnitudes brighter and have light curves 3 times broader than SNe Ibc, but that the intrinsic shapes are similar. There are a number of SLSNe with particularly broad light curves, possibly indicating two progenitor channels, but statistical tests do not cleanly separate two populations. The general spectral evolution is also presented. Velocities measured from Fe II are similar for SLSNe and SNe Ibc, suggesting that diffusion time differences are dominated by mass or opacity. Flat velocity evolution in most SLSNe suggests a dense shell of ejecta. If opacities in SLSNe are similar to other SNe Ibc, the average ejected mass is higher by a factor 2-3. Assuming $\kappa=0.1\,$cm$^2\,$g$^{-1}$, we estimate a mean (median) SLSN ejecta mass of 10\,\M~(6\,\M), with a range of 3-30\,\M. Doubling the assumed opacity brings the masses closer to normal SNe Ibc, but with a high-mass tail. The most probable mechanism for generating SLSNe seems to be the core-collapse of a very massive hydrogen-poor star, forming a millisecond magnetar.

\end{abstract}

\begin{keywords}Supernovae: general -- Supernovae: LSQ14mo -- Supernovae: LSQ14bdq -- Supernovae: SN 2013hx
\end{keywords}

\section{Introduction}\label{sec:intro}

By the end of the last century, the diversity in observed supernovae (SNe) could be explained in terms of well-understood physical differences: explosions due to iron core collapse, or thermonuclear runaway;  progenitor stars being  hydrogen-rich, or -poor;  the presence or absence of a dense circumstellar medium \citep[e.g.][]{fil1997}. However in the era of expansive, untargetted sky surveys, such as the Palomar Transient Factory \citep[PTF;][]{rau2009}  Pan-STARRS1 \citep[PS1;][]{kai2010}, 
the La Silla QUEST survey \citep[LSQ;][]{balt2013} and the Catalina Real-Time Transient Survey \citep[CRTS;][]{dra2009} thousands of supernovae are being discovered each year, revealing unexpected new types of explosions. 

Perhaps the most mysterious of these are the ``super-luminous'' supernovae \citep[SLSNe;][]{gal2012}, so called because of peak luminosities over 2 magnitudes brighter than the bulk of the SN population \citep[of which they make up only $\sim$0.01\%; e.g.][]{qui2013,mcc2015}. Although outliers had been noted previously, \citet{qui2011} were the first to define the SLSN class, and showed that the light curves of these objects are difficult to explain using only the release of energy deposited by the SN shock wave and the decay of synthesised \Ni, which are the energy sources for the known SN population. Because of their high lumosities, SLSNe have been spectroscopically confirmed up to redshift $z\ga 1.5$ \citep{ber2012}, and candidates have even been detected photometrically at $z=2-4$ \citep{coo2012}.

Three main models have been proposed to account for the enormous energy radiated by SLSNe. One is a central engine, such as a magnetised neutron star spinning with a period of order milliseconds \citep[often referred to as the magnetar model;][]{kas2010,woo2010}. Magnetars (albeit with longer periods) have been observed in our Galaxy, and are thought to originate from stars with main-sequence masses $M_{\rm ZAMS}=30-40\,$\M~(\citealt{gae2005}; but see also \citealt{dav2009}). The compact object spins down and heats the ejected SN gas through high-energy emission, though it is an open question as to exactly how this energy can thermalise in the ejecta \citep{met2014}. Another model is a collision with (or shock breakout from) a highly opaque circumstellar medium (CSM), releasing shock energy at a large radius \citep{woo2007,ofek2010,che2011,gin2012}. Comparisons with data have shown that magnetar engines and circumstellar interaction can both reproduce the range of shapes of SLSN light curves and distinguishing between them has been problematic \citep{ins2013,cha2013,nic2014}.

The final mechanism is thermonuclear runaway in a star with a main-sequence mass above 130\M, triggered by pair-production in the hot carbon-oxygen core \citep[a pair-instability supernova, or PISN;][]{bar1967,rak1967}. A small number of SLSNe have been proposed to be PISNe based on their slowly fading light curves, which have decline rates that approximately match the decay of $^{56}$Co \citep[e.g.~SN 2007bi;][]{gal2009,you2010}. This physical scenario may be responsible for only a fraction of the SLSN population, since the slowly fading types appear to be rarer than those which decay too rapidly to be radioactively powered \citep[][estimate the slowly fading SLSNe to be around 10\% of the total SLSN population]{nic2013,mcc2015}. However the physical reality of pair-instability explosions is not firmly established, and the rise times of well-observed events do not satisfactorily match the predictions of quantitative light curve and spectral modelling \citep{nic2013,mcc2014,des2012}. 

Extensive studies of SLSNe in the last few years illustrate a diversity of spectral and photometric properties. Most display no signs of hydrogen \citep{pas2010,ins2013} and occur in low-metallicity dwarf galaxies -- perhaps similar to the hosts of long-duration gamma-ray bursts \citep{nei2011,chen2013,lun2014}, and possibly in even more extreme star-forming environments \citep{chen2014,lel2015}. These objects have been termed SLSNe Ic, by analogy with normal-luminosity SNe from stripped progenitors. Within this group, objects can show nearly identical spectral evolution and yet have very different light curves \citep{ins2013,nic2014}. 
Some SLSNe Ic (SN 2007bi-like objects) clearly evolve on very long timescales \citep{gal2009,nic2013,mcc2014}, but it  is unclear whether these form a distinct subclass (and arise from a different physical mechanism, such as the pair instability) or are part of a continuous distribution. For example, \cite{ins2014} found that their standardisation of SLSN peak magnitudes could encompass the slowly fading objects as well as the more typical ones. 

Another group, SLSNe II, do have hydrogen in their spectra. This is sometimes in the form of strong, multi-component emission lines, almost certainly indicating interaction with CSM. The prototypical example here is SN 2006gy \citep{ofek2007,smi2007b}, and the physical mechanism (the conversion of kinetic energy to radiative energy by shocks at the ejecta-CSM collision) is well established. Quite a few such objects are now known -- e.g. SNe 2006tf \citep{smi2008}, 2008fz \citep{dra2010}, 2008am \citep{cha2011} and 2003ma \citep{rest2011} -- and these are quite correctly dubbed `SLSNe IIn', by analogy with the fainter SNe IIn, which are hydrogen-rich SNe showing narrow spectral lines from shocked CSM.
However, a few SLSNe have much weaker hydrogen lines visible, which are not obviously multi-component and do not unambiguously point to interaction being the dominant power source of the radiative energy. The earliest example of this class is SN 2008es \citep{gez2009,mil2009}. Although they are classed as SLSNe II, they resemble SLSNe Ic with H lines superimposed. Their overall light curve and other spectral properties are closer to SLSNe Ic than to the SLSNe IIn.
One object in particular, CSS121015, prompted \citet{ben2014} to propose that the two spectroscopic classes of SLSNe may in fact come from the same underlying physical process, with their observational properties modified by the hydrogen mass in the ejecta/CSM. CSS121015 had one of the highest peak luminosities of any SLSN to date, but its light curve could be well fit with both interaction and magnetar models. The authors favoured the interaction scenario, due to the presence of time-variability in the narrow Balmer emission, indicating slow-moving material close to the SN.

As SLSNe are extremely rare events, and were largely unknown before the discoveries of SNe 2005ap \citep{qui2007} and SCP06F6 \citep{bar2009}, the paucity of observed events has so far restricted the analysis of their properties as a group. This situation is now beginning to change, as transient surveys are becoming better at picking out these objects
and dedicating resources to follow them up. The first studies of SLSN samples were recently conducted by \citet{ins2013,ins2014}, who showed that their properties could be explained by magnetar powered models, and their potential utility as standardisable candles for cosmology.

The Public ESO Spectroscopic Survey of Transient Objects (PESSTO) has a strategy designed to classify unusual types of transients early in their evolution, using light curve information from feeder surveys such as La Silla QUEST \citep[LSQ;][]{balt2013}, Catalina Real-Time Transient Survey \citep[CRTS;][]{dra2009}, SkyMapper \citep{kel2007}, 
OGLE-IV \citep{wyr2014} and Pan-STARRS1 \cite[e.g. as described in][]{ins2013}.
PESSTO  is described in detail in \citet{sma2014} and reduced and calibrated spectra are publicly available through both ESO archive\footnote{For details on how to get the PESSTO Phase 3 data, see www.pessto.org} 
and WISeREP\footnote{http://www.weizmann.ac.il/astrophysics/wiserep/} \citep[Weizmann Interactive Supernova data REPository;][]{yar2012}. In this work, we construct the largest SLSN sample to date: a total of 24 objects, 
composed of 7 from PESSTO \citep[][Inserra et al.~in prep., Nicholl et al.~in prep., Chen et al.~in prep.]{nic2014,ben2014}, and 17 from the literature. As all of the theoretical scenarios invoked to power SLSNe likely require stars more massive than typical SN progenitors, we investigate whether our sample have systematically different ejected mass than normal-luminosity stripped-envelope SNe. In section \ref{obj}, we describe the SLSNe in our sample, including the new PESSTO objects. The construction of our bolometric light curves is outlined in section \ref{bol}. We investigate the light curve timescales in section \ref{cor}, leading to an analysis of generalised light curve shapes in section \ref{shape}, and a search for evidence of a bimodal SLSN Ic population (i.e.~with rapid and slow decline rates after maximum luminosity) in section \ref{pop}. The typical spectral evolution is investigated in section \ref{spec}. Velocity measurements from spectra are described in section \ref{vel}, and these are used to estimate SLSN masses relative to normal hydrogen-poor SNe Ibc in section \ref{mass}. We summarise our main results in section \ref{sum}, and conclude in section \ref{conc}.

\begin{table}

\caption{SLSNe in our sample}
\label{sample}
\begin{tabulary}{\columnwidth}{LLLLL}
\hline
Name & Type & $z$ & \Mgriz$^*$ & Reference\\
\hline
\multicolumn{5}{l}{`Gold' sample: rest-frame \textit{gri}(\textit{z}) coverage}\\
\hline
SN2007bi		&	Ic$^\dagger$	&	0.127	&	-20.20	&	\citet{gal2009}\\
SN2008es		&	II		&	0.205	&	-21.43	&	\citet{gez2009},\\
				&			&			&			&	\citet{mil2009}\\
SN2010gx		&	Ic		&	0.230	&	-20.64	&	\citet{pas2010},\\
				&			&			&			&	\citet{qui2011}\\
SN2011ke		&	Ic		&	0.143	&	-20.69	&	\citet{ins2013}\\
SN2011kf		&	Ic		&	0.245	&	-20.80	&	\citet{ins2013}\\
SN2012il		&	Ic		&	0.175	&	-20.73	&	\citet{ins2013}\\
SN2013dg		&	Ic		&	0.265	&	-20.30	&	\citet{nic2014}\\
SN2013hx		&	II		&	0.130	&	-20.84	&	Inserra et al.~(in prep)\\
LSQ12dlf		&	Ic		&	0.255	&	-20.68	&	\citet{nic2014}\\
LSQ14mo		&	Ic		&	0.253	&	-19.95	&	Chen et al.~(in prep) \\
LSQ14bdq	&	Ic		&	0.347	&	-21.68	&	\citet{nic2015b}\\
PTF10hgi		&	Ic		&	0.100	&	-19.61	&	\citet{ins2013}\\
PTF11rks		&	Ic		&	0.190	&	-20.01	&	\citet{ins2013}\\
PTF12dam	&	Ic$^\dagger$	&	0.107	&	-20.56	&	\citet{nic2013}\\
CSS121015	&	II		&	0.287	&	-22.00	&	\citet{ben2014}\\
SSS120810	&	Ic		&	0.156	&	-20.45	&	\citet{nic2014}\\
PS1-11ap		&	Ic$^\dagger$	&	0.524	&	-20.54	&	\citet{mcc2014}\\
\hline
\multicolumn{5}{l}{`Silver' sample: rest-frame $g$-band with bolometric correction}\\
\hline
SN2005ap		&	Ic		&	0.283	&	-21.22	&\citet{qui2007}\\
SCP06F6		&	Ic		&	1.189	&	-21.56	&\citet{bar2009}\\
PTF09cnd	&	Ic		&	0.258	&	-21.34	&\citet{qui2011}\\
PTF09cwl		&	Ic		&	0.349	&	-21.15	&\citet{qui2011}\\
PS1-10ky		&	Ic		&	0.956	&	-21.24	&\citet{chom2011}\\
PS1-10bzj	&	Ic		&	0.650	&	-20.32	&\citet{lun2013}\\
iPTF13ajg	&	Ic		&	0.740	&	-21.50	&\citet{vre2014}\\
\hline
\end{tabulary}

$^*$Pseudobolometric magnitude at maximum light; $^\dagger$Described in the literature as a slowly-declining event
\end{table}

\section{The sample}\label{obj}

In this work we focus on the Type Ic SLSNe. However, we also include three SLSNe II. While SLSNe Type IIn, such as SN 2006gy, show prominent multicomponent Balmer lines indicating circumstellar interaction, the three objects used in our sample showed only weak and/or broad hydrogen lines. As the power source for these objects is ambiguous, they may be related to SLSNe Ic. A full summary of the sample is given in Table \ref{sample}. A comparison sample of normal-luminosity stripped-envelope SNe (Types Ib, Ic and broad-lined Ic) is listed in Table \ref{ics}. This contains a compilation of SNe Ibc from the literature that have good photometric data in \textit{griz}, as well as the homogeneous SDSS II sample of \citet{tad2014}

\subsection{Published PESSTO objects}

The first batch of SLSNe Ic classified by PESSTO were presented and analysed by \citet{nic2014}. These were LSQ12dlf, SSS120810 and SN 2013dg. Each object exhibited spectral evolution typical of the class, despite their light curves being quite diverse. Another PESSTO SLSN, CSS121015, was studied by \citet{ben2014}. This object was an extremely luminous SLSN II, but bore resemblance to SLSNe Ic in both the spectrum and overall light curve shape. Fitting by \citet{nic2014}, with magnetar- and CSM-powering, showed that for a given power source, the CSS121015 models occupied a similar region of parameter space to the SLSNe Ic. Including CSS121015 in our sample may help to clarify the existence of a link between normal SLSNe Ic, and some SLSNe II.

\subsection{LSQ14mo}

LSQ14mo was discovered rising steadily in LSQ observations taken from 2014 Jan 12.2 UT. The transient is located at RA=10:22:41.53, Dec=-16:55:14.4 (J2000.0). A spectrum taken by PESSTO on 2014 Jan 31.2 UT was dominated by a blue continuum and O II absorption at around 4000 \AA, revealing it to be a SLSN Ic at a phase of $\sim$1 week before peak luminosity. The spectrum was an excellent match to PTF09cnd \citep{qui2011} at a redshift $z\sim0.25$. A precise redshift of $z=0.253$ was subsequently determined from narrow Mg II $\lambda \lambda$ 2795,2802 absorption \citep{14mo_atel}. PESSTO has collected extensive data on this target, which will be presented in full in a future publication (Chen et al., 2015, in prep.).

\subsection{LSQ14bdq}

LSQ14bdq was also discovered by LSQ during a long rising phase ($>$40 days), with the first confirmed detection occurring on 2014 April 5.1 UT, at a position RA=10:01:41.60, Dec=-12:22:13.4. A spectrum was obtained by PESSTO on 2014 May 4.9 UT, showing it to be a SLSN Ic before maximum light. The redshift was determined to be $z=0.347$, initially by comparison with other pre-maximum SLSNe such as PTF12dam and PTF09cnd, and then more precisely through the detection of Mg II absorption \citep{nic2015b}.

\subsection{SN 2013hx}

A hostless transient was first detected by SkyMapper on 2013 Dec 27 UT at coordinates RA=01:35:32.83, Dec=-57:57:50.6. It was given the survey designation SKYJ1353283-5757506. PESSTO observed the object on 2014 Feb 20 UT after it had risen in luminosity for $\ga$30 days. The spectrum showed \Ha\ emission at $z=0.13$, at which redshift the absolute magnitude was $\sim-22$, as well as broad features in the blue. It showed similarity to both SN 2010gx \citep[a prototypical SLSN Ic;][]{pas2010} and CSS121015 \citep[SLSN II;][]{ben2014}. The SN has been followed up by PESSTO and given the IAU name, SN 2013hx. A separate followup paper will present the full dataset (Inserra et al., 2015, in prep.)

\begin{table}
\caption{Comparison sample}
\label{ics}

\begin{tabulary}{\columnwidth}{LLLL}

\hline
Name & Type  & \Mgriz & Reference\\

\hline
\multicolumn{4}{l}{Well-observed SNe in the literature}\\
\hline
SN1994I		&	Ic		&	$-16.79$	&	\citet{fil1995},\\
			&			&		&	\citet{ric1996}\\
SN1998bw	&	IcBL$^\ddagger$		&	$-16.84$	&	\citet{pat2001}\\
SN1999ex	&	Ic		&	$-16.84$	&	\citet{str2002}\\
SN2002ap	&	IcBL		&	$-16.49$	&	\citet{maz2002}\\
			&			&			&	\citet{gal2002}\\
SN2003jd		&	IcBL		&	$-18.19$	&	\citet{val2008b}\\
SN2004aw	&	Ic		&	$-17.24$	&	\citet{tau2006}\\
SN2007gr		&	Ic		&	$-16.36$	&	\citet{val2008}\\
SN2008D		&	Ib		&	$-16.24$	&	\citet{sod2008},\\
			&			&			&	\citet{mod2009}\\
SN2009jf		&	Ic		&	$-17.34$	&	\citet{val2011}\\
SN2010bh	&	IcBL$^\ddagger$	&	$-16.97$	&	\citet{cano2011}\\
SN2011bm	&	Ic		&	$-17.63$	&	\citet{val2012}\\
SN2012bz	&	IcBL$^\ddagger$	&	$-18.82$	&	\citet{sch2014}\\
\hline
\multicolumn{4}{l}{SDSS II sample from \citet{tad2014}}\\

\hline
SN2005hl	&	Ib		&	$-17.55$\\
SN2005hm	&	Ib		&	$-15.85$\\
SN2006fe	&	Ic		&	$-17.04$\\
SN2006fo	&	Ib		&	$-17.26$\\
14475	&	IcBL		&	$-19.46$\\
SN2006jo	&	Ib		&	$-18.25$\\
SN2006lc	&	Ib		&	$-17.31$\\
SN2006nx	&	IcBL		&	$-19.71$\\
SN2007ms	&	Ic		&	$-16.95$\\
SN2007nc	&	Ic		&	$-16.82$\\
\hline

\end{tabulary}

$^\ddagger$ Associated with observed gamma-ray burst

\end{table}

\begin{figure*}
\center
\includegraphics[width=13cm,angle=0]{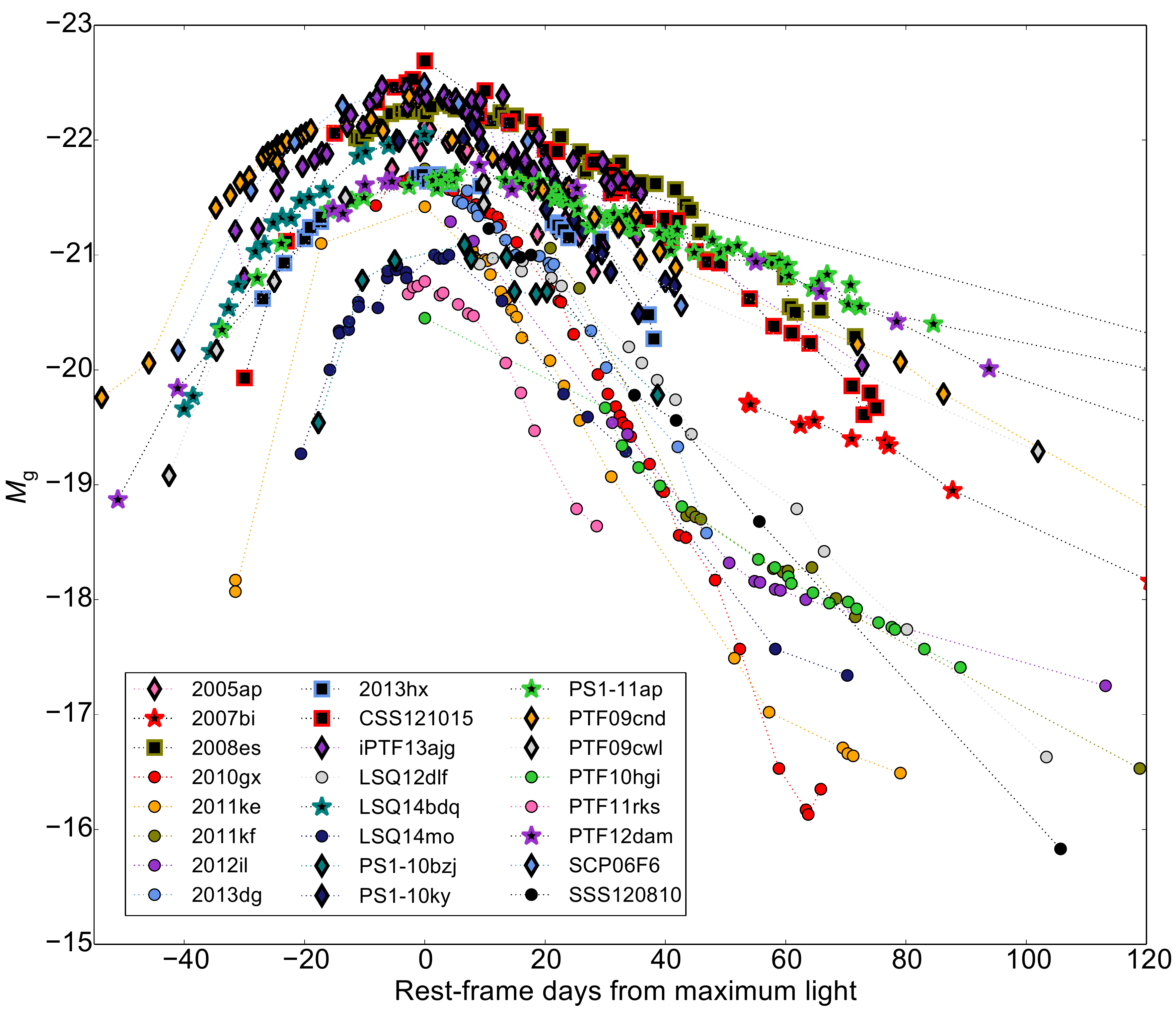}
\includegraphics[width=13cm,angle=0]{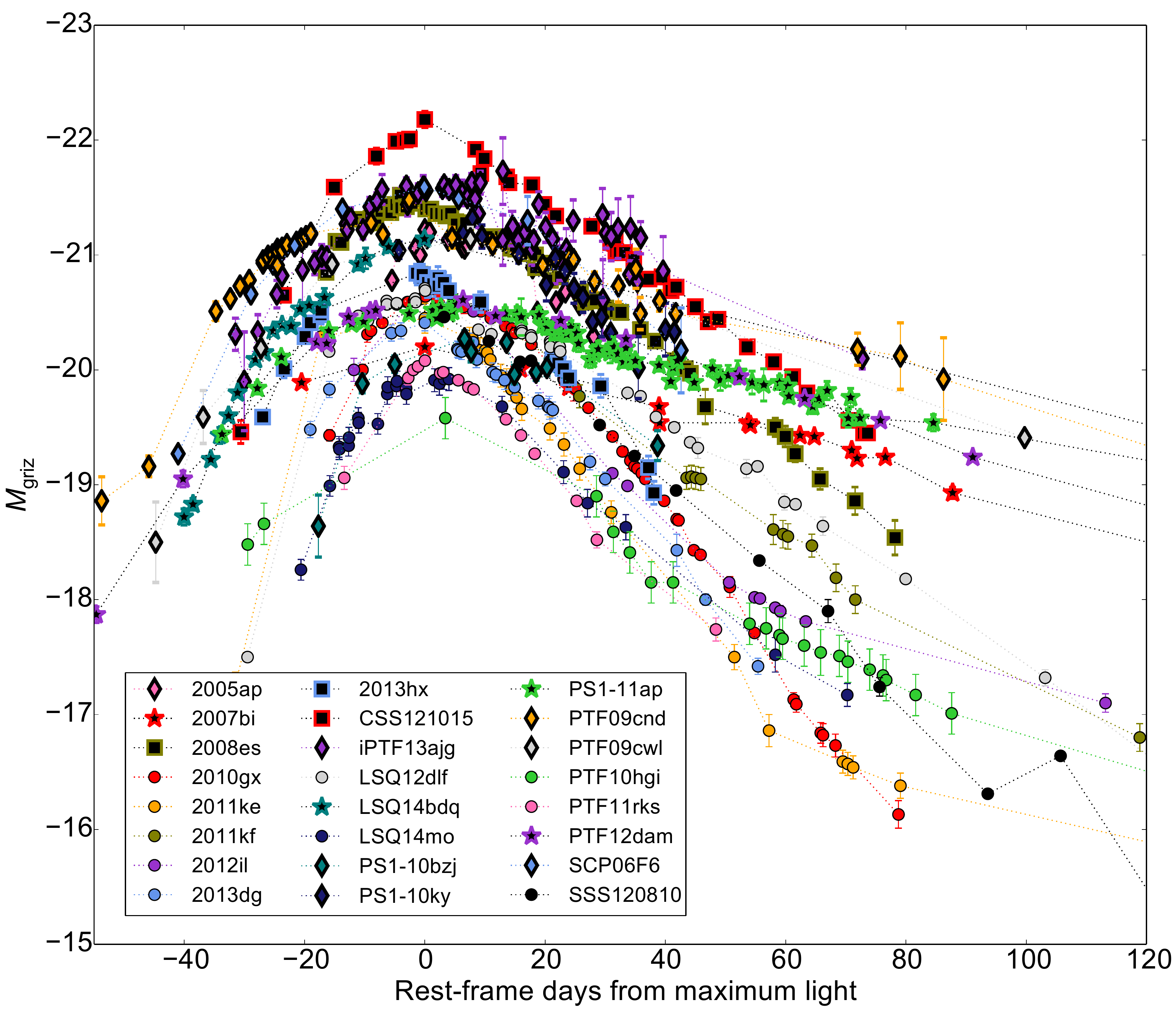}
\caption{Complete set of rest-frame $g$-band (\textit{top}) and $griz$ pseudobolometric (\textit{bottom}) light curves, after extinction and $K$-corrections.}\label{all_lc}
\end{figure*}

\subsection{SLSNe from the literature}

The amount of data available for objects in the literature is highly variable. In some cases, they have only been observed in one or two filters; in others, they are at high redshift and the observed optical light corresponds to ultraviolet (UV) emission in the supernova rest-frame. High-redshift SNe also tend to have sparse spectral data, as they are fainter for observers.

The objects in our sample have therefore been divided into two bins (`Gold' and `Silver' samples), depending on whether they have good coverage at rest-frame optical wavelengths. This can be seen in Table \ref{sample}. The mean redshift for Gold objects is $\langle z \rangle=0.22$. All but two of the SLSNe at $z<0.3$ have extensive photometry in observer-frame $g,r,i$ and in most cases $z$ filters, which at this redshift covers the rest-frame optical regime. This includes all of the PESSTO objects, the 5 low-$z$ SLSNe Ic from \citet{ins2013}, the prototypical SN 2010gx \citep{pas2010}, 3 slowly declining SLSNe Ic \citep[classified as SN2007bi-like;][]{gal2009,nic2013,mcc2014} and one further type II event \citep[SN2008es][]{gez2009}. PS1-11ap, at $z=0.524$, falls in this group because of photometry in the NIR Pan-STARRS1 $y$ filter, which corresponds to rest-frame $i$-band at this redshift \citep{mcc2014}.

In the Silver sample (with a mean of $\langle z \rangle=0.63$), we have two more objects from the Pan-STARRS1 Medium Deep Survey -- PS1-10ky at $z\sim0.9$ \citep{chom2011} and PS1-10bzj at $z\sim0.6$ \citep{lun2013} -- while most of the others featured in the original \citet{qui2011} sample that defined the SLSN Ic class. Although several of these objects are at redshifts $0.25<z<0.3$, the reddest available photometry is in the $R$-band, which corresponds to rest-frame $B$/$g$. SCP06F6 ($z=1.189$) has HST $i$ and $z$ photometry \citep{bar2009}, which corresponds to rest-frame emission between the $u$- and $g$-bands. The final Silver object is iPTF13ajg \citep{vre2014}, which has excellent photometric and spectroscopic coverage, but at $z=0.74$ this mostly probes rest-frame UV. This means that for these objects we must rely on an estimated correction to obtain bolometric light curves.

\section{Bolometric light curves}\label{bol}

To analyse the light curves of our SLSNe in the most homogeneous way possible, we constructed two sets of light curves: rest-frame $g$-band magnitudes, \Mg, and pseudobolometric light curves covering rest-frame SDSS \textit{griz} filters. The first step was to apply $K$-corrections to the observed magnitudes, to transform them to the rest-frames of our objects. These were determined as follows. We calculated synthetic photometry, using the \textsc{iraf}\footnote{Image Reduction and Analysis Facility (IRAF) is distributed by the National Optical Astronomy Observatory, which is operated by the Association of Universities for Research in Astronomy, Inc., under cooperative agreement with the National Science Foundation.} package \textsc{calcphot}, on all available spectra, for each filter in which the SN was observed. Spectra were then corrected to rest-frame using \textsc{dopcor}, including a correction to the flux per unit wavelength by a factor of $1+z$, and new synthetic magnitudes were calculated. The $K$-correction at the epoch of a given spectrum is simply the difference between the rest-frame and observed synthetic magnitudes. These corrections were then linearly interpolated to the epochs with photometry. For most cases we simply corrected $g_{\rm obs}\rightarrow g_{\rm RF}$ etc., aside from the following:  LSQ14bdq ($r_{\rm obs}\rightarrow g_{\rm RF}$); SN 2005ap, PTF09cnd, PTF09cwl ($R_{\rm obs}\rightarrow g_{\rm RF}$); PS1-11ap \citep[$i_{\rm obs}\rightarrow g_{\rm RF}$; $K$-corrections taken from][]{mcc2014}; PS1-10bzj, iPTF13ajg ($i_{\rm obs}\rightarrow g_{\rm RF}$); PS1-10ky ($z_{\rm obs}\rightarrow g_{\rm RF}$); SCP06F6 ($i_{\rm obs},z_{\rm obs}\rightarrow g_{\rm RF}$). For the objects of \citet{qui2011}, sometimes only one spectrum was available -- in this case we also used the spectra of SN 2010gx from \citet{pas2010} (a spectroscopically typical event, with good temporal coverage), after artificially placing them at the desired redshift. Magnitudes were also corrected for Milky Way extinction according to the dust maps of \citet{schlaf2011}, though host reddening was assumed to be negligible.

For the Gold sample, the bolometric light curve was then calculated using this rest-frame photometry from the $g$ to $z$ filters. For LSQ14mo, the $z$-band was estimated using the colours of SN 2013dg, to which it has a very similar light curve in the $gri$-bands, while the $z$-band magnitudes for PS1-11ap and LSQ14bdq were taken from PTF12dam, as in \citet{mcc2014}. Most objects have good coverage before/around peak in one filter only. In these cases, the colours were assumed to be constant, with values from the first epoch that had multi-colour photometry available. This is a reasonable assumption, as the colours show little evolution until 1-2 weeks after maximum light \citep[][]{ins2013}. A spectral energy distribution (SED) was then constructed by converting these magnitudes into flux at the effective wavelength of each filter. For all objects, the flux was set to zero blue-wards of the $g$-band and red-wards of $z$. The luminosity, \Lgriz, was then calculated by integrating this SED over wavelength, and correcting for distance to the SN assuming a cosmology $H_{\rm 0}=72\,$km$\,$s$^{-1}$, $\Omega_{M}=0.27$, and $\Omega_{\Lambda}=0.73$. This follows the procedure of \citet{ins2013}. We express this in terms of pseudobolometric magnitude, using:
\begin{equation}
M_{griz} \equiv -2.5 \log_{10} \left( \frac{L_{griz}}{3.055\times 10^{35}} \right),
\end{equation}
based on the standard definition of bolometric luminosity. Our complete set of $g$-band and pseudobolometric light curves are shown in Fig.~\ref{all_lc}.

\begin{figure}
\includegraphics[width=8.5cm,angle=0]{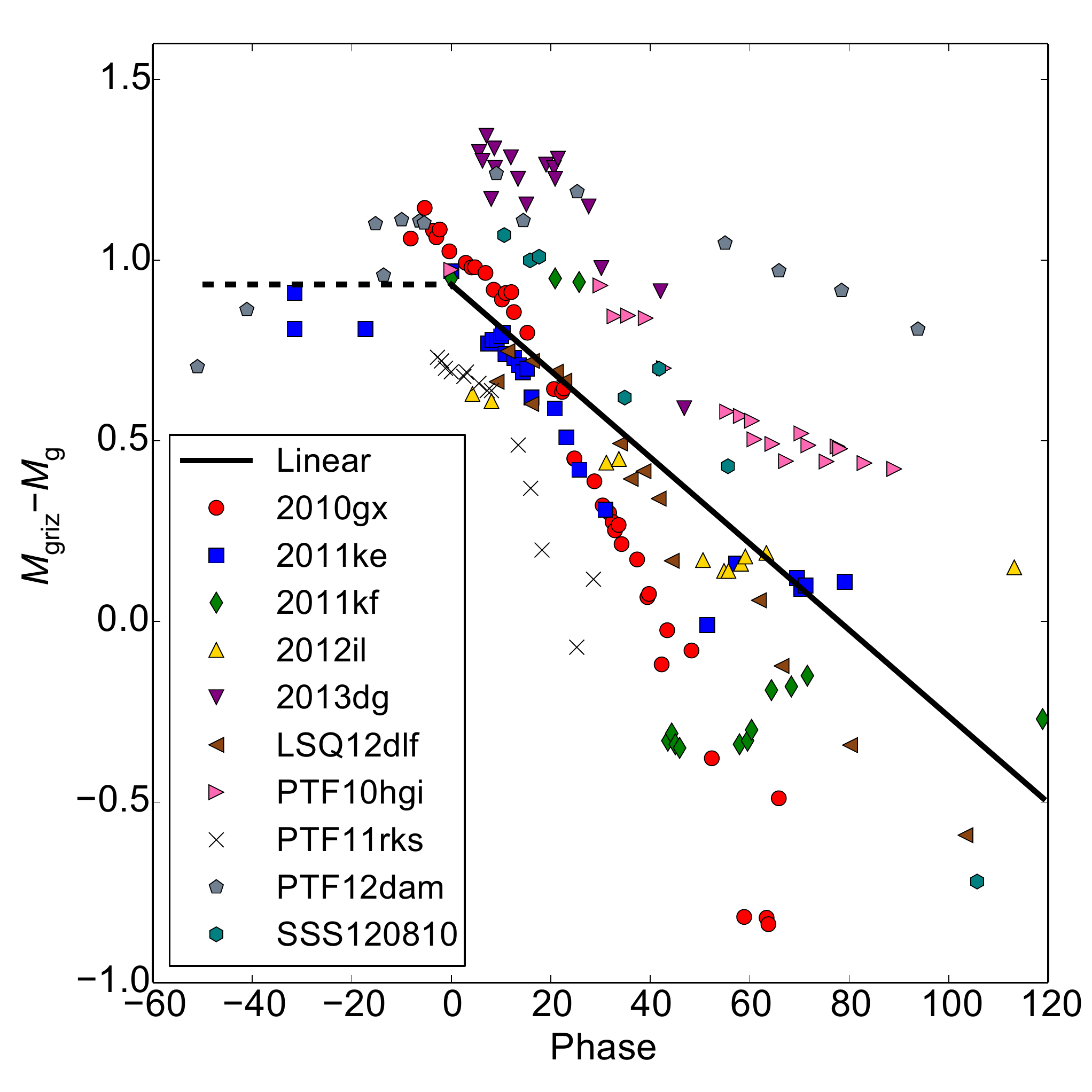}
\caption{Estimating the time-dependent bolometric correction for a typical SLSN. Our best fit is \Mgriz$-$\Mg$\,=0.90 - 0.012 t$ for $t>0$, though the gradient can vary from this by a factor $\sim$2 for individual objects. Uncertainty in the $y$-intercept term has no effect on our analysis.}\label{bol_corr}
\end{figure}

However, a different process was needed to derive the \M~light curves of the Silver objects. It was possible to find the average bolometric correction, \Mgriz$-$\Mg, as a function of time (we define $t=0$ as the epoch of maximum luminosity), for the Gold sample, and apply this correction to our other objects. This correction is shown in Fig.~\ref{bol_corr}. Unfortunately, there is significant scatter, but it is clear that the bolometric correction becomes more negative as a function of time. This is expected; as the SNe cool, bluer wavelengths fade faster at late times. We fit only points where $t>0$, assuming a constant correction before this. Our best fit is \Mgriz$-$\Mg$\,=0.90 - 0.012 t$ for $t>0$ and \Mgriz$-$\Mg$\,=0.90$ for $t<0$. The uncertainty in this correction is $\sim$0.5 magnitudes by 50d after peak, hence at late phases the \M~light curves of Silver objects may become unreliable. However, note that only two Silver objects have data at this phase: PTF09cnd and PTF09cwl. Both of these objects closely resemble PTF12dam in rest-frame $g$ (but were not classified as being 2007bi-like), and this resemblance is preserved in the pseudobolometric light curves. Hence we feel justified in including these objects in our sample.

\section{Light curve timescales}\label{cor}

\subsection{Measurements}

Having constructed our pseudobolometric light curves, we proceeded to measure the rates at which these supernovae brighten to their maximum luminosity, and subsequently decline. We define:

\begin{itemize}
\item{\tr~(the rise timescale): the time ($t<0$) relative to maximum light (\Lp) at which \Lgriz$\,=\,$\Lp$/e$}
\item{\td~(the decline timescale): the time ($t>0$) relative to maximum light (\Lp) at which \Lgriz$\,=\,$\Lp$/e$.}
\end{itemize}

\begin{figure}
\includegraphics[width=8.5cm,angle=0]{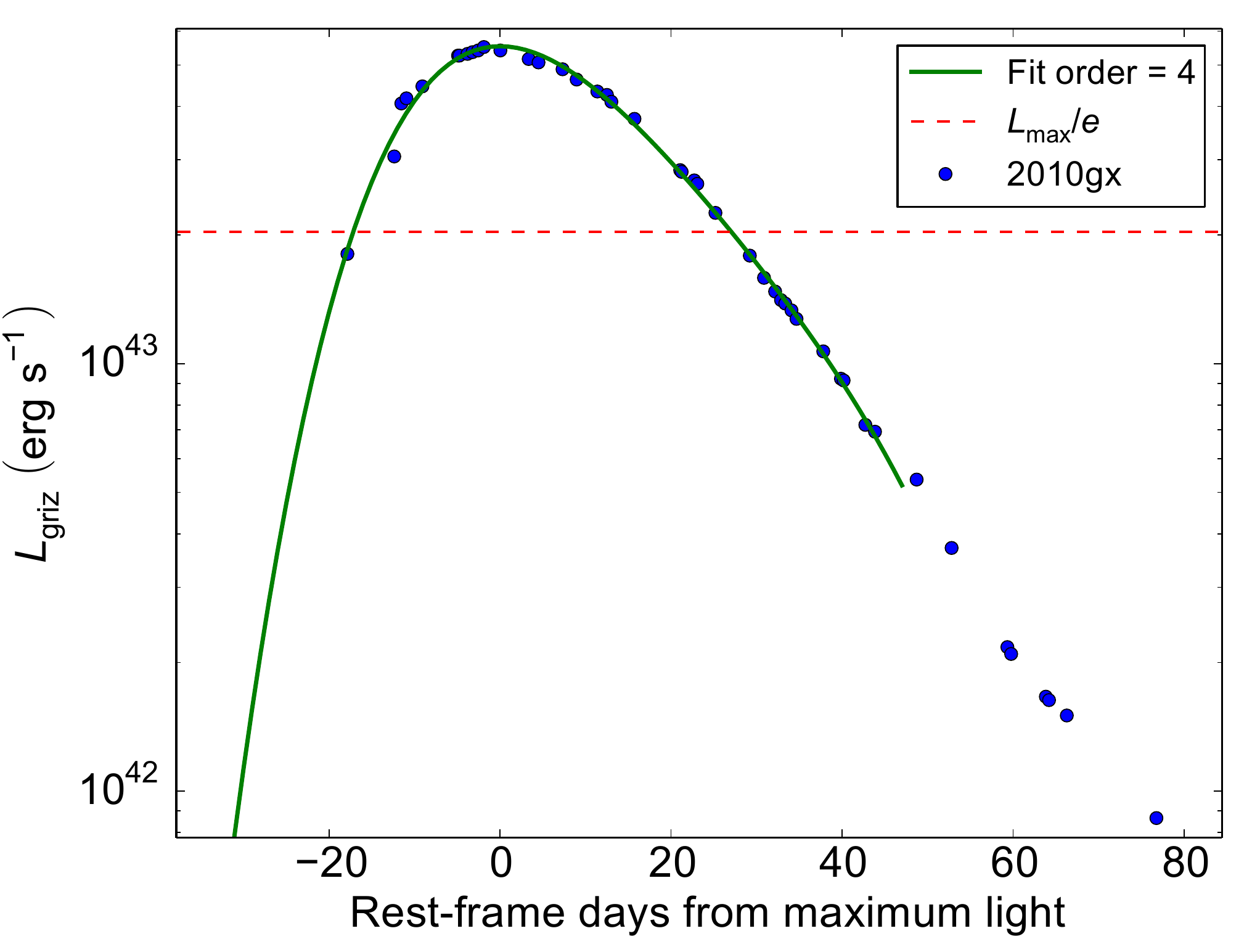}
\caption{Interpolating the light curve of SN 2010gx. The times at which the dashed line intersects the polynomial fit give the exponential rise and decline times.}\label{10gx_fit}
\end{figure}

We make our measurements by fitting the light curves with low-order polynomials. Order four polynomials were found to give a good fit to all of our light curves for epochs $t\la50\,$d. The fits were used to make a new estimate of the date when the pseudobolometric luminosity peaks. This tends to be later than the peak in $g$-band, as one would expect since the ejecta cool over time. We define the new peak, and then measure the quantities described above by interpolating the light curves with the polynomial fits. The method is demonstrated in Fig.~\ref{10gx_fit}. In some cases, the rise time had to be estimated by extrapolation using our polynomials. We consider this to be reliable for most objects, where we extrapolate only by a few days, but the rise time is poorly constrained for SNe 2007bi, 2005ap, and PS1-10ky. For slowly declining objects, the fourth order fit to the peak was not always a good fit at late epochs; for these objects, we made one fit to $t\la50\,$d to estimate the peak and the rise time, as for the rest of our sample, and then measured the decline time by fitting another polynomial to only the post-maximum data points (fourth order and linear fits gave similar results). Our measurements are given in Table \ref{props}. The rise time of SSS120810 could not be constrained from the available data.

\begin{figure*}
\includegraphics[width=18cm,angle=0]{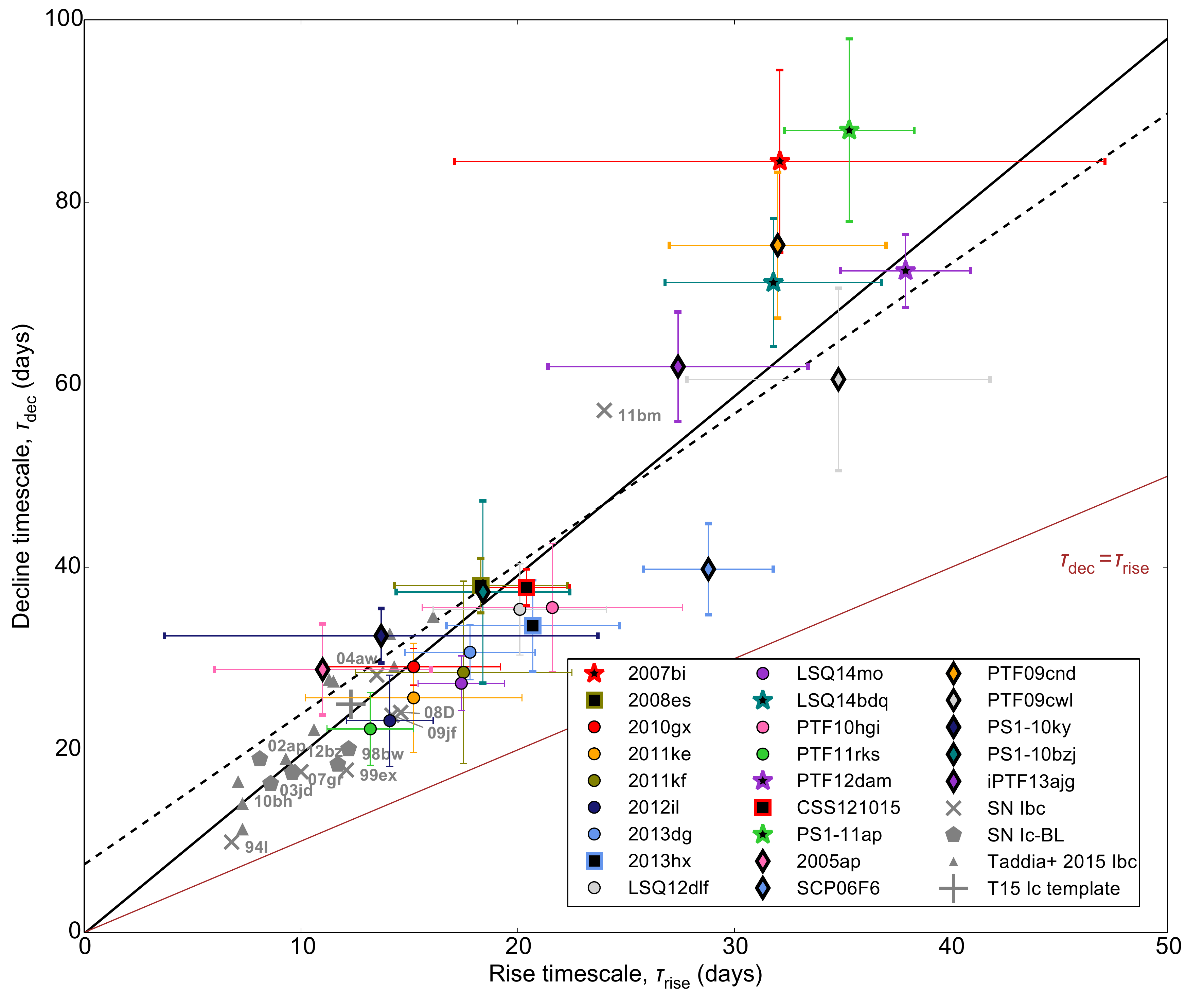}
\caption{Rise vs decline timescales for SLSNe and normal stripped-envelope supernovae. The rise and decline times are clearly correlated, and in a similar way for both samples. The dashed black line gives the best linear fit to the entire SLSN sample ($y = 1.65 x + 7.38$), and the solid black line to the Gold sample only ($y = 1.96 x - 0.10$). The T14 Ic light curve was constructed by integrating the $griz$ templates for SDSS SNe Ic from \citet{tad2014}.
}
\label{corr_plot}
\end{figure*}

\begin{figure*}
\centering
\includegraphics[width=18cm,angle=0]{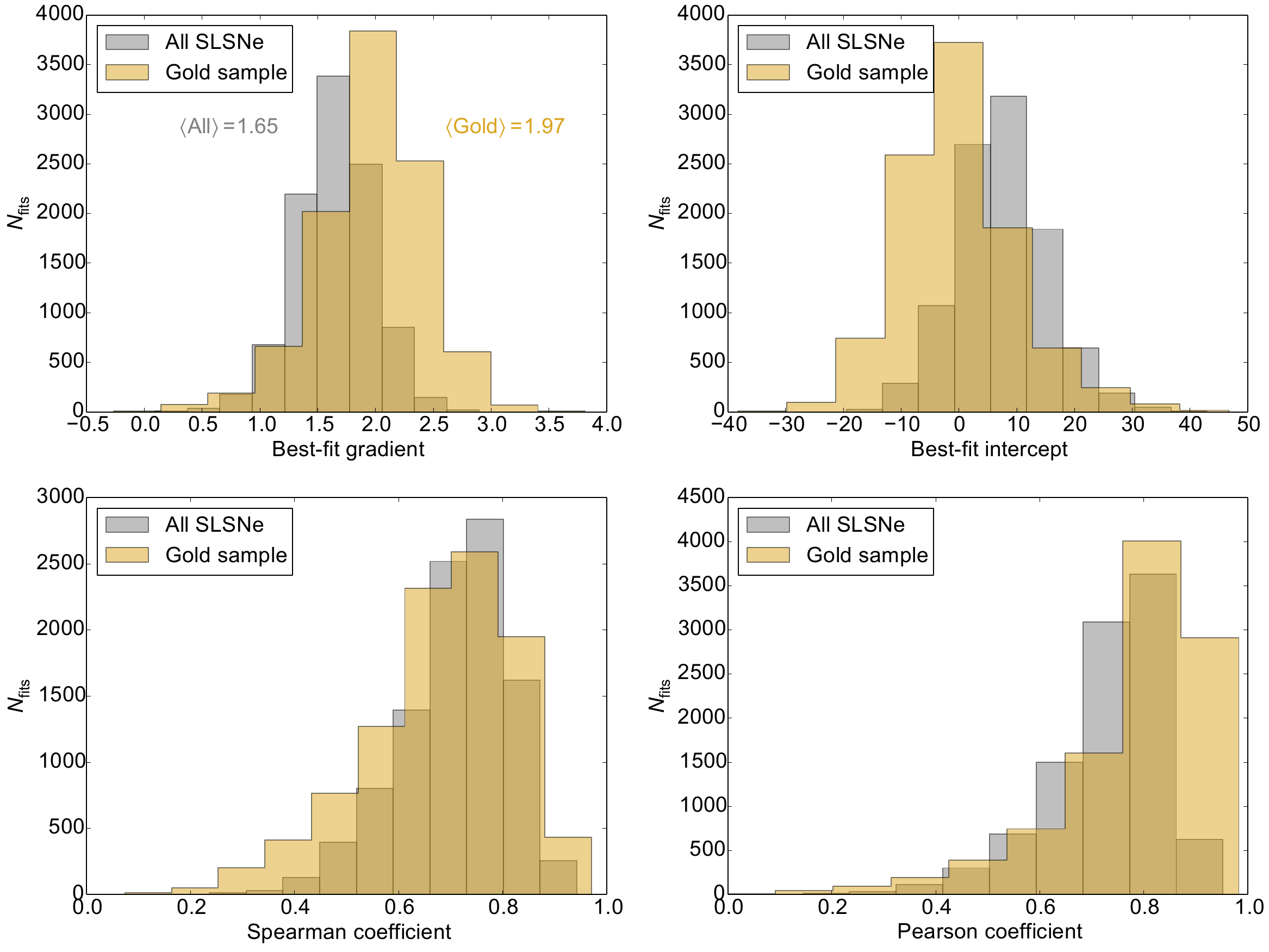}
\caption{Fit parameters for the rise-decline timescale correlation. 
}
\label{signif}
\end{figure*}

\subsection{Correlation}

We plot the rise times vs decline times for our sample in Fig.~\ref{corr_plot}. The best-fit lines to our Gold and complete samples are calculated as follows: we represent each data point by a two-dimensional Gaussian, with a mean given by our measured values of \tr~and \td, and standard deviations by the error bars. A Monte-Carlo method is then employed. A point is drawn at random from each probability distribution defined by these Gaussians, and we use standard {\sc Python} routines to calculate a straight-line fit, as well as Spearman and Pearson correlation coefficients, for the resulting rise-decline relation. This is repeated 10000 times. As can be seen in Fig.~\ref{signif}, the data are clearly correlated with high significance: Spearman's rank correlation coefficient is $0.77\pm 0.10$ for the entire SLSN sample ($0.75\pm 0.11$ for Gold sample only). Pearson's test gives $0.81\pm 0.15$ ($0.84\pm 0.14$). The best-fit straight line to the data is $\tau_{\rm dec} = (1.65\pm 0.33) \tau_{\rm rise} +(7.38\pm 7.79)$ for the full sample, and $\tau_{\rm dec} = (1.96\pm 0.46) \tau_{\rm rise} +(-0.10\pm 10.19)$ for the Gold objects. Although the formal best fit is different for the full sample compared to the Gold SLSNe only, the entire population is clearly consistent with a straight-line, and the two fits agree within the errors. It is no surprise that the gradient is greater than unity, as the light curves of other SN types (both Type Ia and core-collapse) rise to maximum more quickly than they decline. We also measure rise and decline timescales for the SNe Ibc in Table \ref{ics}. They obey a similar correlation as for SLSNe, but with shorter timescales than their more luminous cousins.

\begin{figure*}
\includegraphics[width=18cm,angle=0]{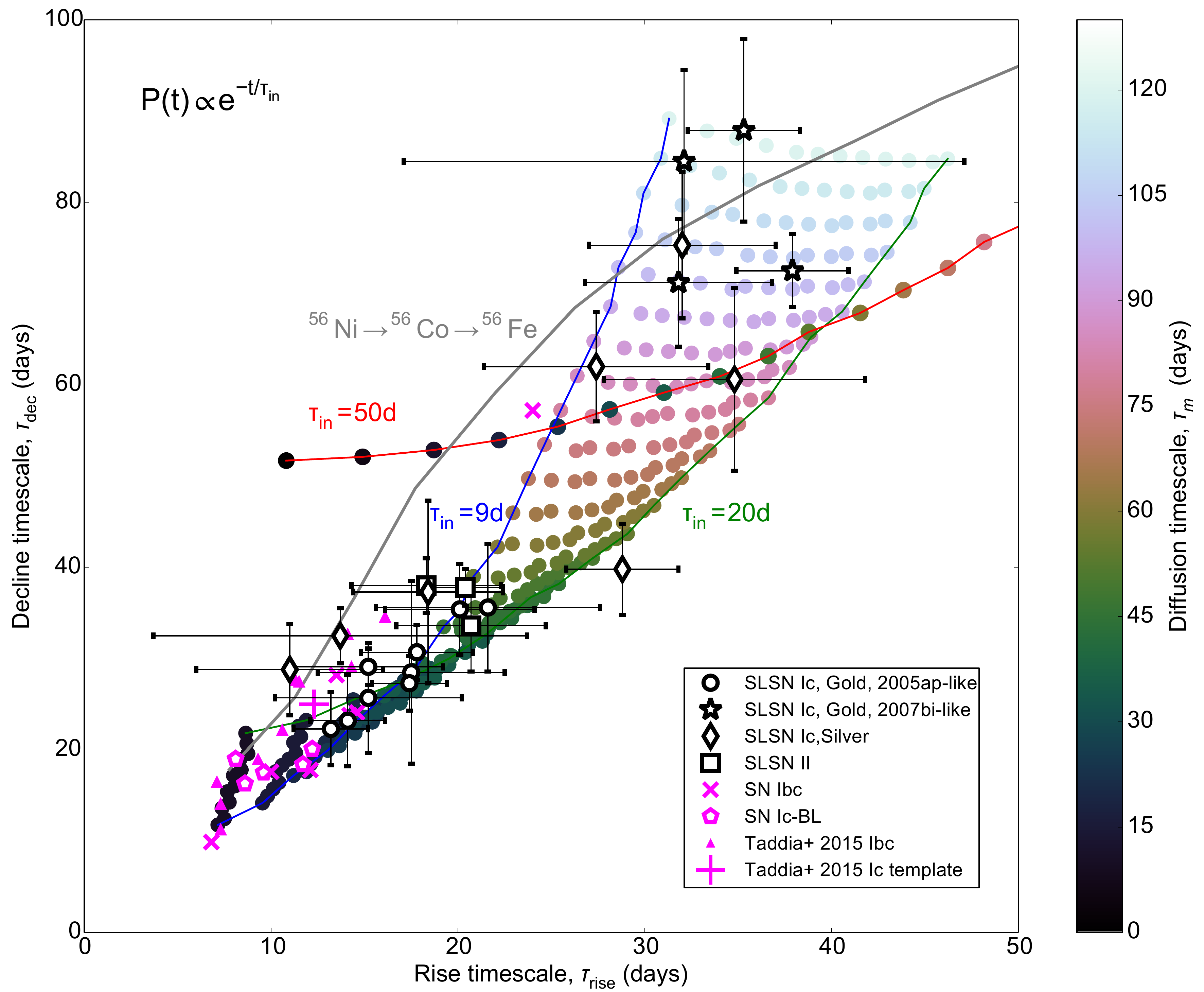}
\caption{The observed rise and decline timescales of our sample overlaid on a grid of diffusion models \citep{arn1982}, with a central heating term that decays exponentially with time. Input times of 10-20 d are needed to reproduce the SLSN correlation, which is then driven by variation in the diffusion timescale, as shown by the colour-bar. \Ni/\Co~models are also shown (grey curve). This roughly matches the normal-luminosity SNe Ibc, although there is a slight offset, which depends on the efficiency of gamma-ray trapping (our model uses the trapping formalism of \citet{arn1982}, with a velocity of 10000 \kms).}
\label{exp}
\end{figure*}

\subsection{Models: overview}

To interpret our correlation, we use the synthetic light curve code described by \citet{ins2013}. This code allows us to model the luminosity from a homologously expanding spherical ejecta with constant opacity and a centrally located power source, $L_{\rm in}(t)$. The light curve equation, re-derived following the original \citet{arn1982} paper but adapted for arbitrary $L_{\rm in}$, is
\begin{equation}
L_{\rm SN}(t) = e^{-\left( t/\tau_m \right)^2} \int_0^t 2 \, L_{\rm in}(t') \frac{t'}{\tau_m} e^{\left( t'/\tau_m \right)^2} \frac{dt'}{\tau_m},
\end{equation}
where \tm~is the diffusion timescale \citep[formally, the geometric mean of the expansion and diffusion timescales;][]{arn1982}.

In the most basic case (for a fixed form of the power input term, $L_{\rm in}$, e.g. an exponentially declining term for heating by radioactive decay, or a central engine with a power-law decline), our model takes three parameters: a diffusion timescale, a power input timescale (which we call \ti), and an overall energy scale (which affects the luminosity of the light curve, but not the shape). In general, the diffusion timescale is a function of ejecta mass, opacity and expansion velocity:
\begin{equation}
\tau_m =\left(\frac{2 \kappa M_{\rm ej}}{\beta c v}\right)^{1/2},
\label{diff}
\end{equation}
where $\kappa$ is the opacity, \Mej~is the ejected mass, $\beta\approx13.8$ (for a wide range of plausible density profiles) is an integration constant, $c$ is the speed of light and $v$ is a scaling velocity for homologous expansion \citep{arn1980,arn1982}. For a given opacity and velocity, the diffusion timescale thus allows us to derive the mass. Other authors have taken the observed rise times of SNe as an estimate of \tm~\citep[most recently][]{whe2014}. This is a reasonable approximation for \Ni-powered light curves, where the decay time is well known, and is closely matched to the typical diffusion times (a coincidence that results in the high peak luminosities in Type Ia SNe). It is not surprising that the normal-luminosity hydrogen-poor sample shown in Fig.~\ref{corr_plot} obey a tight rise-decline correlation: the power input time is the same for all these \Ni-powered SNe, and hence the diversity in both rise and decline times is driven by only one parameter: the diffusion time. The fact that the SLSNe obey such a similar correlation suggests that the diffusion time may also drive the correlation in these objects.

However, for SLSNe the power input time, \ti, is unknown, and may span a wide range of values. If \ti~is very different from \tm, it can have a large influence on the observed rise time, which is no longer a reliable proxy to \tm. A better method here is to use the light curve width: we estimate that the diffusion time through the ejecta is $\tau_m \sim\left(\tau_{\rm rise}+\tau_{\rm dec}\right)/2$. This is explored in detail for the following models, and will be important when we later attempt to estimate masses, in section \ref{mass}.

\begin{figure}
\includegraphics[width=9cm,angle=0]{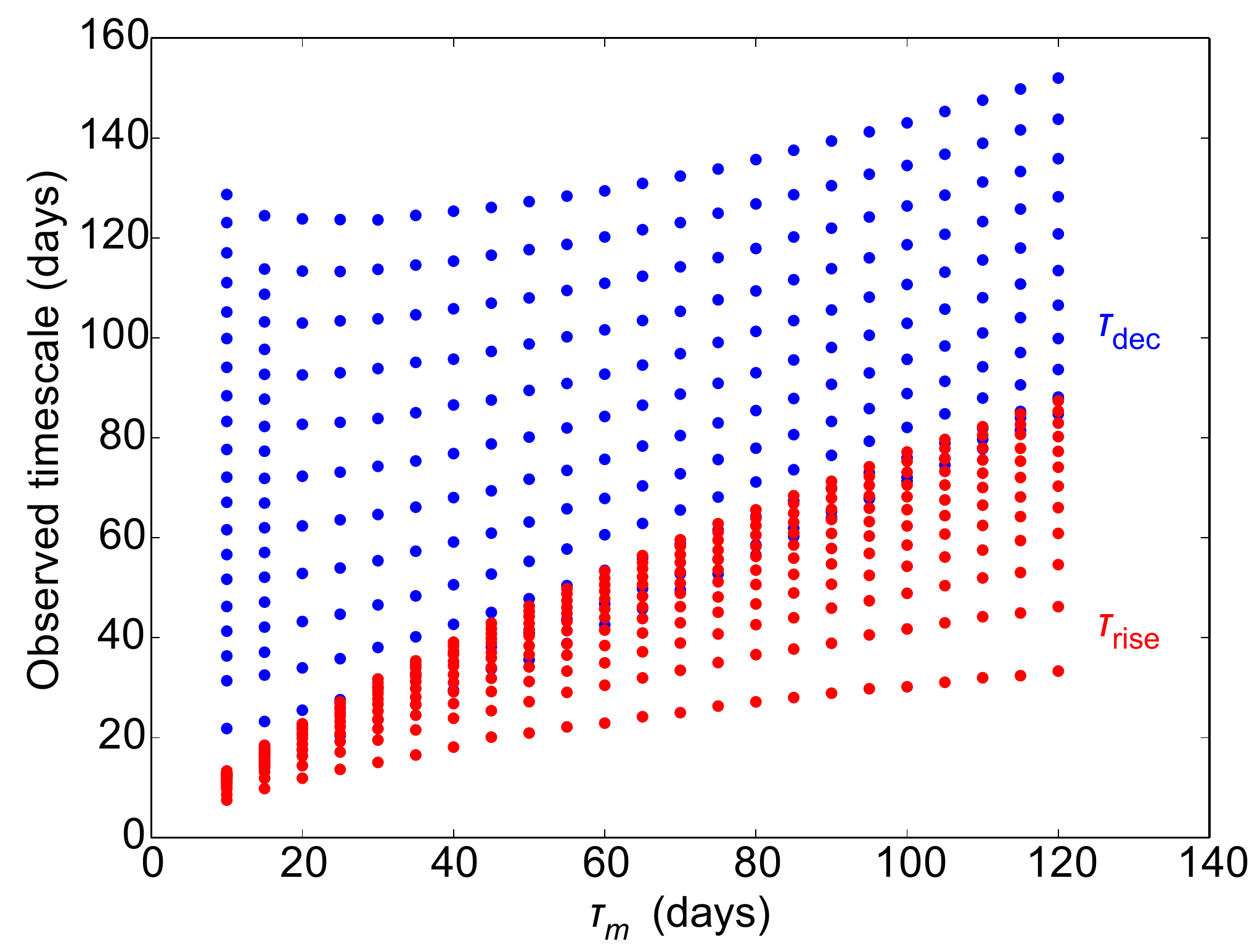}
\includegraphics[width=9cm,angle=0]{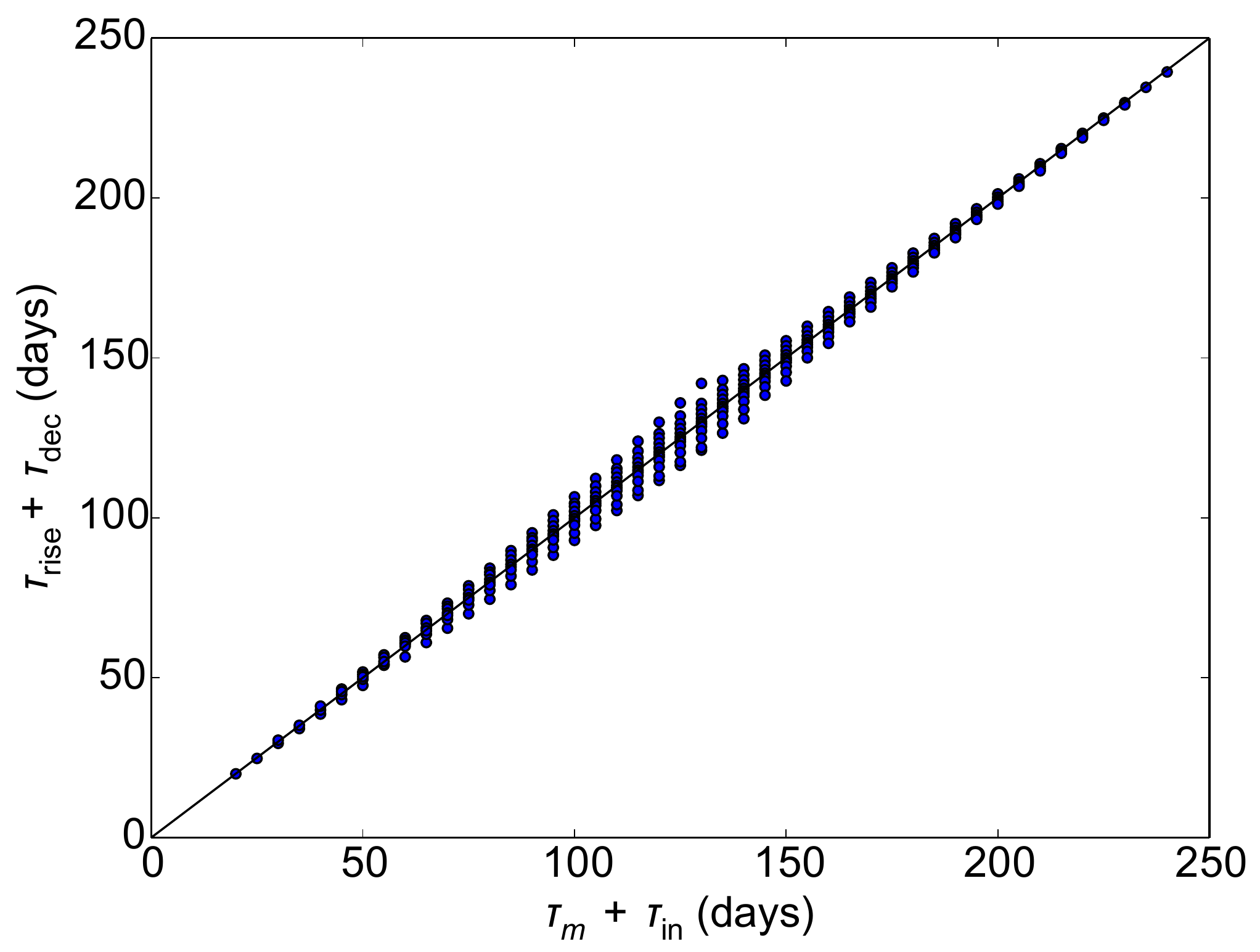}
\caption{The relationships between various timescales in the simple exponential model. While neither the rise nor decline timescale is a good tracer of the diffusion time (\textit{top}), the light curve width is determined by the sum of the power input and diffusion times, such that \tm~can be deduced from observed \tr~and \td~for a given \ti~(\textit{bottom}).}
\label{exp_corr}
\end{figure}

\subsection{Models: \Ni ~and generalised exponential models}
\label{models:ni}

As a first step towards investigating the SLSN parameter space, we generated an array of models with a hypothetical exponentially decaying power source (i.e., with the same functional form as \Ni~decay, but for a variable lifetime). While this power source is not motivated by any proposed physical model, it aids in understanding the relevant timescales, by virtue of being the simplest possible scenario. This model takes only two parameters: the diffusion time (\tm) and the input time (\ti), with $L_{\rm in}=L_0 \exp(-t/\tau_{\rm in})$ ($L_0$ is arbitrary). We varied the diffusion time between 10 and 120 days, in steps of 10 days. A subset of these models are shown on Fig.~\ref{exp}. We see that, if SLSNe are powered by some universal, exponentially declining process, such a process must have a lifetime of $\sim$10-20 days (the smaller points shown are spaced by 1$\,$d in \ti). However, if the timescale for power input is variable, SLSNe in the top right can have longer timescales (shown by the red curve and larger points, with \ti$\,=50\,$d). Most importantly, this figure shows that for this simple model, the correlation in rise and decline timescales is driven by \tm, as seen in the colour scale.

Figure \ref{exp_corr} shows the relationship between diffusion time and measured rise/decline times for a full grid of models, with \tm,\ti$\,=10$-120$\,$d, in steps of 10 days. Clearly, there is no straightforward way to deduce the diffusion time directly from either the rise or the decline. However, we find a strong correlation when the 
quantities are combined: \tr$\,+\,$\td$\,\approx\,$\tm$\,+\,$\ti. This holds over the full range in \tm~and \ti. Hence if we know the timescale of our exponential power source, we can accurately recover the diffusion time by measuring the rise and decline timescales. Measurement of the ejected mass through light curve fitting has been applied for many years to SNe Ibc. Since the input timescale is known for \Ni-powered SNe, the light curve width is typically a measurement of the diffusion time \citep[e.g.][]{arn1982,dro2011,val2008b}. For SLSNe, most of the best-fitting exponential models have \ti$\,\sim\,$\tm. Therefore, from our relation between the 4 important timescales, an estimate of the diffusion time is given by \tm$\,\approx($\tr$\,+\,$\td)$/2$.

Figure \ref{exp} also shows the expected rise-decline curve for models powered by \Ni-decay. In this case, the only important variable is the diffusion time (\Ni~mass sets the overall luminosity, but not the light curve width). We see that this model does predict the correlation exhibited by SNe Ibc, which must be controlled by the diffusion time as mentioned in the previous subsection. Although some of fast rising and fast decaying SLSNe Ic lie close to this \Ni\ decay curve, that power source has already been ruled out for these. As discussed 
in \cite{chom2011}, \cite{qui2011} and \cite{pas2010}, the peak luminosity means that the \Ni~mass would have to be greater than or similar to the total ejecta mass. Such expanding balls of \Ni~are unphysical and ruled out by the observed spectra.

\begin{figure*}
\includegraphics[width=18cm,angle=0]{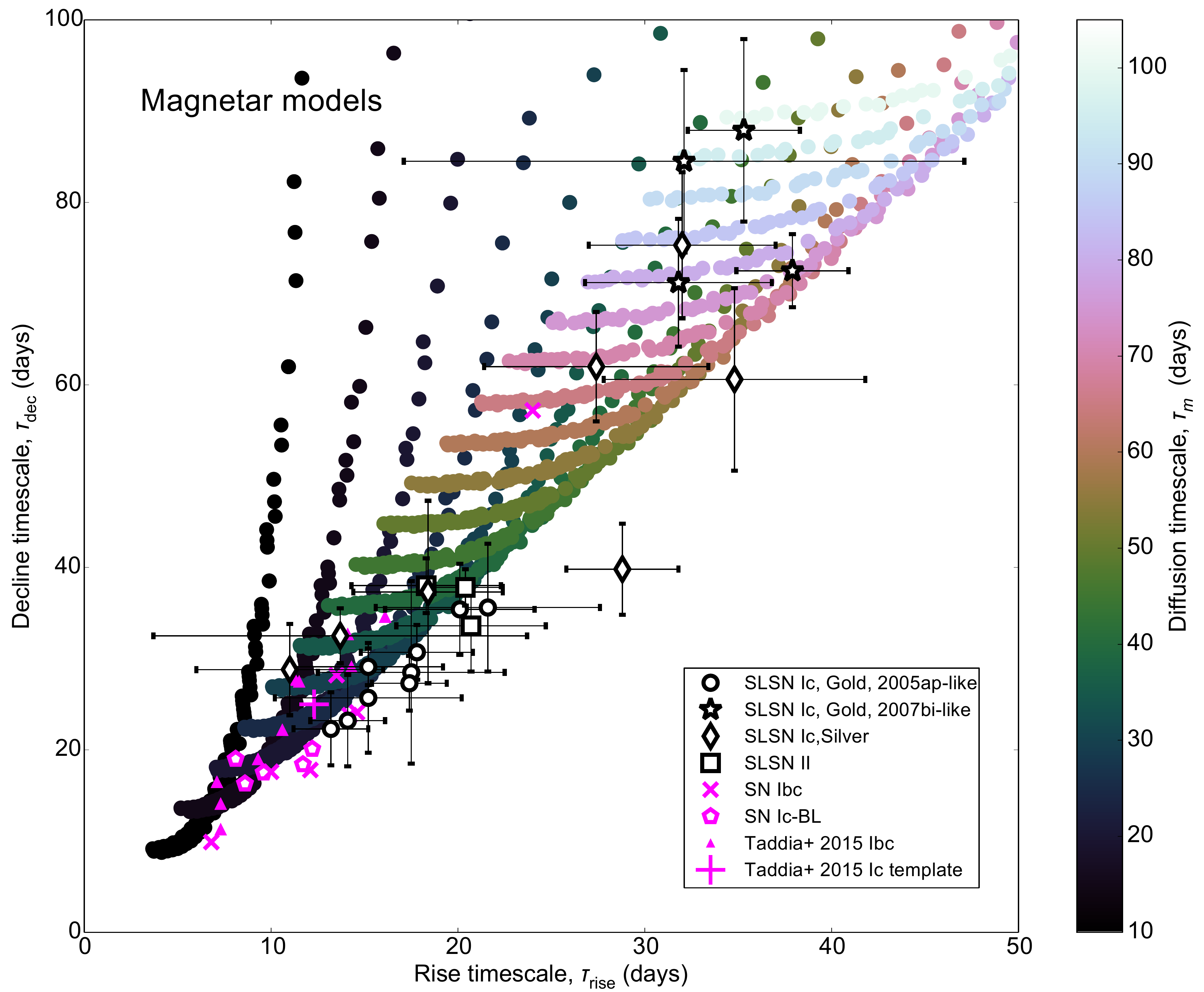}
\vspace{-1em}
\includegraphics[width=18cm,angle=0]{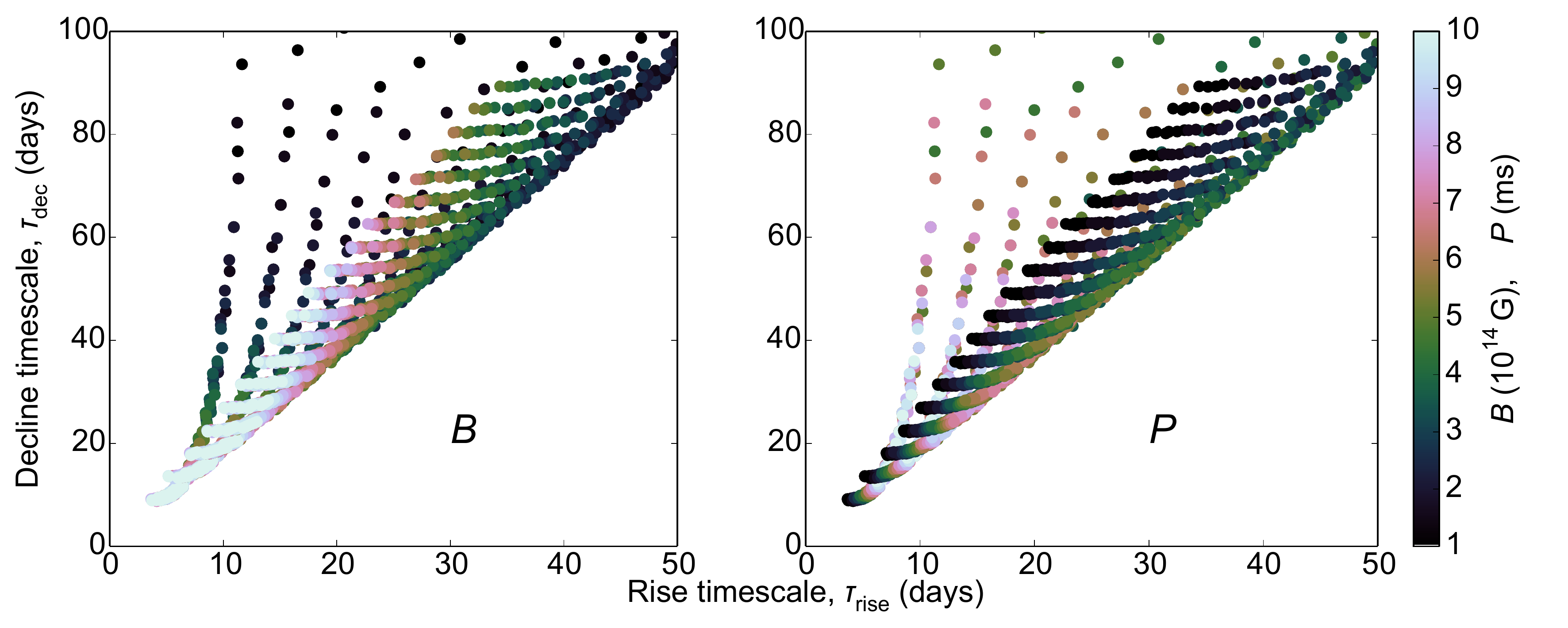}
\caption{We overlay a grid of magnetar-powered diffusion models with different spin period ($P$), magnetic field ($B$) and diffusion time (\tm). All are varied in uniform steps. Only models with $L_{\rm peak}>3\times 10^{43}\,$erg$\,$s$^{-1}$ are plotted. The top panel shows how increasing the diffusion timescale, \tm, traces out the correlation we see in our data, and that \tm~is the parameter most strongly driving the diversity in our light curves. The lower panels show the effect of varying $P$ and $B$. The bottom left region (where normal SNe Ic reside) is difficult to reach with magnetar models, as very high $B$ is required. SLSN lie along the lower right edge of the magnetar distribution, where $P$ and $B$ are least extreme (for given \tm).}
\label{mag}
\end{figure*}

\subsection{Models: magnetar}

\begin{figure}
\includegraphics[width=9cm,angle=0]{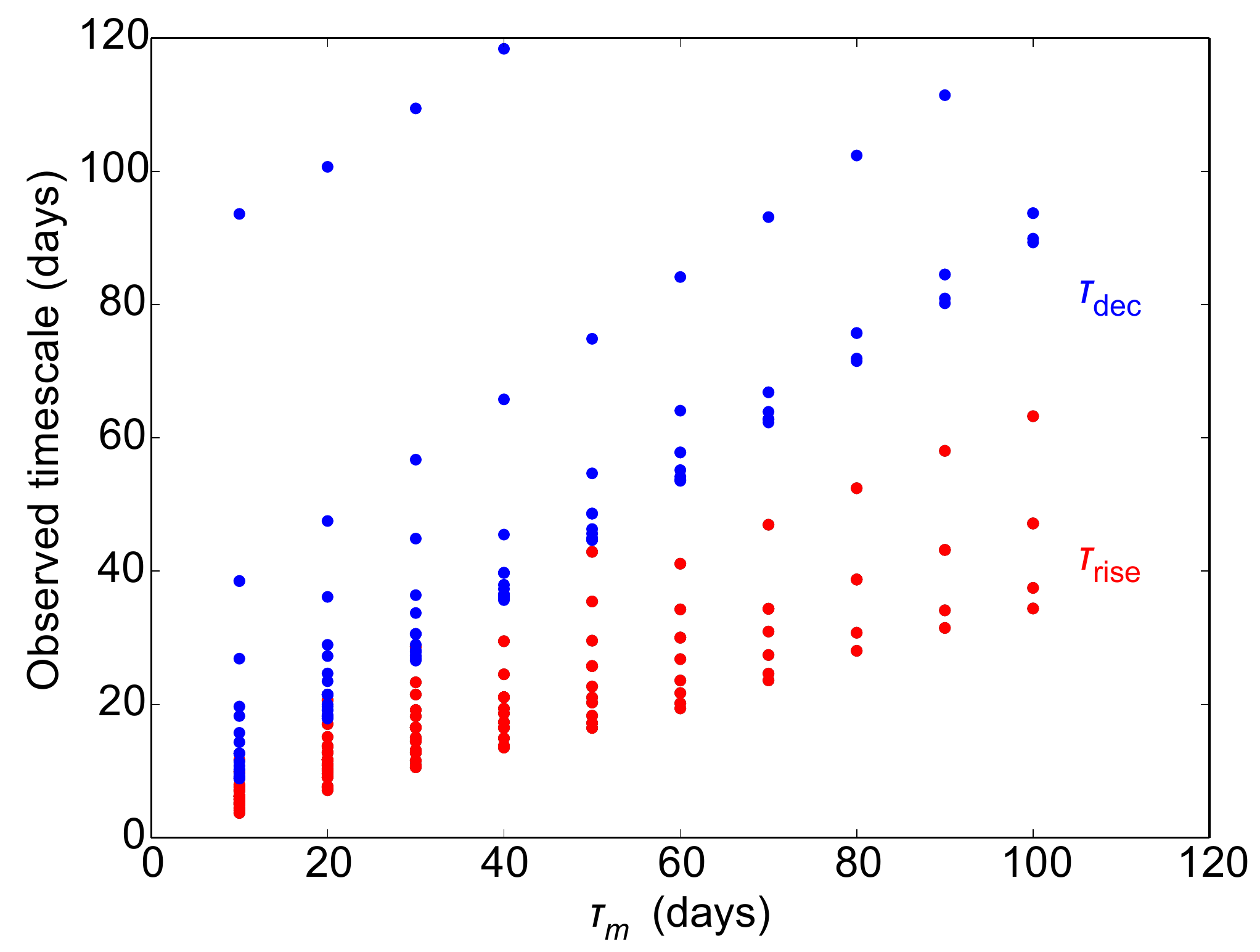}
\includegraphics[width=9cm,angle=0]{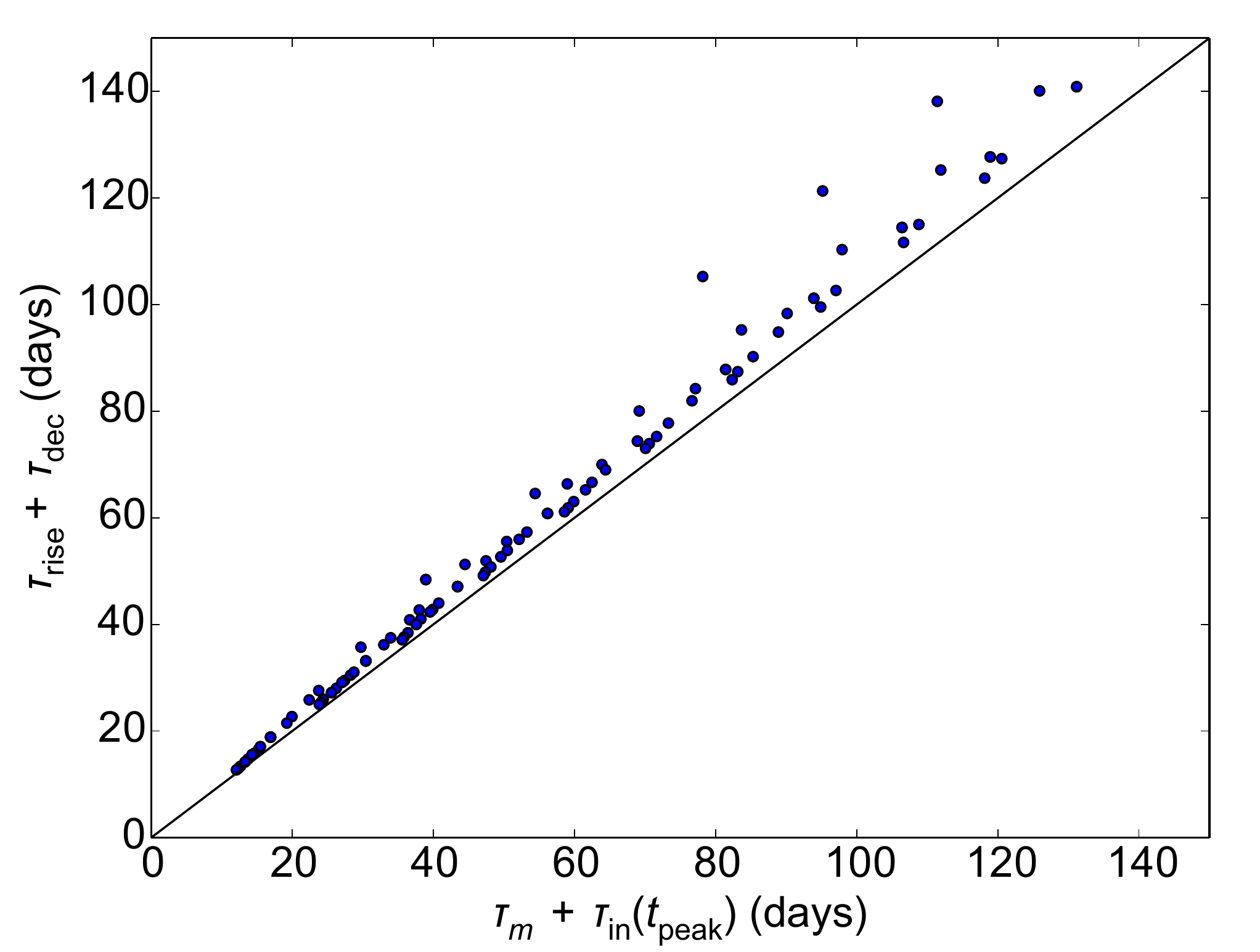}
\caption{Same as Fig.~\ref{exp_corr}, but for our magnetar model grid. The models shown have diffusion times of 10-100 d (steps of 10 d), $B=\,$\{1,3,5,7,9\}$\,\times10^{14}\,$G, and $P=\,$\{1,3,5,7,9\}$\,$ms. Only models with peak luminosity greater than $3\times10^{43}\,$erg$\,$s$^{-1}$ are plotted.}
\label{mag_corr}
\end{figure}

\begin{figure}
\includegraphics[width=9cm,angle=0]{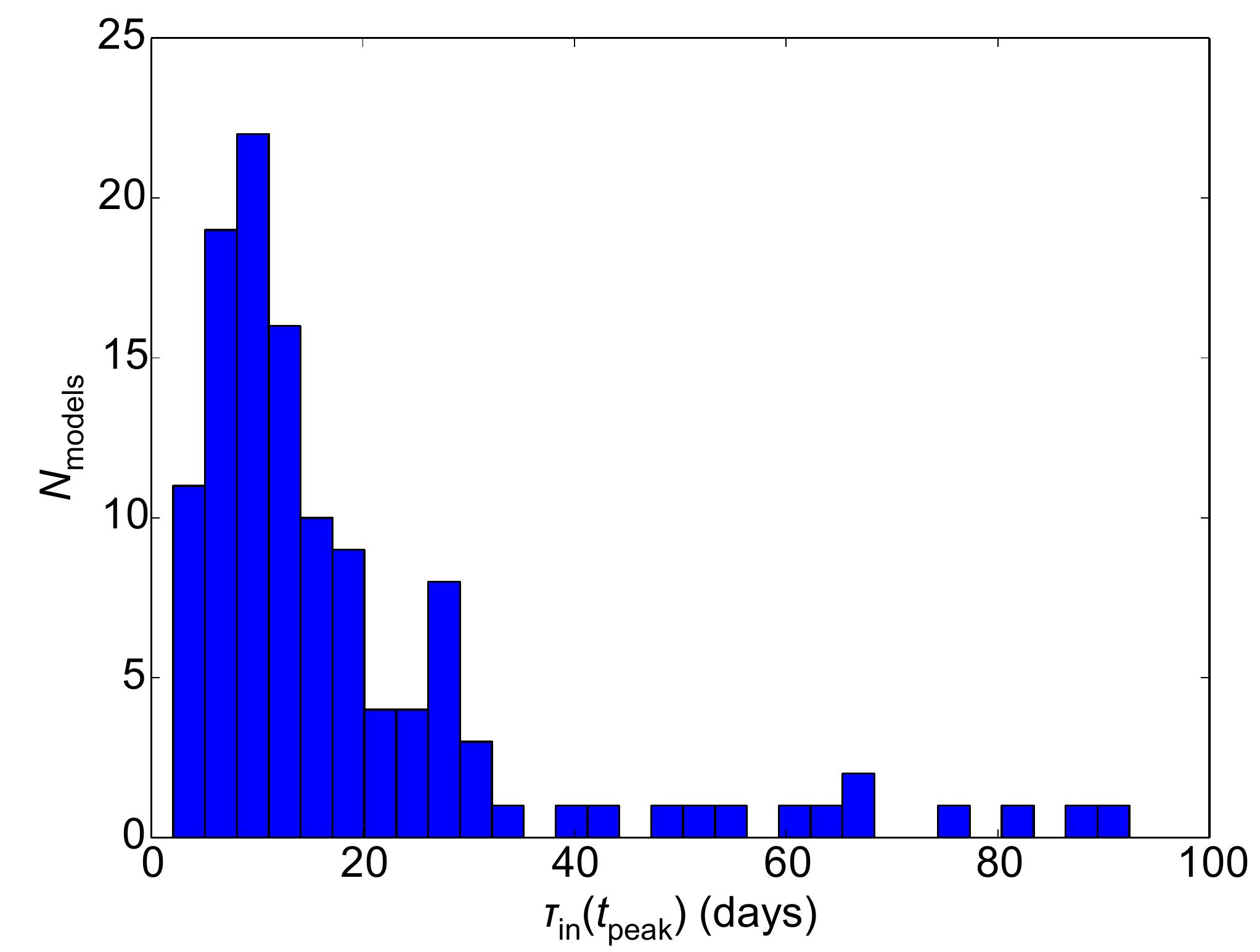}
\caption{The distribution of the power input timescales at maximum light, for the magnetar models in Fig.~\ref{mag_corr}. We caution that this is for a uniform distribution in $B$ and $P$, whereas we do not know what initial spins and magnetic fields magnetars are likely to form with. In particular, models with \ti$<10\,$d require high magnetic field and fast spin.}
\label{tau_in}
\end{figure}

In one of the most popular models, SLSNe are powered by a central engine which re-shocks the ejecta after it has expanded to large radius, thus overcoming adiabatic losses. In the magnetar spin-down model, the energy source is the rotational energy of a millisecond pulsar, which is tapped via a strong magnetic field. It is generally assumed to radiate with the functional form of a magnetic dipole \citep{kas2010}:
\begin{equation}
L_{\rm magnetar}(t) = \frac{E_{p}}{\tau_p} \frac{1}{\left(1 + t/\tau_p\right)^2}~\mbox{erg s}^{-1},
\end{equation}
where $E_p$ (the rotational energy, $I_{\rm ns}\Omega^{2}/2$) and $\tau_p$ are determined by the spin period, $P$, and magnetic field, $B$, of the magnetar. The energy input timescale for magnetar spin-down is given by $\tau_p = 4.75 B_{14}^{-2} P_{\rm ms}^2\,$d, where $B_{14}$ is the magnetic field in units of $10^{14}\,$Gauss and $P_{\rm ms}$ is the initial spin period in milliseconds. The shortest possible rotation period (corresponding to the largest energy reservoir) is $P\sim1\,$ms; any shorter and centrifugal forces would lead to breakup. Galactic magnetars have $B<10^{15}\,$G \citep[e.g.][Table 3]{dav2009}. This combination makes it difficult to achieve spin-down timescales $\tau_p \ll 0.1\,$d.

We ran a grid of magnetar models, uniformly varying \tm~(in steps of 10 days), $B$ and $P$ (in unit steps of $10^{14}\,$G and milliseconds, respectively), which we compare to the data in Fig.~\ref{mag}. However, it is not obvious that we can vary $B$ and $P$ independently. The high magnetic field is likely generated by a dynamo mechanism during core-collapse \citep{dunc1992}, as a primordial $B$-field in the progenitor core would couple its angular momentum to the envelope, braking the core and likely precluding formation of a millisecond pulsar at collapse. The simplest assumption for the dynamo mechanism is that a constant fraction of the rotation energy, $E_p \propto P^{-2}$, is converted to magnetic energy, $E_{\rm mag} \propto B^2$. In this case, we would have $B \propto 1/P$. Using this approach (with uniformly distributed $P$), rather than uniformly distributed $B$ and $P$, results in no significant effect on the distribution of rise and decline timescales in Fig.~\ref{mag}. Bearing in mind that we do not know the initial distribution of spin periods (and hence $B$-fields), we will continue our analysis assuming that $B$ and $P$ are independent parameters for simplicity.

Only models that are brighter than the faintest SLSN in our sample (PTF10hgi) are plotted. We produced many more models of course, but the only ones that are relevant are those producing luminosities of the same order as the SLSNe. The models shown display a correlation in rise and decline timescales similar to that observed in SLSNe, although a small number of models have very slow declines relative to the rise time. The colour map shows that increasing \tm~drives the models to longer rise and decline times, tracing out the observed correlation. We also investigate the effect of $B$ and $P$. The $B$-field influences the rise time, but has little effect on the decline. $B\sim10^{15}\,$G is needed to reach \tr$\, \la10\,$d. This could explain why no SLSNe are seen with such short rise times. The corollary also holds: if we were to observe SLSNe with rise times less than about 10 days, it would preclude the magnetar model would struggle to explain them. Models with long $P$ tend to rise quickly, but this is not a necessary condition, unlike the constraint on $B$.

Most of our SLSNe lie below the magnetar grid, but are consistent (within the errors) with lying along the locus of points on the sharp lower-right edge of the model distribution. This locus corresponds to weaker $B$, and $P\sim$2-5$\,$ms. If SLSNe are powered by millisecond magnetars, the less-extreme magnetic field may account for why the data preferentially lie at the lower edge of the model distribution. Overall, magnetar models satisfactorily reproduce the basic trend we see in the data. For a sensible range of parameters, diffusion time dominates the diversity in timescales, just as we saw previously for the simplest models with a central (exponential) power source.

Although most of the energy injection in this model takes place within a time $\tau_p$, the power input timescale is not a constant, unlike in exponential models. Using the definition \ti$\,= | L_{\rm in}(t)/(dL/dt) | $, we find that for magnetars, \ti$(t)=\frac{1}{2}(\tau_p + t)$, whereas for radioactive sources, \ti~is simply the lifetime of the nucleus. In setting the peak width, the important timescale (in addition to \tm) is \ti($t_{\rm peak}$)$\,=\frac{1}{2}(\tau_p + t_{\rm peak})$. Using this definition, we recover almost exactly the same correlation between the 4 timescales: \tr$\,+\,$\td$\,\approx\,$\tm$\,+\,$\ti($ t_{\rm peak})$. This is shown in Fig.~\ref{mag_corr}. Does this mean that we can assume \tm$\,\approx($\tr$\,+\,$\td)$/2$ in this case also? Fig.~\ref{tau_in} shows the distribution of \ti($t_{\rm peak}$) for our magnetar grid. The mean is 19 days (standard deviation: 18 d). Of all the models, 76\% have \ti~between 5 and 30 days at peak; however those with \ti($t_{\rm peak})<10\,$d need both fast rotation and strong magnetic field. This corresponds to the bottom left region in Fig.~\ref{mag}, where we do not have any observed SLSNe. Therefore the relevant models have timescales mostly in the range 10-30 d, with a tail extending to many tens of days. This is good news, as the assumption that \ti$\,\sim\,$\tm~is thus also reasonable for magnetar models. Hence we conclude that \tm$\,\approx($\tr$\,+\,$\td)$/2$, for a range of sensible models with central power sources.

\begin{figure*}
\includegraphics[width=18cm,angle=0]{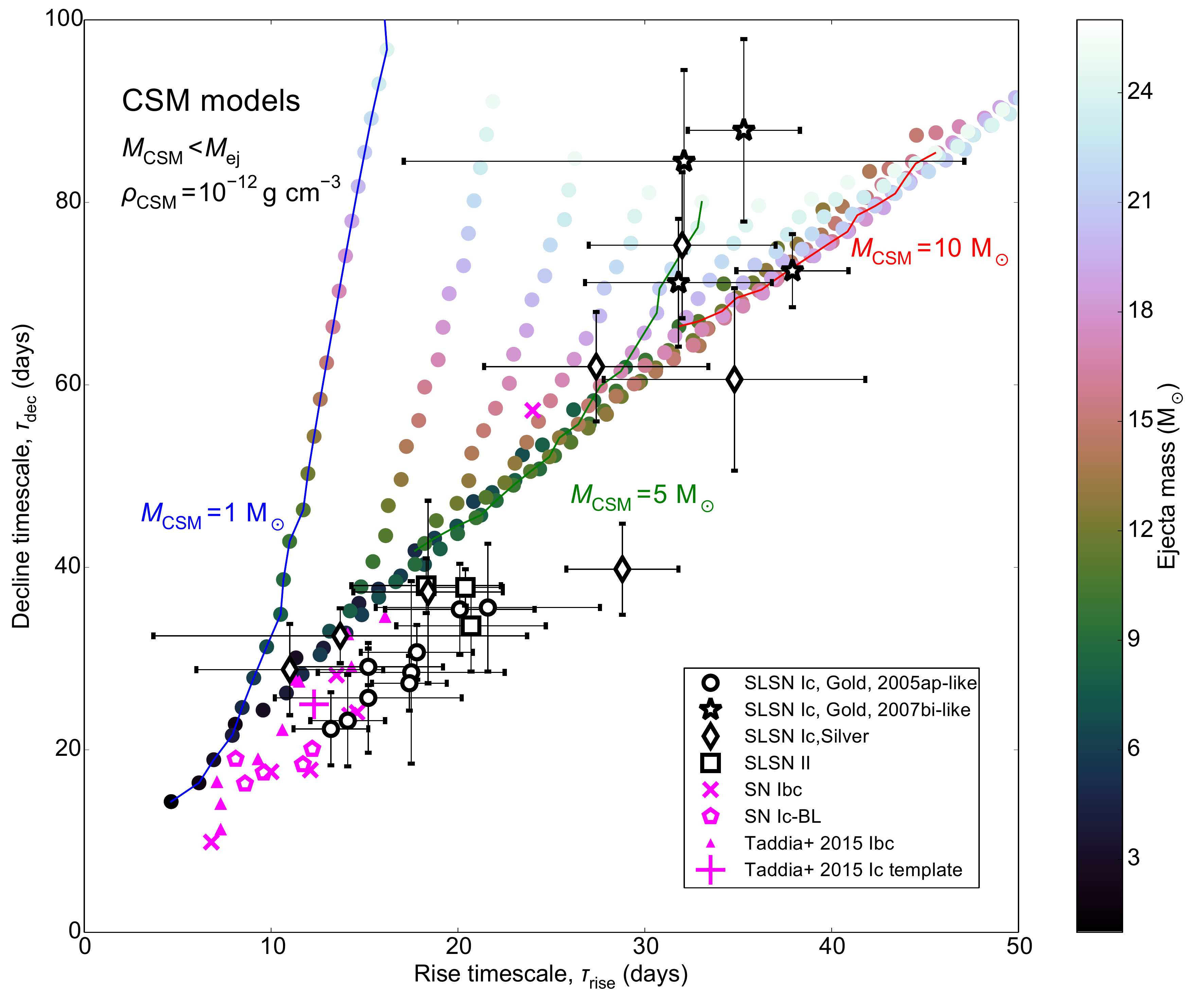}
\caption{Same as Fig.~\ref{mag}, but for interaction-based models. In this case, we vary the ejected mass and CSM mass, and CSM density. We find that the shape of the rise-decline distribution is highly sensitive to the CSM density -- the observed correlation is approximately recovered for $\rho_{\rm CSM} = 10^{-12}\,$g$\,$cm$^{-3}$. All of the light curve fits by \citet{nic2014} \citep[and many by][]{cha2013} found $\rho_{\rm CSM} \sim 10^{-12}\,$g$\,$cm$^{-3}$. Increasing the ejecta mass moves light curves along our correlation, while increasing the CSM mass primarily affects the rise time. It can be seen that the correlation is best reproduced by the subset of models with \Mcsm$\sim$\Mej/2 -- this was a general property of the light curve fits by \citet{nic2014}. Models with lower CSM mass rise too quickly for a given decline rate. If ejecta-CSM interaction does power all SLSNe, our rise-decline correlation puts tight constraints on the progenitor systems.}
\label{csm}
\end{figure*}

\begin{figure}
\includegraphics[width=8.5cm,angle=0]{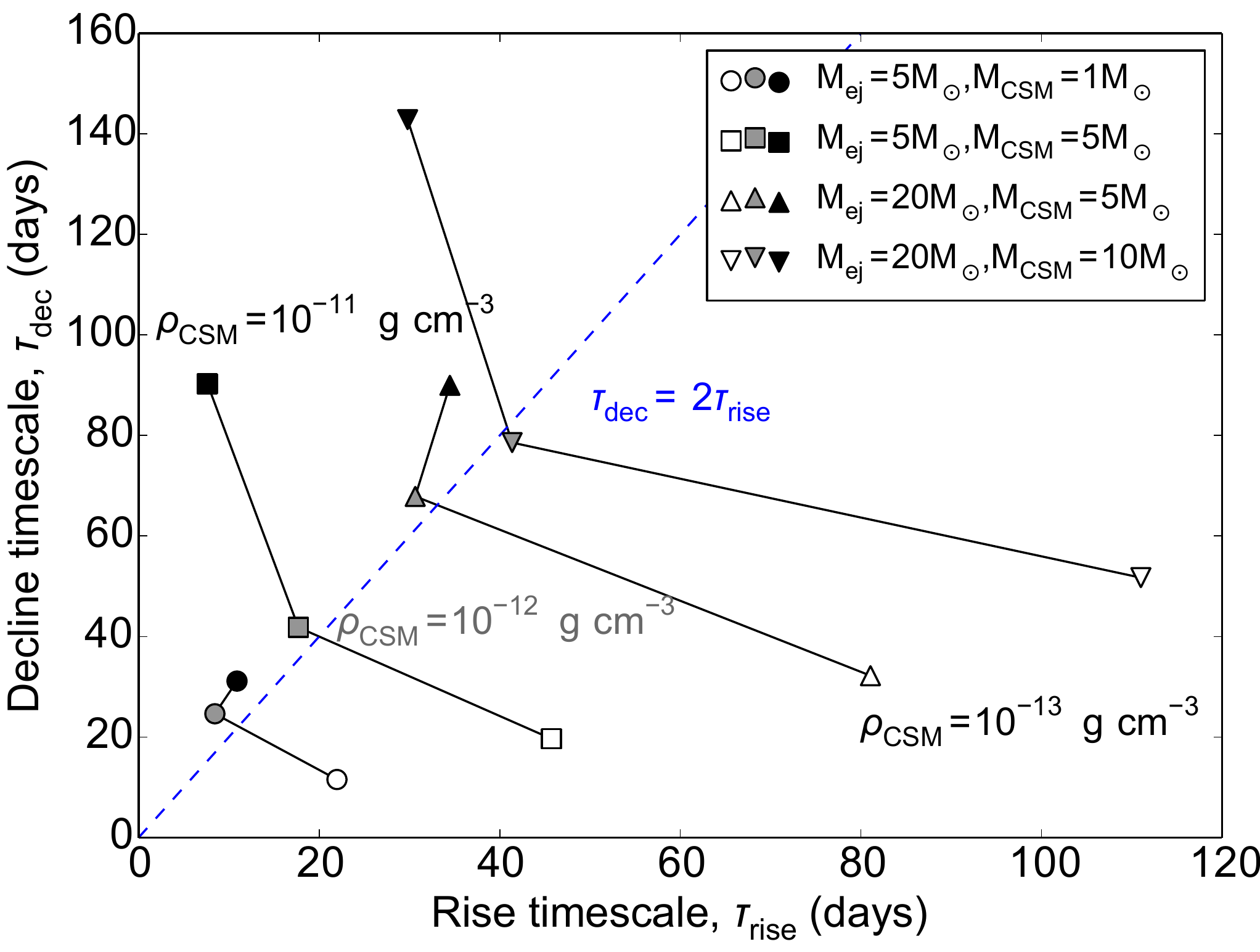}
\includegraphics[width=8.5cm,angle=0]{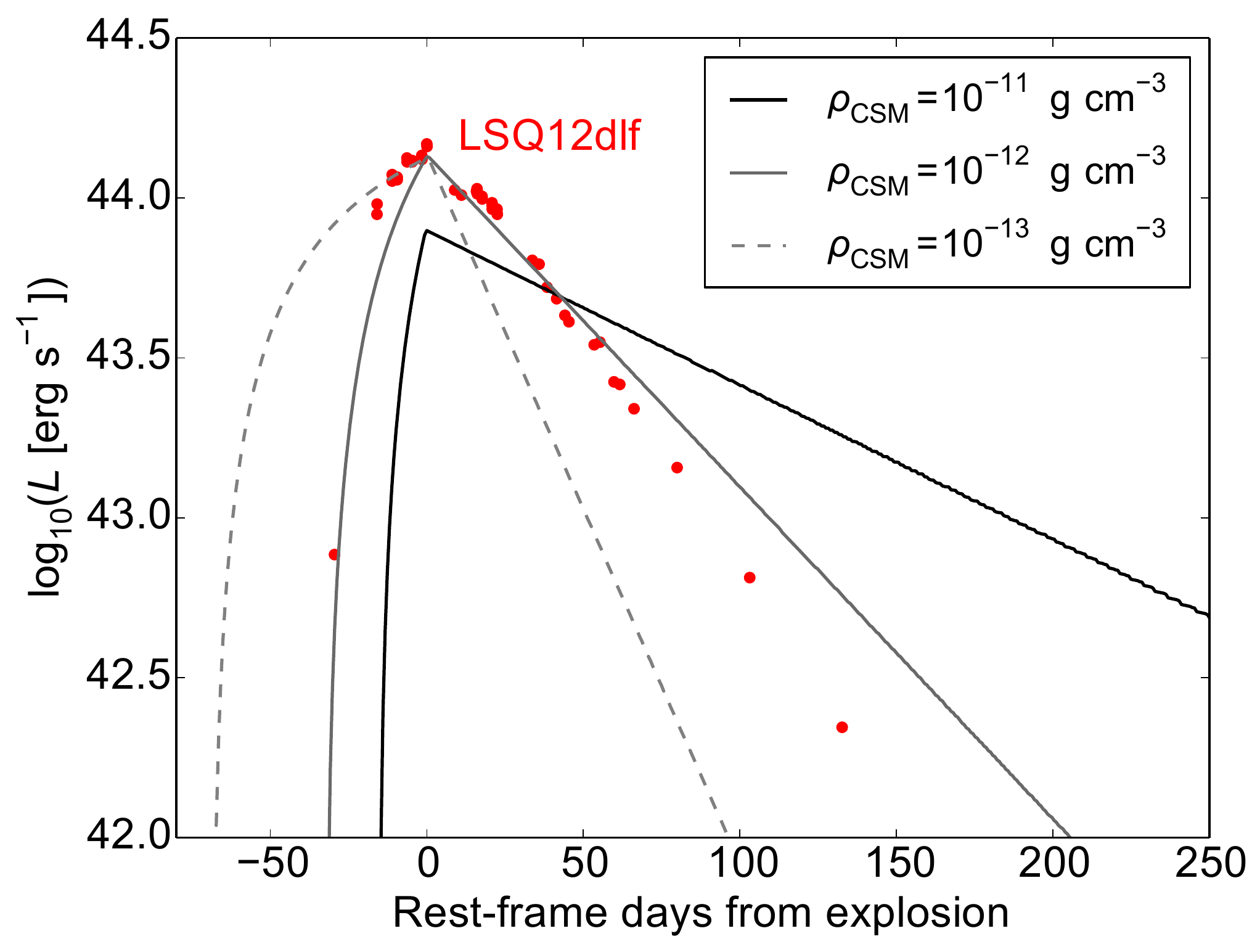}
\caption{Rise vs decline timescales for synthetic light curves powered by ejecta-CSM interaction. The relation is strongly dependent on the density of the CSM. Models with $\rho_{\rm CSM} \sim 10^{-12}\,$g$\,$cm$^{-3}$ (shown in Fig.~\ref{csm}) correspond best to our data. This can also be seen here in the lower panel, where we compare the SLSN LSQ12dlf to 3 models with \Mej$\,=\,$\Mcsm$\,=5\,$\M, and varying CSM density. The light curves peak when the forward shock from the ejecta-CSM collision breaks out of the CSM, and the subsequent decline is controlled mainly by the diffusion time in the CSM. Denser CSM results in faster shock propagation (shorter rise) and slower diffusion (longer decline), giving the inverse relationship between rise and decline times apparent in the top panel.}
\label{rho}
\end{figure}

\subsection{Models: CSM interaction}

Alternatively, we can attempt to fit the observed correlations with parameterised models in which the ejecta collide with a dense CSM. This has been another popular model invoked to explain SLSNe\citep{gin2012,che2011,mor2012}. As in \citet{nic2014} \citep[which followed the physical treatment of][]{cha2012}, we consider the limit of zero expansion velocity and luminosity input by strong external shocks. This model has many free parameters: ejected mass; CSM mass, radius and density (and density profile); explosion energy; \Ni~mass. We ran a grid of models with fixed explosion energy ($10^{51}\,$ erg; this mostly just affects the peak luminosity) and inner CSM radius ($10^{12}\,$cm; the light curve is quite insensitive to this parameter), and assume no \Ni. The CSM is taken to be a spherically symmetric shell of constant density -- its mass and density therefore determine its radial extent. We varied the ejected mass (\Mej) in unit steps from 1-25\,\M, and CSM mass (\Mcsm) in similar steps from 1-15\,\M, with the additional restriction \Mcsm$\,<\,$\Mej. The CSM density was initially set to $\rho_{\rm CSM} = 10^{-12}\,$g$\,$cm$^{-3}$.

The results are shown in Fig.~\ref{csm}. At a given \Mcsm, increasing \Mej~increases both \tr~and \td, in the same way as the diffusion time, \tm, did for our magnetar models. Illustrative lines of fixed \Mcsm~have been marked -- the observed rise-decline correlation is best reproduced by models with \Mcsm$\ga$\Mej/2. This was a common feature of models presented by \citet{nic2014}, and in fact is probably a requirement for SLSNe powered by interaction, as the CSM mass must be an appreciable fraction of the ejecta mass to thermalise the bulk of the expansion kinetic energy.

More restrictive, but perhaps more interesting, is the effect of CSM density on our light curves. The rise-decline relation is similar to our data for $\rho_{\rm CSM} = 10^{-12}\,$g$\,$cm$^{-3}$. In Fig.~\ref{rho}, we show the effect of varying $\rho_{\rm CSM}$ for a few representative models. Marked on the figure is the approximate slope of the observed correlation for SLSNe. Changing $\rho_{\rm CSM}$ to $10^{-11}\,$g$\,$cm$^{-3}$ or $10^{-13}\,$g$\,$cm$^{-3}$ moves our models to regions of the rise-decline plot far from where our data reside.

Why does this density have such a strong effect on the ratio of rise and decline times in our models? The dominant energy source is the forward shock from the interaction, which deposits heat as it propagates through the CSM. At some point, the shock breaks out of the CSM shell, and can no longer contribute energy (a similar effect occurs with the reverse shock in the ejecta, but the forward shock turns out to be dominant in most cases); at this point our light curve usually peaks. The time taken for the forward shock to propagate through the CSM decreases with increasing density \citep[][equation 15]{cha2012}. The model subsequently declines, as the stored energy diffuses out of the CSM, on the characteristic CSM diffusion time. This timescale $increases$ with the CSM density, so that models with earlier peaks fade more slowly. Thus, for all other parameters fixed, there is an inverse relationship between rise and decline timescales as we vary $\rho_{\rm CSM}$, as seen in Fig.~\ref{rho}.

Clearly, if SLSNe are powered by interaction with a dense CSM, our observed rise-decline relationship can place narrow constraints on the range of CSM densities present. Of the 6 SLSN light curves fit with CSM models by \citet{nic2014}, 5 had densities in the range $-12.54<\log_{10}\rho_{\rm CSM}<-11.74$ (no convincing fit was found for the final object, SN 2011ke). It seems contrived that virtually all H-poor SLSNe would have such similar circumstellar densities, particularly when modelling indicates that a range of densities can generate the observed peak magnitudes \citep[e.g.][who also fit H-rich events]{cha2013}. Three possibilities exist. The first and most obvious is that ejecta-CSM interaction is not the power source in SLSNe Ic. 
Alternatively, our simple model may not be a good description of interacting SLSNe (for example, the shape of the CSM density profile may be important, and not a uniform shell). One very important weakness in this analysis is that the interaction models of \citet{cha2012} and \citet{nic2014} treat the shocks following \citet{che1994}, whose derivation was for \Mcsm$\,\ll\,$\Mej, and it is unclear how the picture changes for massive CSM.  Finally, some process in the evolution of SLSN progenitors might somehow be capable of consistently producing circumstellar environments within this density range. The homogeneity in the spectral properties of SLSNe would then result from the similar physical conditions in the CSM. The last of these possibilities is intriguing, and determining this process could prove an important clue to understanding what kinds of stars produce SLSNe. However, observations of SNe known to be powered by CSM interaction (SNe IIn) show huge diversity and variation in their observed characteristics and inferred physical configurations. In any case, CSM models will have to be able explain our observed correlation, if we are to continue to accept them as valid model for SLSNe.

\section{Generalised Light Curves and Peak Luminosity}\label{shape}

\begin{figure}
\includegraphics[width=8.9cm,angle=0]{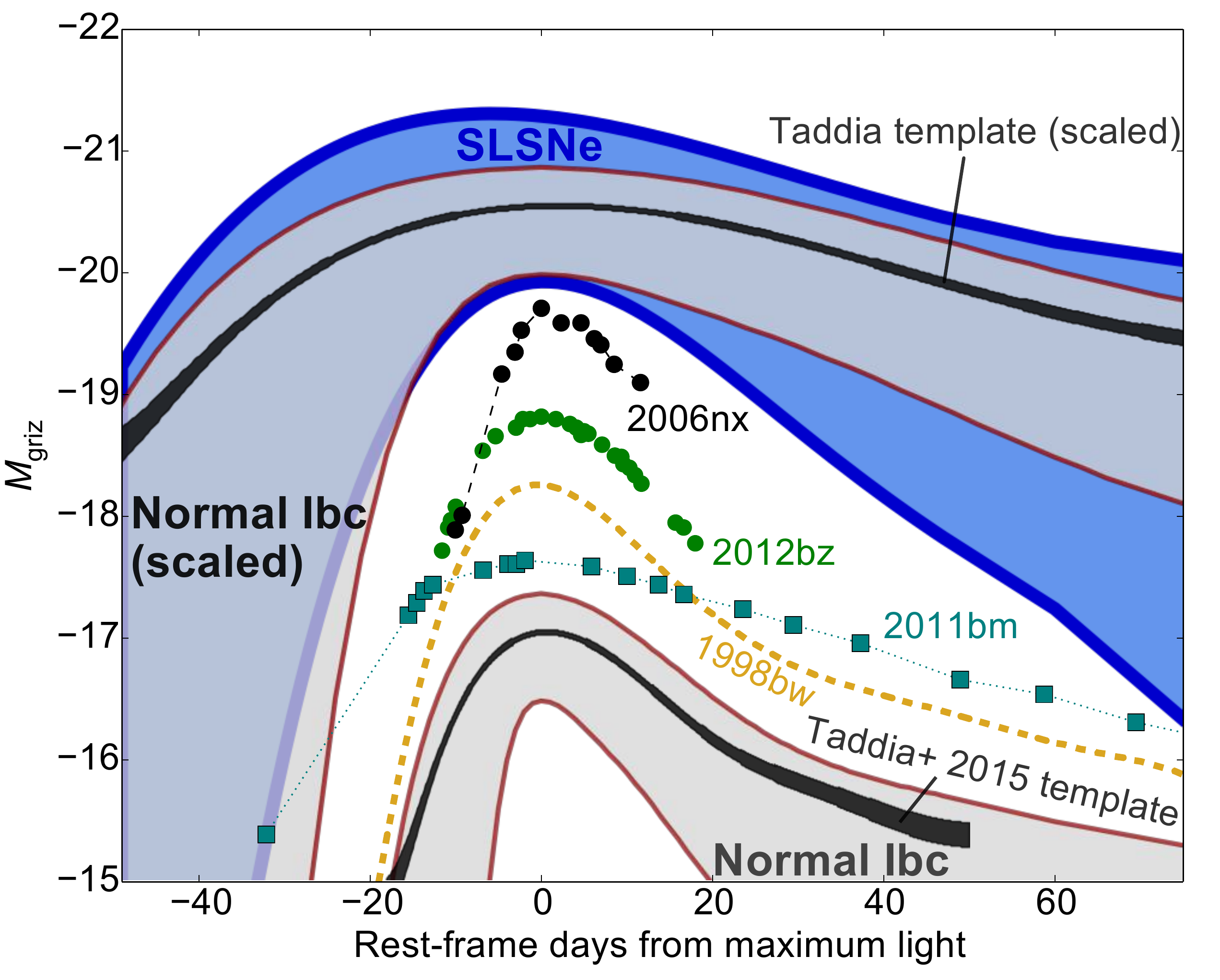}
\caption{
{ Generalised light curves for SLSNe compared with lower luminosity SNe Ic (normal and broad-lined). The blue area represents normal SNe Ic in our well-observed literature sample. and those of \citet{tad2014}. SN Ic light curves can be mapped onto the SLSNe by a 3.5 magnitude increase in brightness and a stretch along the time axis by a factor of 3. Some broad-lined SNe Ic lie in the magnitude gap, but  tend to have narrower light curves than SLSNe. SN 2011bm \citep{val2012} shows a normal Ic spectral evolution but has a light curve width comparable to SLSNe.}
}
\label{gen}
\end{figure}

The correlation in rise and decline timescales, presented in the previous section, was found to be the same for SLSNe Ic and for normal-luminosity, hydrogen-poor core-collapse SNe. This suggested that the two populations have the same basic light curve shape (and we inferred that the slower evolution of SLSNe was due to longer diffusion timescales). To further investigate the relationship between SLSN and normal SN Ic light curves, we construct a generalised light curve for each type of SN. We do this simply by taking the area in magnitude-time parameter space that contains all of the light curves in each sample (SLSNe of Type II are excluded from this analysis). This is shown in Fig.~\ref{gen}, where we also include the SN Ibc light curve template from \citet{tad2014}, and a number of SNe Ibc with unusually high luminosity. As expected, the SLSN light curves are much brighter and broader than typical SNe Ibc. In the unscaled SN Ibc light curve, the decline rate starts to change at around 30 days after peak. This is because the ejecta are becoming optically thin, and the luminosity begins to track the decay of \Co. We do not typically see his behaviour in SLSNe, for two reasons. 
For most SLSNe, the decline after peak is too fast to be compatible with realistic \Ni-driven models with such a high peak luminosity, as discussed in Sect.\,\ref{models:ni}. If they do contain some small amounts of \Co~(comparable to that in SNe Ibc), this is masked by the bright luminosity source that powers the peak. The more slowly declining SLSNe (SN 2007bi-like), on the other hand, do match the \Co~decay rate. If they were radioactively powered, the long rise times and broad peaks would allow them to join smoothly onto the \Co~decay tail after maximum.

We find that a very simple transformation maps the SN Ic light curves onto those of the SLSNe: an increase along the $y$-axis by 3.5 magnitudes (a multiplicative factor of $\approx 25$ in luminosity), and a broadening in time by a factor 3.  The most obvious interpretation of this correspondence is that the two sets of light curves are determined by the same underlying physics: the rapid expansion of shocked gas with small initial radius, heated by some internal power supply. The broader light curves of the SLSNe are indicative of a longer diffusion timescale (higher mass and/or lower velocity) compared to the SNe Ic. The higher peak luminosity tells us that some additional energy source is heating the ejecta, compared to $\sim0.1-1\,$\M~of \Ni~in SNe Ic (this could be, for example, a millisecond magnetar). Indeed, this is the established theory explaining the diversity within the SN Ic class, where higher ejecta mass and \Ni~mass result in broader and brighter light curves, respectively.

If SLSNe Ic were powered by interaction of SN Ic ejecta with dense (and H-poor) circumstellar material, the light curve physics would be quite different: a combination of forward and reverse shocks in the ejecta and CSM, with a strong dependence on the various density profiles. Indeed, for the massive, optically thick CSM needed to generate super-luminous peak magnitudes, we should not actually see direct emission from the supernova ejecta until well after maximum light \citep{ben2014}. It would therefore be surprising to recover such a trivial transformation between normal and super-luminous Ic light curves. Circumstellar interaction can generate a range of light curve shapes (as shown in Fig.~\ref{rho}). For instance, the most conclusive example of a Type Ic SN interacting with H-deficient CSM is SN 2010mb \citep{ami2014}, which had an extremely unusual light curve shape with a plateau lasting for hundreds of days.

One interesting question is whether there is a continuum of peak luminosities between normal and super-luminous Type Ic SNe. Since the discovery and characterisation of SLSNe, an apparent gap has been recognised. 
\cite{rich2014} have compiled large samples of SNe Ibc to determine absolute magnitude distributions in the standard Johnson $B$-band. Their study attempted to correct for bias and derive volume limited absolute magnitude distributions, however the targets do not all have enough data to determine bolometric luminosities ($M_{griz}$ at maximum). There are some typing inaccuracies in the Richardson et al. sample (e.g. 2006oz as  a Ib, and 2005ap as a type II), however the normal and broad-lined Ibc population does appear to have an upper limit to their peak brightness of around $-18$ to $-19$, with the SLSNe sitting 3 magnitudes brighter.

\begin{figure}
\includegraphics[width=8.5cm,angle=0]{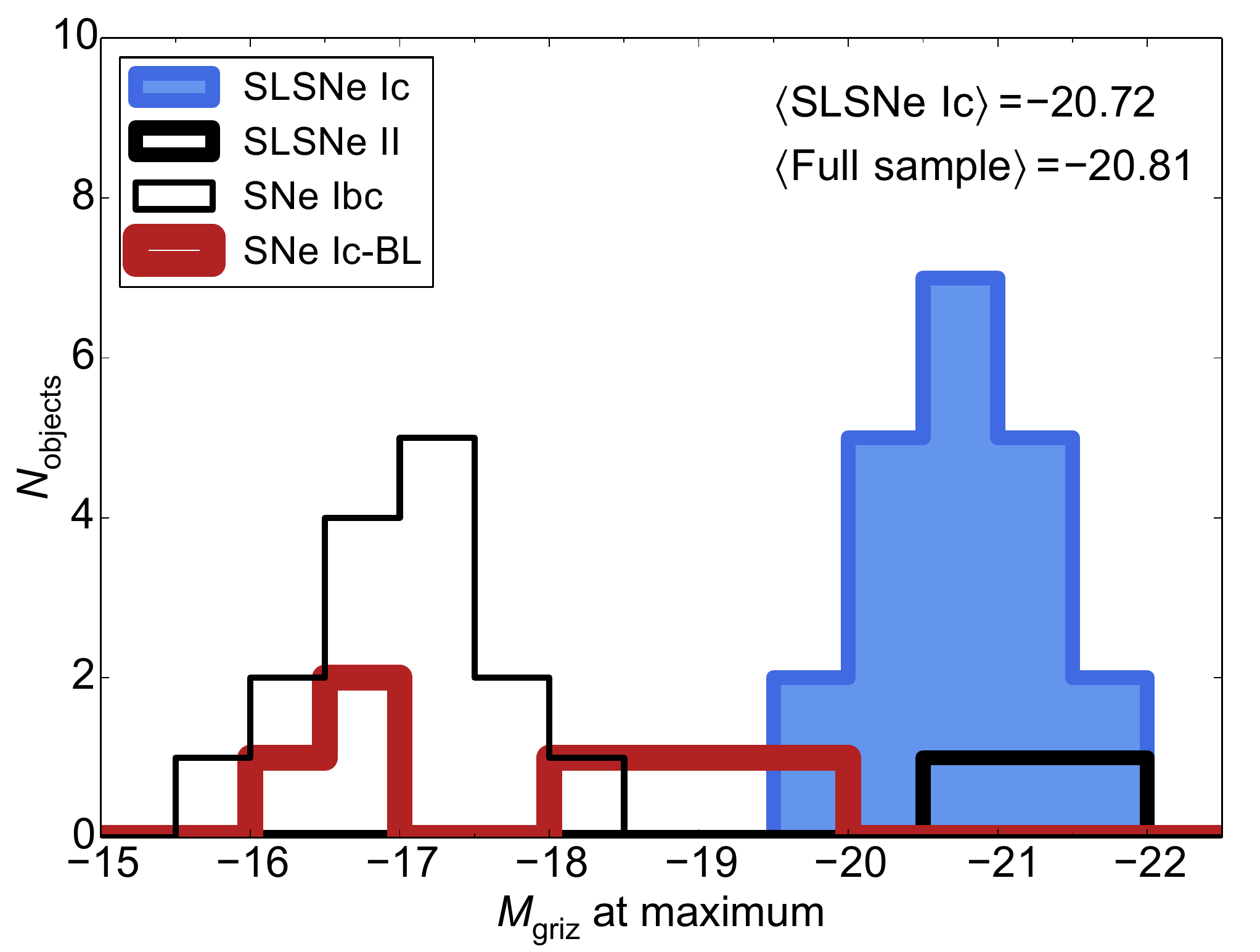}
\caption{The distribution of magnitudes at maximum light. The scatter is quite low, with a standard deviation of only 0.50 magnitudes (0.58 if we include the two SLSNe II). The mean peak magnitude for normal SNe Ic is $-17.03$. A few broad-lined events have luminosity comparable to the fainter SLSNe. The data are binned in 0.5\,mag intervals.}
\label{peak}
\end{figure}

We compare the brightness distributions of our SLSNe and other SNe in Fig.~\ref{peak}. The SLSNe Ic peak at \Mgriz$\,=-20.72 \pm 0.59$ magnitudes, while the normal SNe Ibc tend to peak at \Mgriz$\,=-17.03 \pm 0.58$. 
We note here that \citet{ins2014} derived peak absolute magnitudes for a sample of 16 SLSNe (with a large overlap with our sample) in a synthetic bandpass centred on $400\,$nm. They found $\langle M_{400}\rangle=-21.86\pm0.35$. This seems to be more uniform than \Mgriz, though it is worth bearing in mind that in some cases our derived \Mgriz~has been estimated using assumed colours, introducing additional uncertainty, whereas \citet{ins2014} had observations fully covering their synthetic bandpass for all of their objects at maximum light.

Selection effects and bias in our sample mean we cannot definitively say whether there is an excess of hydrogen-poor supernovae with peak absolute magnitudes brighter than \Mgriz$\,\sim-20$ (i.e. SLSNe), or whether such events are the bright tail of a continuous magnitude distribution. The plot in Fig.~\ref{peak} is not meant be 
be a representative luminosity function, as one would require accurate relative numbers in either a 
magnitude or volume limited survey. We know definitively that SLNSe are rare and their relative rate with respect
to normal SNe Ibc is of order 1 SLSNe per 3000-10000 SNe Ibc \citep{qui2013,mcc2015}. 
The SLSNe are obviously over-represented in our sample, relative to their rates of occurrence in nature, as we have compiled similar numbers of SLSNe and comparison SNe Ibc. To prove statistically that SLSNe comprise a separate population of events from the brightest `normal' SNe Ic would require careful consideration
of the selection factors in surveys such as PTF, PS1 and LSQ+PESSTO to determine unbiased relative numbers of SLSNe, SNe Ibc, and SNe Ic-BL and construct meaningful luminosity functions \citep[e.g. as done for normal SNe in][]{li2011}

Nevertheless we can make some comments on the luminosity differences observed if we assume that  
the SNe Ibc we have compiled are fairly representative of the general population of such stripped envelope SNe.
None of our spectroscopically normal SNe Ibc are observed to peak at \Mgriz$\,\approx-19$. However, some broad-lined SNe Ic do have peak magnitudes spanning the gap between normal and super-luminous SNe. Broad-lined SNe Ic are often high \Ni-producers, and some have been shown to be associated with observed gamma-ray bursts (GRBs) 
\citep{woo2006}. The large \Ni~mass makes them brighter than typical SNe Ic. Fig.~\ref{gen} shows two GRB-SNe, SN 1998bw \citep{pat2001} and SN 2012bz \citep{sch2014}.
 The brightest Ic-BL in the sample of \citet{tad2014} is also shown. The object, SN 2006nx, was discovered at redshift $z=0.137$, but if there was an associated GRB, it was not seen. SN 2006nx actually has a similar peak magnitude to SLSNe, suggesting that it may in fact be a member of that class. However, its light curve is quite narrow not only compared to SLSNe, but also for such a luminous \Ni-powered SN. Its unknown spectral evolution precludes a robust answer to the question of whether it is physically related to the SLSN population.

It is interesting that some SNe Ic-BL/GRB-SNe do  lie in the gap between SNe Ic and SLSNe, as some authors have suggested that SLSNe and GRBs are in fact related. \citet{lun2014} argued that the two types of explosion occur in similar low-metallicity environments. \citet{lel2015} also claimed that their host galaxies are similar, but also that those of SLSNe are more intensely star-forming than those of GRBs, perhaps implying more massive progenitors for SLSNe. Magnetar/engine-powered models have been invoked to explain both the high luminosities in SLSNe and the relativistic jets in GRB-SNe\citep{tho2004,met2011}. However, the required magnetar parameters in GRB models are more extreme than for SLSNe, with spin periods $\sim$1\,ms and magnetic fields 10-100 times stronger, in order to drive a jet that punches through the stellar envelope \citep{buc2009}.
\citet{gal2012} notes that some broad-lined SNe Ic, such as SN 2007D \citep{dro2011} and SN 2010ay \citep{san2012}, reached peak magnitudes close to those of SLSNe, and likely required an additional energy source on top of the inferred $\sim1\,$\M~of \Ni. Perhaps bright SNe Ic-BL represent events where the central engine enhances the luminosity as well as the kinetic energy of the explosion, but not to the same degree as in SLSNe, where the additional power source overwhelms the input from \Ni~decay. In this framework, there could thus be a continuum in luminosity between the various subclasses of SNe Ic, depending on the properties of the engine.
 
If, however, there is a significant gap in peak brightness, with very few objects intermediate between normal and super-luminous SNe Ic, it may be difficult to explain for circumstellar interaction models. For example, Type II supernovae show a broad but continuous distribution in peak magnitudes: from SNe II-P through Type II-L's to bright Type IIn and SLSNe IIn. Some authors have argued that this hierarchy is driven by varying degrees of circumstellar interaction \citep{che1994,ric2002,gal2012}. A gap between SNe/SLSNe Ic could be more indicative of a threshold process -- for example, only progenitors above some critical mass can form magnetars, or undergo a particular instability \citep[e.g.~pulsational pair instability;][]{woo2007}. In the magnetar model, there may also exist an observational `desert' between cases where the spin-down power is sufficient to drive a stable jet (GRB), and cases where the spin-down is weaker and slower, forming a wind nebula and enhancing the late-time luminosity instead (SLSN). If the magnetar spins down very quickly but the jet does not break the stellar envelope, there may be neither a GRB nor an enhancement in optical brightness (B.~Metzger, private communication.) In summary, peak magnitude distributions of SNe from large, homogeneous samples are needed, and will provide an important constraint on the possible relationships between SLSNe and SNe Ic.

\begin{table}
\caption{Measured properties and derived masses}
\label{props}

\begin{tabulary}{\columnwidth}{LLLLL}
\hline
SN & \tr (d)$^a$ & \td (d)$^b$ & $v_{5169}$ (\kms)$^c$ & \Mej/\M \\

\hline
\multicolumn{5}{l}{Super-luminous SNe}\\
\hline

SN2007bi		&	32.1	&	84.5	&	11900		&	31.1$^{+34.3}_{-21.7}$		\\
SN2008es$^{*}$	&	18.3	&	38.0	&	--			&	3.7$^{+3.0}_{-2.1}$	\\
SN2010gx		&	15.2	&	29.1	&	10900		&	4.4$^{+3.2}_{-2.3}$	\\
SN2011ke		&	15.2	&	25.7	&	10200		&	3.3$^{+1.9}_{-1.5}$	\\
SN2011kf		&	17.5	&	28.5	&	9000		&	3.7$^{+2.0}_{-1.5}$	\\
SN2012il		&	14.1	&	23.2	&	9100		&	2.4$^{+1.3}_{-1.0}$	\\
SN2013dg		&	17.8	&	30.7	&	14000		&	6.3$^{+3.8}_{-2.9}$	\\
SN2013hx		&	20.7	&	33.6	&	6000		&	3.4$^{+1.8}_{-1.4}$	\\
LSQ12dlf		&	20.1	&	35.4	&	13700		&	8.1$^{+5.1}_{-3.9}$	\\
LSQ14mo		&	17.4	&	27.3	&	10200	&	3.9$^{+1.9}_{-1.5}$	\\
LSQ14bdq$^\dagger$	&	31.8	&	71.2	&	--	&	20.4$^{+18.6}_{-12.6}$	\\
PTF10hgi		&	21.6	&	35.6	&	4800		&	3.0$^{+1.7}_{-1.3}$	\\
PTF11rks		&	13.2	&	22.3	&	18100		&	4.4$^{+2.5}_{-2.0}$	\\
PTF12dam	&	37.9	&	72.5	&	10500		&	27.0$^{+19.6}_{-14.3}$		\\
CSS121015	&	20.4	&	37.8	&	10000$^{\ddagger}$		&	6.5$^{+4.5}_{-3.3}$	\\
SSS120810	&	--	&	30.2$^{\dagger \dagger}$	&	11200		&	5.7$^{+3.6}_{-2.1}$	\\
PS1-11ap$^\dagger$	&	35.3		&	87.9	&	--	&	29.2$^{+30.3}_{-19.6}$	\\
SN2005ap$^\dagger$		&	$>$11	&	28.8	&	--	&	3.0$^{+3.3}_{-2.1}$	\\
SCP06F6$^\dagger$	&	28.8		&	39.8	&	--	&	9.1$^{+3.1}_{-2.7}$	\\
PTF09cnd$^\dagger$	&	32.0		&	75.3	&	--	&	22.2$^{+21.5}_{-14.3}$	\\
PTF09cwl$^\dagger$	&	34.8		&	60.6	&	--	&	17.5$^{+10.8}_{-8.2}$	\\
PS1-10ky$^\dagger$	&	13.7		&	32.5	&	--	&	4.1$^{+4.0}_{-2.7}$	\\
PS1-10bzj	&	18.4	&	37.3	&	13000	&	7.8$^{+6.2}_{-4.4}$	\\
iPTF13ajg	&	27.4	&	62.0	&	9100	&	14.0$^{+12.9}_{-8.7}$	\\

\hline
\multicolumn{5}{l}{Other H-poor SNe}\\
\hline
SN1994I		&	6.8	&	9.9	&	10100		&	0.5$^{+0.2}_{-0.2}$	\\
SN1998bw		&	12.2	&	20.1	&	26600		&	5.3$^{+2.9}_{-2.3}$	\\
SN1999ex		&	12.1	&	17.8	&	9300		&	1.6$^{+0.7}_{-0.6}$	\\
SN2002ap		&	8.1	&	19.0	&	20400		&	2.9$^{+2.8}_{-1.9}$	\\
SN2003jd			&	9.6	&	17.5	&	16400		&	2.3$^{+1.5}_{-1.2}$	\\
SN2004aw		&	13.5	&	28.2	&	12100		&	4.1$^{+3.4}_{-2.4}$	\\
SN2007gr		&	10.0	&	17.6	&	8400		&	1.2$^{+0.8}_{-0.6}$	\\
SN2008D		&	14.6	&	24.1	&	8700		&	2.5$^{+1.4}_{-1.1}$	\\
SN2009jf		&	14.2	&	23.8	&	10100		&	2.8$^{+1.6}_{-1.2}$	\\
SN2010bh	&	8.6	&	16.3	&	35000	&	4.2$^{+3.0}_{-2.2}$ \\
SN2011bm$^\dagger$	&	24.0	&	57.2	&	--		&	12.7$^{+12.5}_{-8.3}$		\\
SN2012bz	&	11.7	&	18.4	&	23000	&	4.0$^{+2.0}_{-1.6}$ \\
\hline
SN2005hl	&	14.3	&	29.1	&	5450$^{\ddagger}$		&	2.0$^{+1.6}_{-1.1}$\\
SN2005hm	&	11.5	&	27.5	&	9470$^{\ddagger}$		&	2.8$^{+2.7}_{-1.8}$\\
SN2006fe	&	11.3	&	27.7	&	5000$^{\ddagger}$		&	1.5$^{+1.5}_{-1.0}$\\
SN2006fo	&	14.1	&	32.7	&	10500$^{\ddagger}$	&	4.4$^{+4.2}_{-2.8}$\\
14475	&	7.3	&	14.1	&	18700$^{\ddagger}$	&	1.6$^{+1.2}_{-0.9}$\\
SN2006jo	&	7.3	&	11.3	&	14400$^{\ddagger}$	&	1.0$^{+0.5}_{-0.4}$\\
SN2006lc	&	9.3	&	19.0	&	9100$^{\ddagger}$		&	1.4$^{+1.1}_{-0.8}$\\
SN2006nx	&	7.1	&	16.5	&	15400$^{\ddagger}$	&	1.7$^{+1.6}_{-1.1}$\\
SN2007ms	&	16.1	&	34.6	&	11400$^{\ddagger}$	&	5.6$^{+4.9}_{-3.4}$\\
SN2007nc	&	10.6	&	22.2	&	12700$^{\ddagger}$	&	2.6$^{+2.2}_{-1.5}$\\
\hline

\end{tabulary}

Masses derived using equation \ref{mass_eq} with $\kappa=0.1\,$cm$^2\,$g$^{-1}$ and \tm$\,=\,($\tr$\,+\,$\td$)/2$. Error bars correspond to estimates with \tm$\,=\,$\tr~(lower) and \tm$\,=\,$\td~(upper). 
$^a$Characteristic time for SN to rise to maximum light, defined using \tr$\,\equiv t(L_{\rm peak}/e)$ for $t<t_{\rm peak}$; $^b$characteristic time to fade, \td$\,\equiv t(L_{\rm peak}/e)$ for $t > t_{\rm peak}$; $^c$Velocity from minimum of Fe II $\lambda$5169 absorption, 20-30 d after maximum light;
$^{*}$Assumes $v\sim6000\,$\kms, based on SN 2013hx spectrum; $^{\ddagger}$Velocity from the literature, not necessarily Fe II; $^\dagger$Assumes $v=10000\,$\kms; 
$^{\dagger \dagger}$For mass estimate, we take \tr$\,=\,$\td$/1.6$ (see Fig.~\ref{corr_plot})

\end{table}

\section{Two types of SLSN Ic?}\label{pop}

\begin{figure}
\includegraphics[width=8.5cm,angle=0]{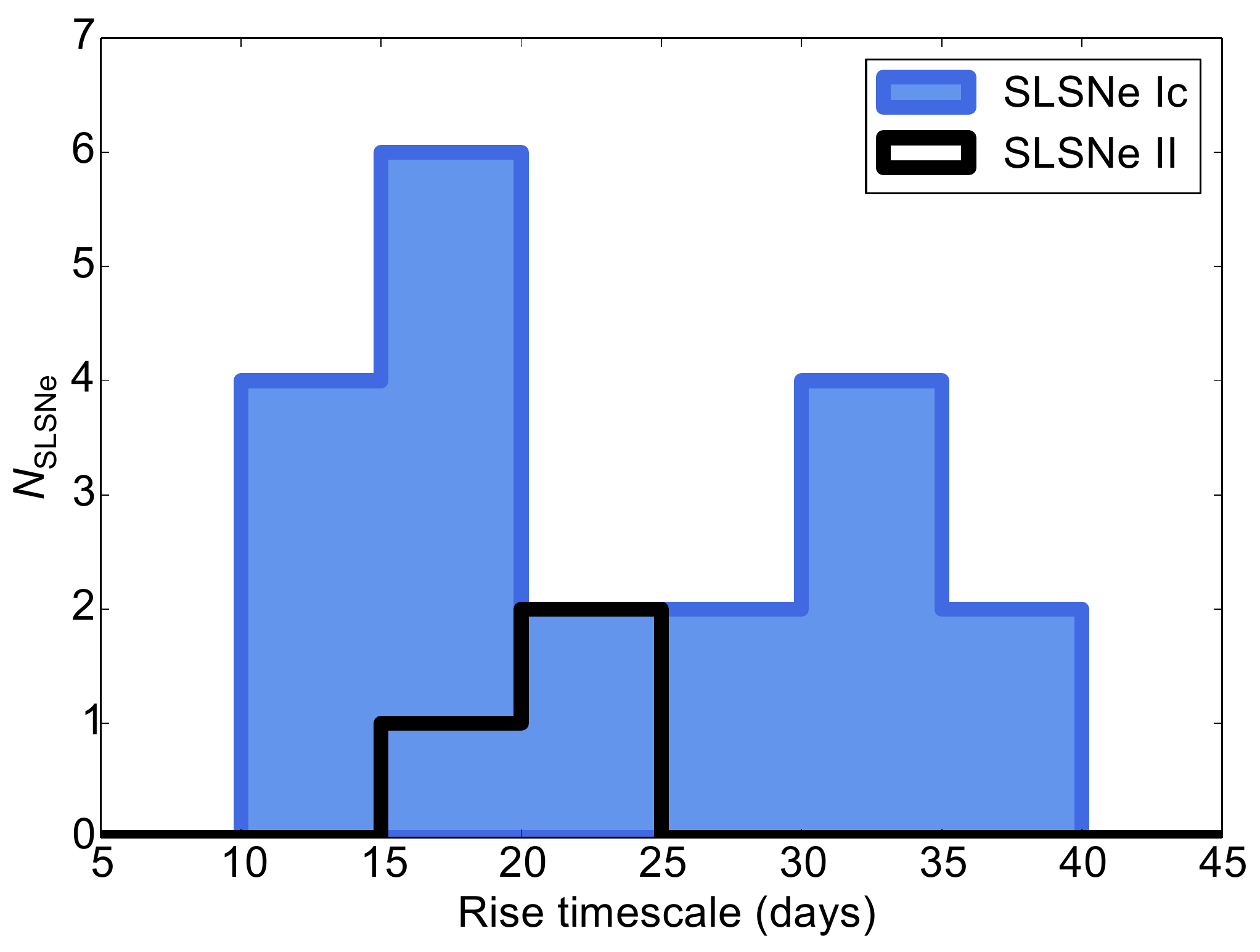}
\includegraphics[width=8.5cm,angle=0]{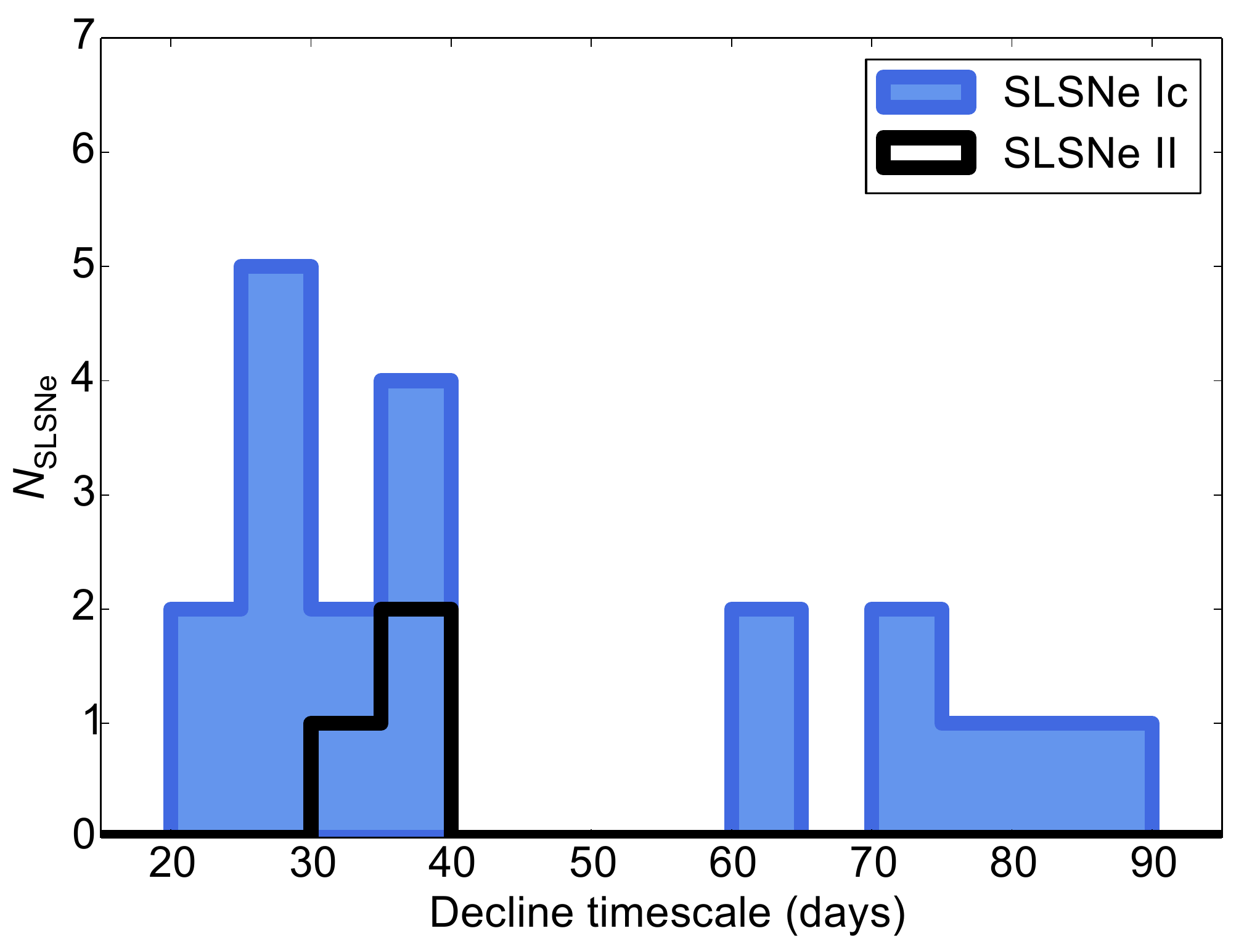}
\includegraphics[width=8.5cm,angle=0]{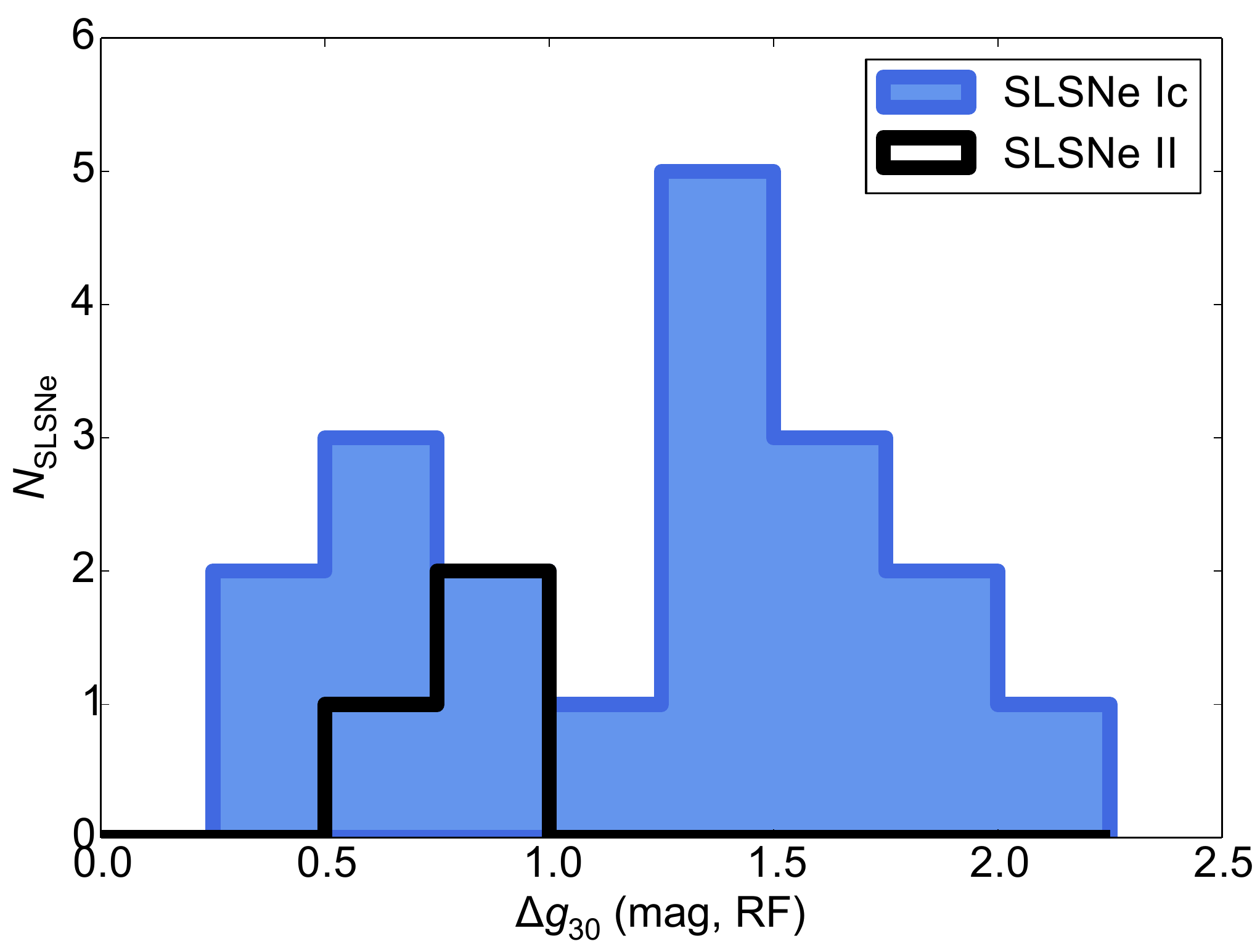}
\caption{Histograms showing the rise (\textit{top}) and decline (\textit{middle}) timescales (binned in 5 day intervals), and the $g$-band decline (binned in 0.25\,mag intervals) in 30 days after maximum (\textit{bottom}). While the distributions show some indication of bimodality by eye, applying Hartigan's Dip Test \citep{har1985} shows that this is not statistically significant.}
\label{hist}
\end{figure}

For most SLSNe, the data unambiguously exclude \Ni~and \Co~decay as the main power source around light curve maximum: the \Ni-mass needed to power the peak is of the order of the total ejected mass inferred from light curve fitting, and moreover exceeds the limiting \Ni-mass inferred from the late-time luminosity \citep[e.g.][]{ins2013}. However, \citet{gal2012} proposed that the events with broader light curves, such as SN 2007bi, are radioactively powered, and likely exploded as pair-instability SNe. These would be fundamentally different from the other SLSNe, having a different explosion mechanism and power source, and would be characterised observationally by longer light curve timescales. We now look to see if we can distinguish two distinct classes in our data. In fact, Fig.~\ref{corr_plot} shows a relative lack of SLSNe with \tr$\,\sim25$-30$\,$d and with \td$\,\sim40$-60$\,$d. To make this clearer, we plot histograms of \tr, \td, and the $g$-band decline in magnitude 30 days after maximum light. This is a proxy for the decline timescale that is much simper to observationally measure, and particularly useful for parameterising the decline in high-redshift SLSNe, which may not have good rest-frame coverage in the redder bands. The data are shown in Fig.~\ref{hist}. %By eye, it appears plausible that the SLSNe in our sample may fall into two populations, separated in light curve timescales.

We apply the Dip Test of \citet{har1985} to each parameter. The dip statistic, $D$, measures the maximum difference between the empirical distribution and a unimodal distribution (chosen so as to minimise $D$). A larger value of $D$ indicates that the data are not well described by a unimodal probability distribution function (PDF). We test for multi-modality using a bootstrapping method. We construct 5000 random sets of length $n$, where $n$ is the size of our SLSN sample ($n=19$ for our full sample), drawn from a uniform PDF, and calculate $D$ for each set. The probability $p$-value for the null hypothesis to be correct (i.e.~that the data are unimodal) is then given by $p=N(D_{\rm SLSNe}<D_{\rm boot})/5000$ (this is the fraction of random sets, drawn from a uniform PDF, which appear to be less unimodal than our data).

The results of the test ($D$ and $p$) are shown in Table \ref{stats}. In no case do we find statistically significant evidence for bimodality. Therefore we have no confidence in rejecting the null hypothesis, and cannot confirm the existence of two sub-classes (fast/slow) of SLSNe. The similarity in overall light curves (up to some stretch factor) presented in sections \ref{cor}-\ref{shape} may actually be evidence in favour of a single class, and if so, likely a single distribution in timescales. We note, however, that we do see lower $p$-values when we remove the hydrogen-rich events from the sample -- but not low enough to exclude a unimodal distribution of light curve timescales. Many more objects will need to be observed before a stronger statement can be made. If the SLSN population does turn out to be significantly bimodal when a larger sample is constructed, it could indicate either two separate explosion mechanisms, or two progenitor channels for the same explosion mechanism.

\begin{table}
\caption{Significance testing for bimodality}
\label{stats}

\begin{tabulary}{\columnwidth}{LLLLL}

\hline
Property & Sample & Dip Statistic, $D$ & $p$-value \\
\hline

Rise time	&	Full sample	&	0.060	&	0.801 \\
		&	SLSNe Ic	&	0.069	&	0.666 \\
Decline time	&	Full sample	&	0.067	&	0.605 \\
		&	SLSNe Ic	&	0.077	&	0.439 \\
$\Delta g_{30}$	&	Full sample	&	0.069	&	0.722 \\
		&	SLSNe Ic	&	0.078	&	0.357 \\
Estimated mass	&	Full sample	&	0.044	&	0.992 \\
		&	SLSNe Ic	&	0.050	&	0.974 \\
		
\hline
\end{tabulary}

`Full sample' includes the 3 events with hydrogen lines in their spectra.

\end{table}

\section{Spectral evolution}\label{spec}

\begin{figure*}
\center
\includegraphics[width=18cm,angle=0]{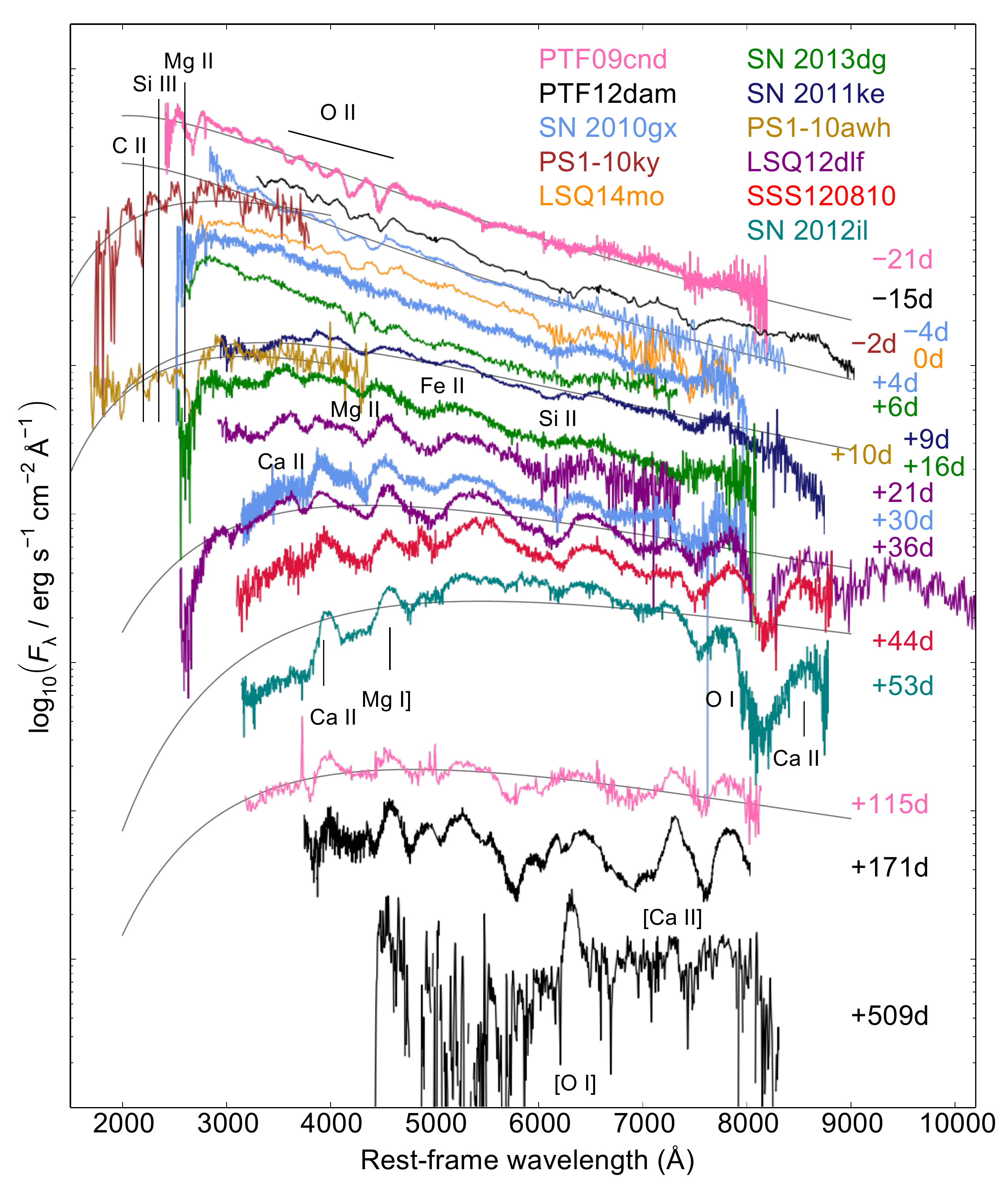}
\caption{Typical spectroscopic evolution of SLSNe Ic. The grey curves are representative blackbody fits to determine the colour temperature (shown in Fig.~\ref{temp}). The main line-forming ions are also marked.}
\label{evol}
\end{figure*}

\citet{qui2011} first presented pre-maximum spectra of a sample of SLSNe from PTF; the high degree of similarity enabled the authors to determine that these objects together formed a new class of supernova. Since \citet{pas2010} showed the spectroscopic evolution of SLSN 2010gx into a more typical SN Ic, all H-poor SLSNe have been seen to follow this path, including slowly-declining objects \citep{nic2013}. Here we compile high signal-to-noise spectra of 11 objects from our sample, to construct the complete spectral evolution, shown in Fig.~\ref{evol}. The data for these are from the references in Table \ref{sample}, and are available from WISeREP \citep{yar2012} or via the PESSTO  data release through the ESO archive 
\citep{sma2014}.

\begin{figure}
\includegraphics[width=9cm,angle=0]{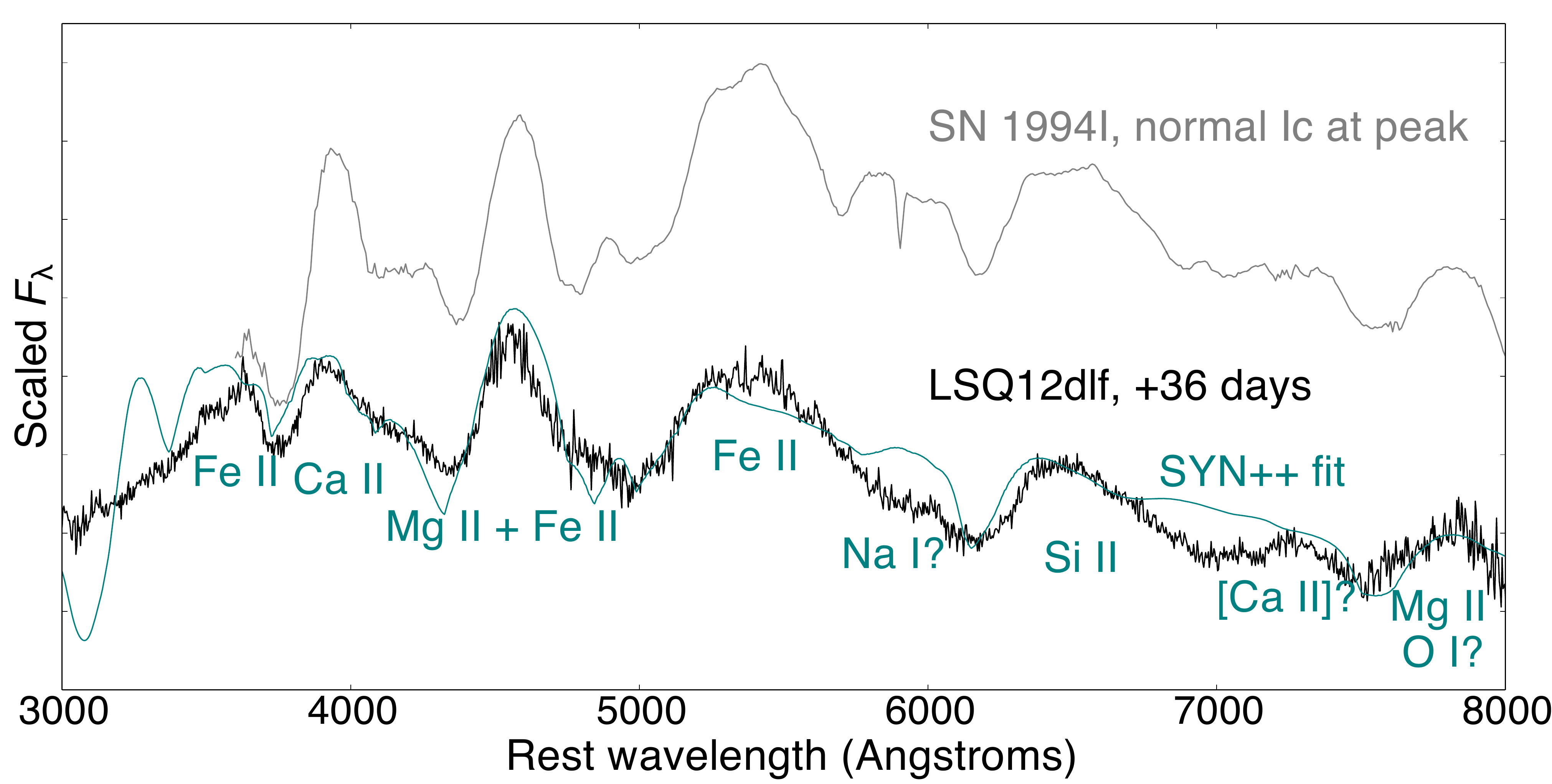}
\caption{SLSN LSQ12dlf at around one month after maximum light, compared to the normal SN Ic 1994I at maximum. The dominant lines in the two spectra are identical. LSQ12dlf shows stronger continuum emission in the blue, despite being older. A SYN++ fit \citep[from][]{nic2014} identifies the strongest lines.}
\label{spec_compare}
\end{figure}

The pre-max spectra are extremely blue, peaking at around 2500\AA, and dominated by high-ionisation lines -- particularly O II at around 4000\AA, seen mainly in absorption. Blackbody fits at this phase give a colour temperature \Tcol$\,\sim15000\,$K. There is a major change at around maximum light: as the ejecta cool, the O II lines disappear (oxygen, with an ionisation potential of 
O\,{\sc i} of  $13.6\,$eV, is mostly neutral at $T<15000\,$K), leaving the optical spectrum largely featureless. Within 10-20 days after the luminosity peaks, broad P-Cygni lines of singly-ionised metals (with lower first ionisation potentials, $\sim\,$6-8$\,$eV) emerge, mainly Ca II H\&K, Mg II, FeII blends (particularly between 5000-5500\AA) and Si II. By 30 days, these lines are strong, and the spectra closely resemble a conventional SN Ic at around maximum light (though even at this phase, SLSNe are often bluer than SNe Ic at peak). This similarity, and the slightly bluer colours, are highlighted in Fig.~\ref{spec_compare}.

The next lines to appear are O I $\lambda$7775 and the Ca II near-infrared triplet, along with the semi-forbidden Mg I] $\lambda$4571 emission. This line becomes dominant over the Mg II/Fe II P-Cygni somewhere around 30-50 days after maximum, and can be difficult to disentangle from the allowed transitions in the region. Strong forbidden emission lines do not appear in the spectra until over 100 days after maximum light (slowly declining objects such as PTF12dam are still not fully nebular at 200d). In a normal Ic, [O I] $\lambda\lambda$6300,6363 and [Ca II] $\lambda\lambda$7291,7323 are already prominent at 60-100 days after peak. These do eventually become the strongest lines in our very late spectra \citep[PTF12dam at 500d;][]{chen2014}. The late appearance of nebular lines in SLSNe may indicate densities higher than those in normal SNe Ic. However, the nebular transition point depends most strongly on the opacity of the ejecta, which is in turn determined by the ionisation of the dominant elements (opacity drops by several orders of magnitude when these recombine). Therefore the higher temperatures in SLSN ejecta could account for the slower evolution towards the nebular phase.

\begin{figure}
\includegraphics[width=9cm,angle=0]{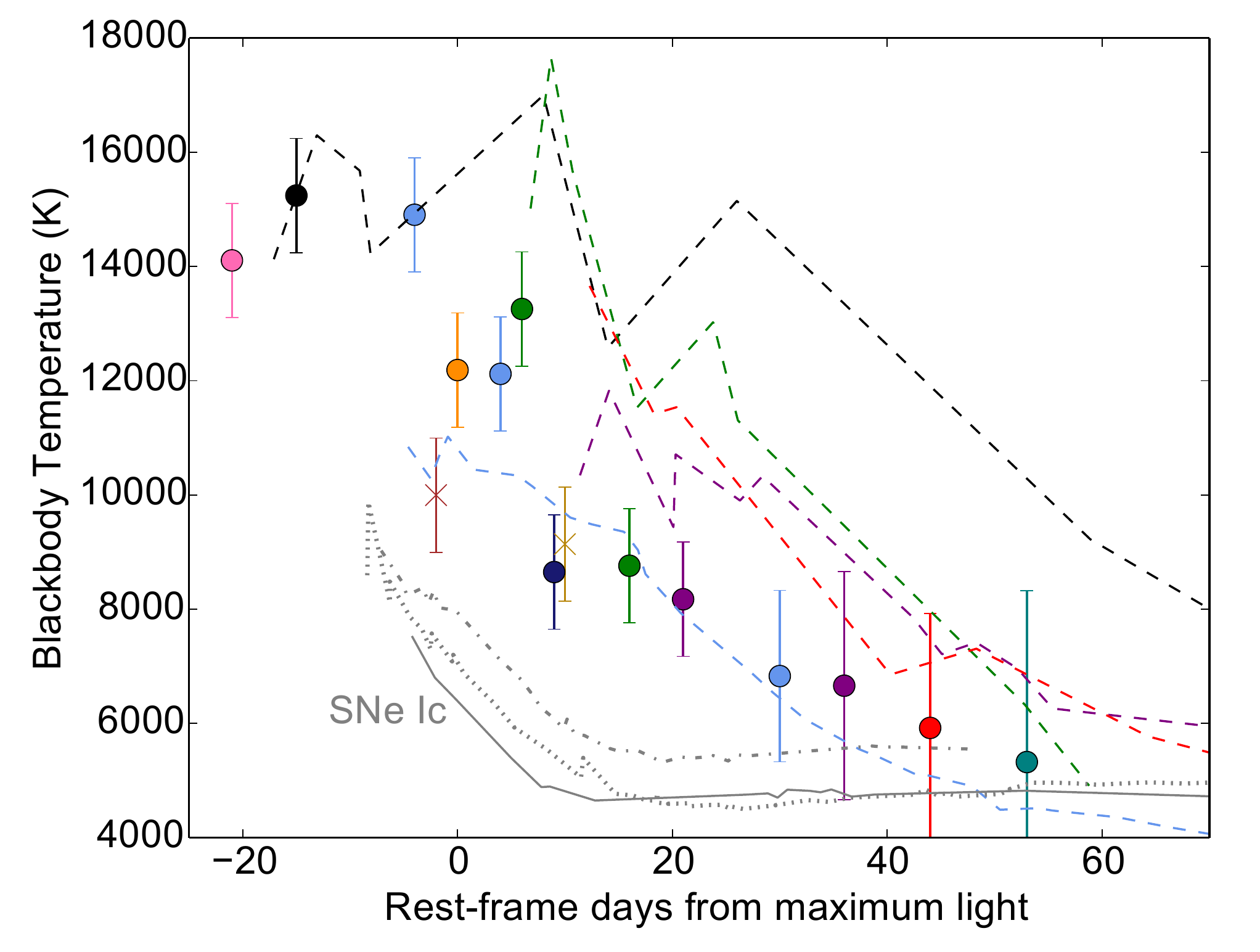}
\caption{The evolution of the colour temperature for SLSNe Ic and normal SNe Ic. Colours are the same as in Fig.~\ref{evol}. Temperatures are derived from blackbody fits to spectra (points) and photometry (lines). Photometric measurements tend to be hotter, as there is a lack of $U$-band data to constrain the fit at $\sim3000\,$\AA, which is not a problem for the spectral fits. Crosses indicate UV spectra ($\lambda<2000\,$\AA). The SNe Ic are 1994I (solid), 2007gr (dotted) and 1998bw (dot-dashed). SLSNe appear to have a very slow colour temperature evolution prior to maximum light, with typical \Tcol$\,\sim15000\,$K. Around maximum light, the colour temperature drops rapidly, reaching $\sim10000\,$K around 10 days later. This is partly due to decreasing gas temperature, but the evolution is accelerated at this phase due to the emergence Mg II, Ca II and Fe II lines, which absorb flux in the blue. The SLSNe then cool at a constant rate for 50-60 days. Normal SNe Ic do most of their cooling soon after explosion, and reach a constant temperature ($\sim5000\,$K) 10-20 days after maximum light.}
\label{temp}
\end{figure}

In Fig.~\ref{temp}, we plot the colour temperatures measured from blackbody fits to our spectra (using the whole observed range in wavelength), as well as fits to the photometry of some well-observed low-$z$ SLSNe Ic. The temperature evolution is markedly different from other Type Ic SNe, including the energetic SN 1998bw. SLSNe display (and maintain) much higher temperatures before and around peak light, and cool much more slowly afterwards. This requires significant heating over an extended period of time (several tens of days); therefore the abnormally high UV emission in SLSNe is a consequence of sustained energy input rather than a shock breakout phenomenon. This could plausibly be explained either by central engine models \citep[as shown by, e.g.][]{ins2013,nic2013}, or by CSM-interaction models, provided the CSM is sufficiently extended that the forward shock continues to propagate and deposit heat for tens of days after the supernova becomes visible. The temperature behind a strong shock is approximately $T_s=3/16\, m v_s^2 k_B^{-1}$, where $v_s$ is the shock speed, $m$ is the average particle mass and $k_B$ is the Boltzmann constant \citep{mck1991}. For singly-ionised, oxygen-rich material, the shock speed needed to reach $T\sim15000\,$K is $v_s\sim12000\,$\kms. In 10 days, such a shock travels $\sim10^{15}$cm, which is approximately the blackbody radius of the photosphere in SLSNe -- thus the temperature evolution does seem to be consistent with interaction models.

However, there are several features of SLSN spectral evolution which may be difficult to reconcile with interaction-powered models. In particular
\begin{itemize}
\item{We do not see narrow lines from slow-moving material in any SLSN Ic (though we do in CSS12105).}
\item{Moreover, the dense, massive CSM required to match the luminosity should be extremely optically thick, so the spectral lines seen at early times should be from the outermost material. These lines are puzzlingly broad ($>10000\,$\kms) if they are not supernova ejecta.}
\item{The spectral evolution is very homogeneous among SLSNe. Supernovae which are known definitively to be interacting with their CSM  (i.e.~SNe IIn) exhibit quite diverse spectra, generally showing gas with a range of velocities.}
\end{itemize}

Thus, the spectral evolution is more easily explained without circumstellar interaction. The most natural interpretation is that the ejecta from SLSNe have a similar composition to that of SNe Ic, but because of the much higher temperatures, due to heating from a power source equivalent to several solar masses of nickel, we do not see normal SN Ic spectral lines until later in the evolution, instead seeing a very blue continuum and high ionisation lines around peak. The slow spectral evolution may be exacerbated by higher densities in SLSNe.

\section{Velocity measurements}\label{vel}

\begin{figure}
\includegraphics[width=9cm,angle=0]{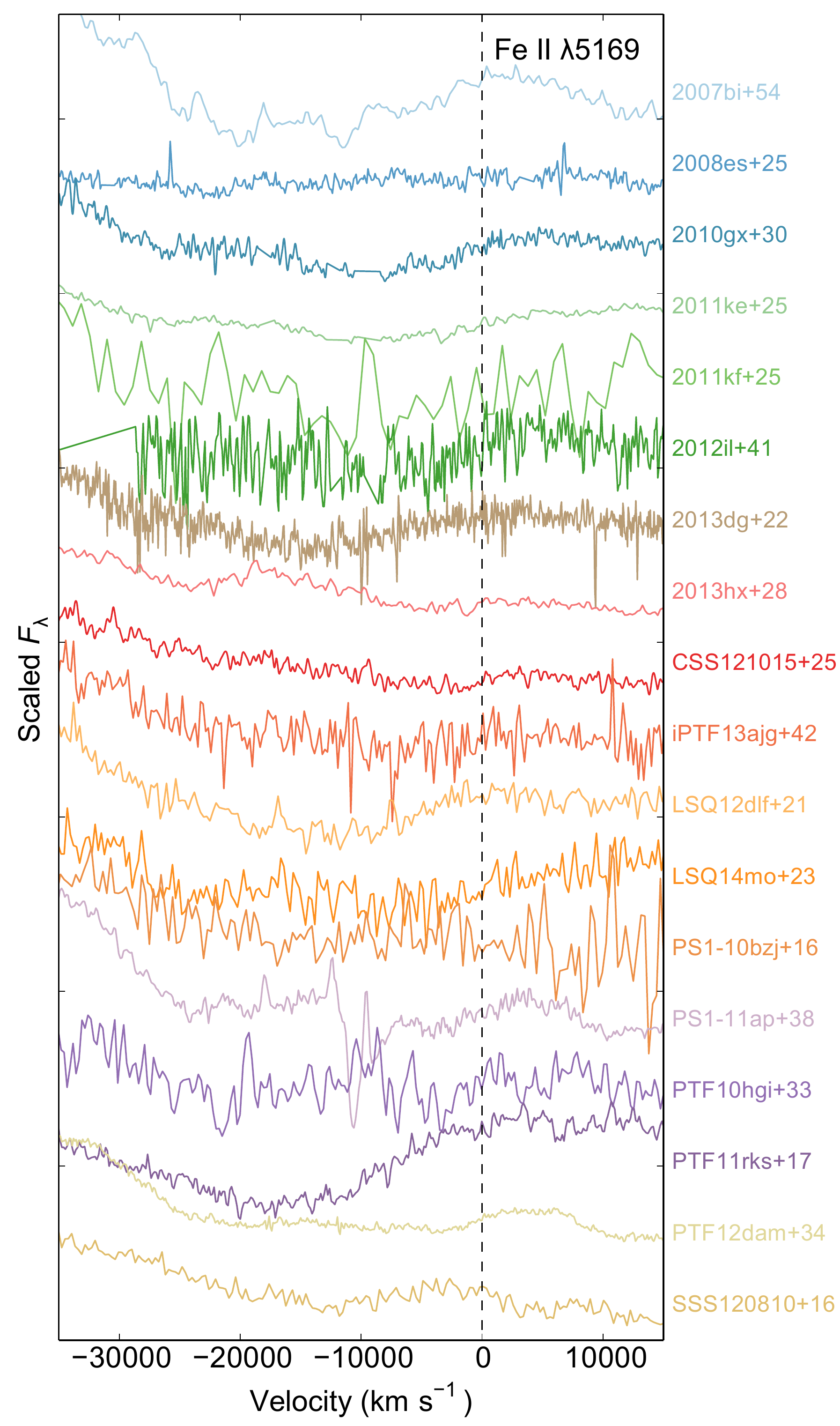}
\caption{The Fe II $\lambda$5169 P-Cygni profile in SLSNe at $\sim$20-30 days after maximum light. The minimum of the absorption trough gives an indication of the photospheric velocity. The line is clearly weaker (or absent entirely) in our SLSNe Type II (2008es and CSS121015) -- it does not become prominent in these objects until $\ga40\,$d after maximum light \citep{ben2014}.}
\label{fe}
\end{figure}

The analysis in section \ref{cor} demonstrated that diffusion time in the ejecta is the most important factor in generating the observed diversity in SLSN light curves. As shown by equation \ref{diff}, this timescale is a function of ejected mass, velocity and opacity. Since all of our super-luminous objects (at least the hydrogen-deficient ones) have very similar spectral evolution, we expect that they have similar compositions and ionisation states, and hence similar opacity. This leaves us with a degeneracy in \Mej~and the expansion velocity, $v$. Velocities can be estimated using the spectra, allowing us to break this degeneracy, and to determine whether it is variations in \Mej~or $v$~ that are most important between the SNe (super-luminous and normal) in our sample. We measure velocities using the absorption minimum of the Fe II $\lambda$5169 P-Cygni profile. This line profile is shown at $t=20$-$30\,$d in Fig.~\ref{fe}

\begin{figure}
\includegraphics[width=8.8cm,angle=0]{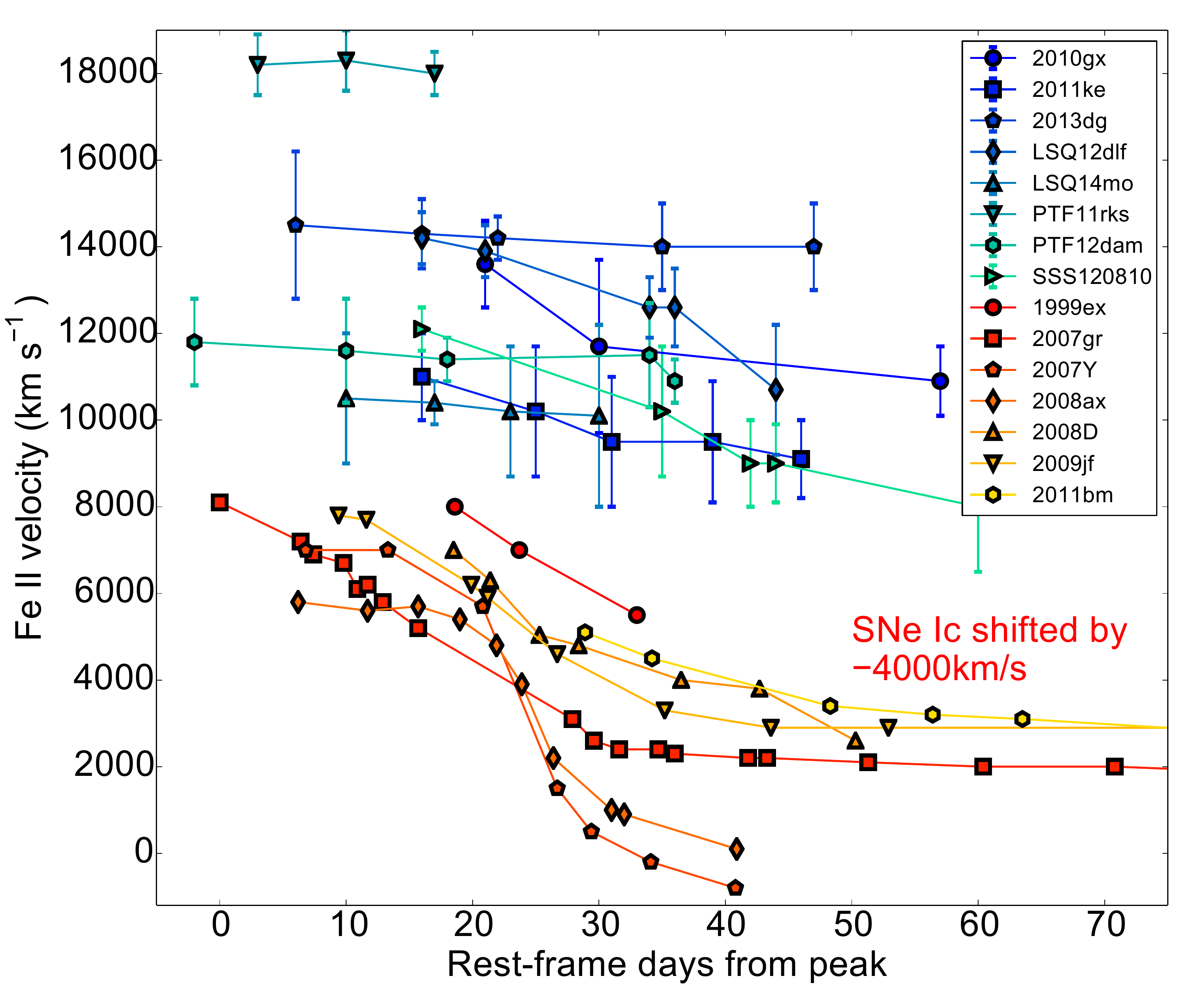}
\caption{Velocity evolution of well-observed SLSNe Ic. All velocities have been measured from Fe II$\lambda$5169 absorption minima. The SN Ic velocities are taken from \citet{val2012}, who used the same method; these have been shifted downwards by 4000\kms for clarity of presentation. The velocity curves for SLSNe are seen to be much flatter than those of SNe Ic, where we see a rapid decline after maximum light.}
\label{v}
\end{figure}

For a few objects, we have sufficient temporal coverage and signal-to-noise to measure this over a period of more than 30 days, and see how the velocity evolves. This is shown in Fig.~\ref{v}. The velocities are remarkably constant in time, declining by at most 2000\kms in the first 30 days after maximum light (and in many cases showing no clear decline at all). This is quite discrepant with the velocity evolution of other SNe Ic, such as the sample from \citet{val2012} shown in the figure. In normal SNe Ic, the ejecta expand in a roughly homologous fashion \citep[a natural consequence of a spherical shock;][]{arn1980}. Thus, as the ejecta expand and the photosphere recedes to deeper layers, we see more of the relatively slow-moving inner material, and hence measure lower velocities. Constant velocities, on the other hand, are predicted by the magnetar models of \citet{kas2010}, who showed that for magnetars with rotational energy greater than the kinetic energy of the supernova ejecta (i.e. $\ga10^{51}\,$erg), essentially all of the ejecta is swept up into a dense shell of uniform velocity. This is an unavoidable consequence of the central overpressure, and thus far the clearest observational test for engine-powered supernovae. The fairly flat velocity curves of our SLSNe are therefore consistent with the \citet{kas2010} models. This was also pointed out by \citet{chom2011}, in their study of SLSNe PS1-10ky and PS1-10awh. Moreover, because of this slow evolution, we can use velocity measurements at 20-30 days after maximum light as a reasonable proxy to the velocity at day 0, which is useful as for most objects we do not have good detections of Fe II until this phase.

\begin{figure}
\includegraphics[width=8.8cm,angle=0]{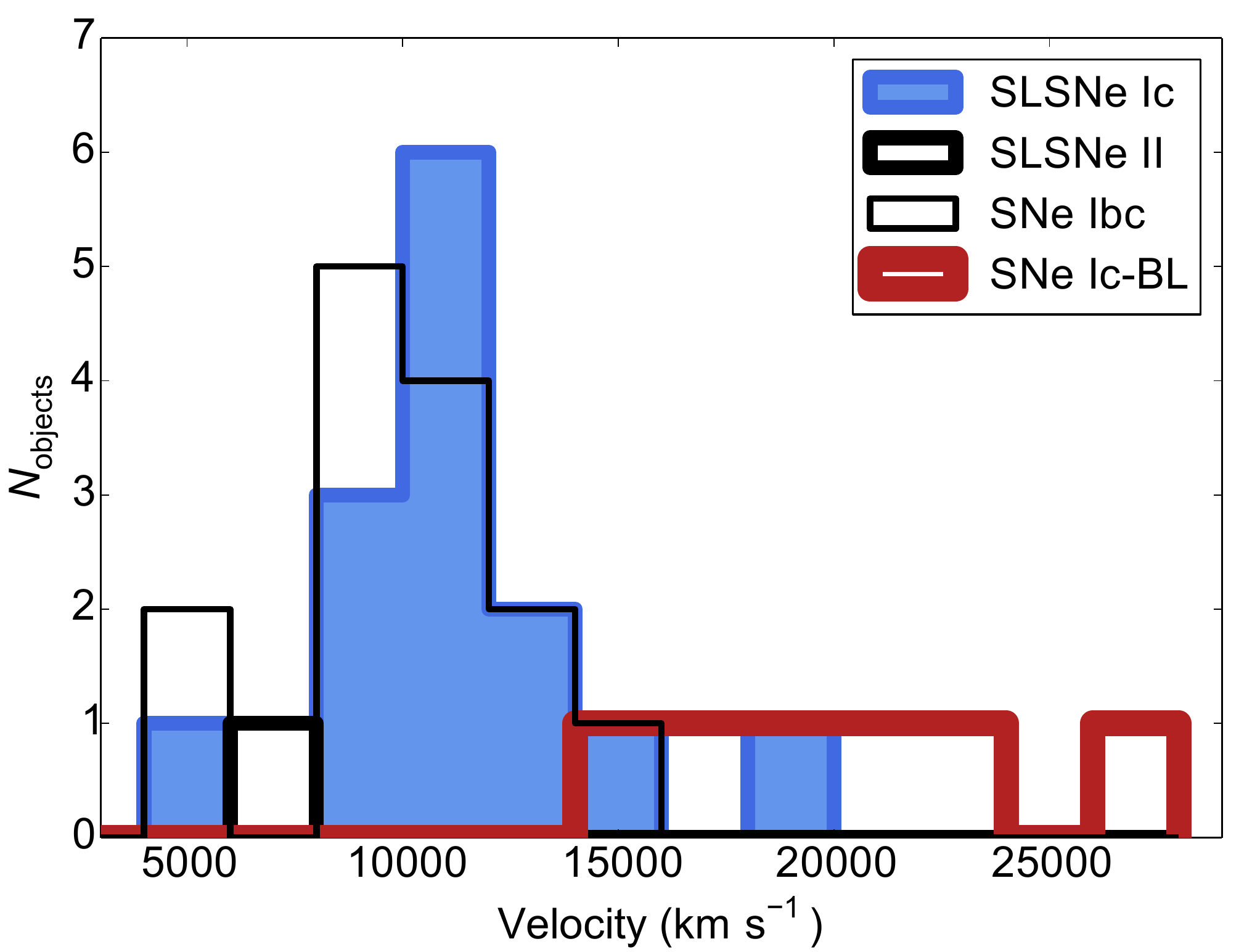}
\caption{Photospheric velocities measured from the lines shown in Fig.~\ref{fe}, compared with the velocities of SNe Ic at maximum light (the same objects in Fig.~\ref{corr_plot}). Bin size is 2000\,\kms. Most SLSNe have velocities close to the mean value, $\approx10800\,$\kms, with a standard deviation of $\approx3000\,$\kms. The outliers are PTF11rks (fast) and PTF10hgi (slow). The other slow object is CSS121015, though the Fe II line is very weak at this epoch (and not robustly detected in SN 2008es or  SN 2013hx, the other SLSNe II). The median velocities are 10500\,\kms, 9800\,\kms~and 20400\,\kms~for SLSNe, normal SNe Ibc and broad-lined SNe Ic, respectively.}
\label{v_hist}
\end{figure}

The velocity measurements from the Fe II line are given in Table \ref{props}, as well as velocities for the SDSS SN Ibc sample from \citet{tad2014}, and the distributions for SLSNe and SNe Ibc are shown in Fig.~\ref{v_hist}. The  velocity distributions of SLSNe and normal SNe Ibc are almost indistinguishable within the errors: the median velocity for SLSNe is 10500 \kms, with a standard deviation of 3100 \kms, while the median for normal SNe Ibc is 9800 \kms, with a standard deviation of 2500 \kms. The broad-lined SNe Ic, on the other hand, all have velocities greater than 15000 \kms. The similarity between typical photospheric velocities at peak light for super-luminous and normal SNe Ic indicates that the broader light curves and slower spectral evolution in SLSNe are not caused by a slower expansion. However, there is a possible caveat to this: since SLSNe have higher temperatures, and therefore higher ionisation, the photosphere may be formed further out in mass coordinate compared to normal SNe Ic, and therefore amidst faster moving ejecta. This would give a high photospheric velocity even if the bulk expansion was slower. \citet{ins2013} and \citet{nic2013} looked at the widths of [Ca II] and Mg I] emission lines in late spectra of SLSNe, which give an independent trace of the expansion velocity, finding typical values of 10000 \kms, fully consistent with our estimates from the Fe II P-Cygni lines. This gives us confidence that SLSNe expand at least as fast as other SNe Ic.

Two of our objects, LSQ12dlf and SSS120810, do show significant velocity evolution. We also note that \citet{ins2013} found evidence of decreasing photospheric velocity in some SLSNe Ic, compatible with the predictions of simple spherical models, rather than dense shells. They presented Fe II velocities close to maximum light for a number of objects, whereas here we had difficulty to reliably determine the Fe II profiles at such early epochs, when the line is very weak. The apparent velocity decline shown in their study also relied on measurements of different lines at different epochs, including emission line widths at late times. In fact, their Fe II velocities after 15-20 days from peak are flatter than at early times, and in reasonable agreement with our measurements here. However, if these objects do show an initial velocity gradient, it may indicate that the process causing the flat velocity evolution (for example, a second shock) has not yet terminated, or may suggest a tail in the ejecta distribution extending to high velocity. Another possibility is that the overpressure from a central engine is anisotropic \citep[stronger along the polar axes;][]{buc2007}. An observer might initially see high velocities from the faster polar region, with lower \tm, before later seeing the contribution from the slower, higher-\tm~equatorial material. In any case, if peak velocities estimated from our measurements at 20-30 days are really lower limits, higher velocities at peak in SLSNe compared to normal SNe Ic would only serve to strengthen our conclusion: the broad light curves of SLSNe cannot be explained by low velocities.

\section{Mass estimates}\label{mass}

Having shown that velocity is not the parameter driving the diversity in SLSN evolution timescales, and if the opacity is similar for all of our objects, we are left with the ejected mass as the most important factor. Equation \ref{diff} can be rearranged to give
\begin{equation}
M_{\rm ej} = 7.7\times10^{-7} \left(\frac{\kappa}{0.1\,\rm{cm}^2\,{\rm g}^{-1}} \right)^{-1} \frac{v}{{\rm km}\,{\rm s}^{-1}} \left(\frac{\tau_m}{{\rm days}}\right)^2,
\label{mass_eq}
\end{equation}
where all variables are as defined in section \ref{cor}, and our velocity measurements are taken from section \ref{vel}. There is an important caveat to our velocity measurements: we have measured the velocity of the photosphere (\vph), whereas the $v$ in equations \ref{diff} and \ref{mass_eq} are the ejecta `scale velocity' \citep{arn1980,arn1982}. There is no simple way to measure this scaling velocity, and it is unclear whether the photospheric velocity is a good representation of it. However, we will proceed with the assumption that $v\sim v_{\rm phot}$.

We take $\kappa=0.1\,$cm$^2\,$g$^{-1}$ as a fiducial value. \citet{ins2013} showed that this is a reasonable approximation for opacity dominated by electron scattering, if the temperature behind the photosphere is $\sim10^5\,$K, since abundant species such as oxygen, carbon and iron will be roughly half-ionised. For fully-ionised gas, $\kappa=0.2\,$cm$^2\,$g$^{-1}$. Shortly after explosion, we expect full ionisation, but this drops as the expanding SN cools. The opacity will therefore vary over the duration of the light curve, but this is not taken into account in the \citet{arn1982} formalism. Most authors use constant (essentially, time-averaged) opacities of  $\kappa=0.07-0.1\,$cm$^2\,$g$^{-1}$ for modelling normal SNe Ic. The ionisation fraction, and hence the electron-scattering opacity, may be higher for longer in SLSNe, due to the additional energy source, but this is unlikely to be by more than a small factor, simply because progressively more energy is required to remove successive electrons from an ion. From equation \ref{mass_eq}, increasing the opacity decreases the derived mass by the same factor. The lower limit on ejecta mass is found by assuming full ionisation -- this is therefore half of the mass estimated using $\kappa=0.1\,$cm$^2\,$g$^{-1}$.

The final parameter to estimate is the diffusion timescale, \tm. This was discussed in detail, both in general and for specific models, in section \ref{cor}, but we recap here for convenience. It is common practice to use the rise time as an estimate of \tm~\citep[for a recent example, see][]{whe2014}. This is approximately true in the original formulation of \citet{arn1982}, but that derivation was only for \Ni-powered SNe (\Ni~having an exponential lifetime of 8.8 days, comparable to the diffusion time in SNe Ia). As the decay time of \Ni~is fixed, the power input is the same for all normal Type I SNe, and any variation in light curve timescales depends only on the diffusion time. However, SLSNe may have a range of power input times, for example magnetars with different spin-down times, which may be much longer or shorter than the diffusion time. While it remains true that the diffusion time should generally be represented in the shape of the light curve around peak, we lack a straightforward mapping between \tm~and either \tr~or \td. As a best estimate, we take \tm$\,\approx\,($\tr$\,+\,$\td$)/2$, which is true for many scenarios, and use the rise and decline timescales as upper and lower limits on \tm. The uncertainty in mass owing to the choice of timescale is a factor of $\la2$.

More precise determinations can only be done through detailed modelling. Our goal here is not to derive the most exact masses, but to provide estimates for our sample in the most general and homegeneous way possible, thus exposing any underlying trends. The masses derived using our method are given in Table \ref{props}. One caveat we should add is that we have referred to the \textit{ejected} mass, but this assumes that the entire diffusion mass associated with the light curve is supernova material; if SLSNe Ic are powered by CSM-interaction, the diffusion mass we are probing is a combination of ejecta \textit{and} CSM. However, most of our analysis in this paper indicates that SLSNe are governed by generally similar physics to SNe Ibc, i.e. rapidly expanding material being heated from the inside. We therefore propose that our derived masses are likely representative of the ejecta.

\begin{figure}
\includegraphics[width=8.8cm,angle=0]{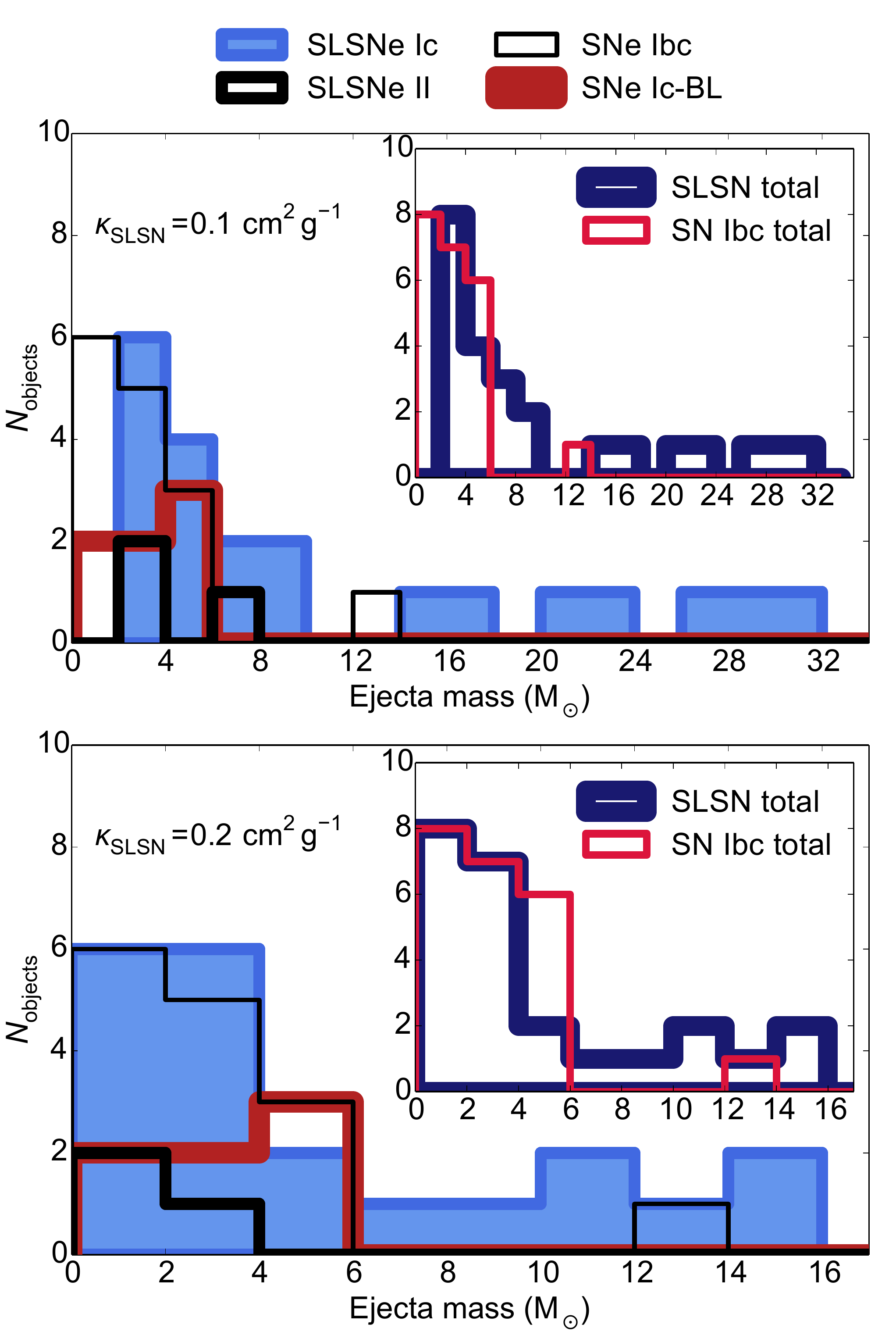}
\caption{The ejecta mass distribution for SLSNe and other H-poor SNe. SLSNe seem to arise from explosions ejecting $\ga2$ times the mass, on average, ejected by normal and broad-lined SNe Ibc, however, the mass distributions actually appear quite similar if the opacity in SLSNe is a factor 2 higher than in other SNe Ibc. A magnetar wind is a good candidate to increase the ionisation (hence opacity) in the ejecta \citep{met2014}}
\label{masses}
\end{figure}

The average ejecta mass in the SLSNe is 10.1\,\M~(standard deviation: 9.0\,\M), compared to 3.1\,\M~(2.9\,\M) and 3.1\,\M~(1.3\,\M) for the normal and broad-lined SNe Ibc, respectively. The median masses are 6.0\,\M, 2.5\,\M, and 2.9\,\M~for the three samples. While we are sensitive to small-number statistics here, the peak of the broad-lined Ic distribution appears to be at higher mass than the normal SNe Ic, especially if the extreme outlier SN 2011bm is neglected (giving a mean/median SN Ibc mass of 2.4/2.3\,\M). Taking only the SNe with an observed GRB counterpart (SNe 1998bw, 2010bh and 2012bz), the mean of 4.5\,\M~is intermediate between SNe Ibc and SLSNe. Fig.~\ref{masses} shows the ejecta mass distributions for SLSNe, SNe Ibc and SNe Ic-BL, with the ejected mass higher for SLSNe by a factor of 2-3. Although the broad-lined SNe Ic appear to be skewed towards higher mass than normal SNe Ibc, overall they seem to eject significantly less mass than the SLSNe. This supports the view of \citet{lel2015}, who suggested that SLSNe have more massive progenitors than LGRBs, but may be in tension with the results of \citet{lun2015}, who found that SLSNe do not trace host galaxy UV light (star formation) as closely as do LGRBs, implying older/lower-mass progenitors.

There are clearly several very massive (\Mej$\,\sim20$-$30\,$\M) SLSNe compared to the normal/broad-lined Ibc sample, for which only SN 2011bm hints at a high-mass tail. Five SLSNe eject over 20\,\M, yet there are only two objects in the 10-20\,\M~regime. To check whether this high-mass tail could be fully explained as a consequence of greater ionisation in SLSNe \citep[e.g.~due to the hard radiation field from a magnetar;][]{met2014}, we rescale the masses to $\kappa=0.2\,$cm$^2\,$g$^{-1}$ in the bottom panel of the figure. In this case, the bulk of the SLSN sample have masses consistent with the normal SN Ic population. However, there is still a clear excess of events with \Mej\,$\ga 10$\,\M. Therefore even if many of the SLSN light curves can be explained as a consequence of high ionisation, there remains a substantial number of events that must eject significantly more mass than normal stripped SNe.

The interesting question, then, is whether this indicates a separate population of high-mass SLSNe, arising from a different progenitor, explosion mechanism, or power source. Effectively, this is the physical  interpretation of the simple observational result illustrated in Fig.\,\ref{corr_plot} (and discussed in Section\,\ref{pop}) that the rise- and decline-time correlation visually picks out two groups. We apply Hartigan's Dip Test, as described in section \ref{pop}, to our distribution of SLSN masses. We find that $D=0.050$ and $p=0.974$, showing that the most massive SLSNe are fully consistent with being the tail of a continuous distribution. However, many more objects will be required in order to test this more robustly. There may also be an observational bias here: broad light curves are easier to detect than fast ones, which could lead to the more massive objects being over-represented.

The very large ejecta masses inferred for a few of our objects may be surprising. For example, \citet{nic2013} fit the light curve of PTF12dam with a magnetar engine in 16\,\M~ejecta (for the same opacity used here). Based solely on the light curve timescales and assuming no particular form for the power input, we have here estimated around 25\,\M. Fortunately, this discrepancy can be easily explained. As described by \citet{ins2013}, this magnetar model assumes a kinetic energy $E_{\rm k} = 10^{51} + 0.5(E_{\rm mag} - E_{\rm rad})\,$erg, where $E_{\rm mag}$ is the total energy input by the magnetar and $E_{\rm rad}$ is that lost due to the radiation emitted from the SN. For SLSNe that evolve quickly, the magnetar spins down rapidly, and overwhelms the original kinetic energy, so the derived parameters are not very sensitive to the initial choice of $10^{51}\,$erg. However, for slow objects like PTF12dam, this is not necessarily the case \citep[the fit by][had $E_{\rm k}=1.69\times10^{51}\,$erg]{nic2013}, so the initial supernova explosion energy is important. In fact, for their kinetic energy and mass, the expected velocity is lower than observations by a factor of two. From this we draw several conclusions:
\begin{itemize} 
\item{The explosion energy in PTF12dam exceeded $10^{51}\,$erg.}
\item{Velocity information can be an important constraint on the magnetar models of \citet{ins2013} \citep[ideally this would be the scaling velocity of][ though it is unclear how to derive this quantity from the spectrum]{arn1982}.}
\item{Again using equation 3, for a fitted diffusion time, doubling the velocity means doubling the mass, so the same magnetar model with the \textit{observed} expansion velocity ($\sim10000\,$\kms) would have ejected twice as much material, consistent with our estimate here.}
\end{itemize}

We also note that \citet{kas2010} reproduced the light curve of SN 2007bi with a magnetar model and 20\,\M~ejecta, using an opacity $\kappa=0.2\,$cm$^2\,$g$^{-1}$. In our estimates, we have used $\kappa=0.1\,$cm$^2\,$g$^{-1}$. For a fixed explosion energy and diffusion time, halving the opacity means increasing the ejecta mass by a factor $\sim1.6$ \citep{nic2014}. Our estimate of 31.1\,\M~for SN 2007bi is thus quite consistent with the hydrodynamical magnetar model of \citet{kas2010}. This gives us confidence in our simple mass derivations.

The masses also place constraints on explosion models. A core-collapse explosion ejecting 20\,\M~might be expected to produce a black hole rather than a neutron star \citep{heg2003}. However, recent theoretical results have shown that the final fate of a massive stellar core (neutron star or black hole) is unlikely to be monotonic in mass \citep[e.g.][]{oco2011,cla2015}. In particular, \citet{des2012c} have shown that in rapidly-rotating stars (potential GRB and SLSN progenitors), magnetars are produced more naturally. Nevertheless, if a black hole is formed in the core collapse, the central engine mechanism could still apply, in this case in the form of fall-back accretion \citep{dex2013} rather than magnetorotational powering. It should be noted that the fall-back models considered by \citet{dex2013} only produced super-luminous light curves if the progenitors retained their hydrogen envelopes, in order to delay fall-back to later times. The progenitors of SLSNe are expected to be stripped of their hydrogen, though \citet{nic2015b} and \citet{piro2015} recently found evidence that they may exhibit an inflated He envelope at the time of explosion.

Taking our upper limits on ejecta mass (from the decline timescales), the most massive objects may eject as much as 40-60$\,$\M, suggesting that pair-instability explosions should be considered as viable models. However quantitative comparisons do not lead to  comfortable agreement between the observed timescales and detailed models. Firstly, as shown by \citet{gal2009}, \citet{kas2011}, \citet{des2012} and \citet{nic2013}, H-poor PISN models with \Mej$\,\ga 100\,$\M~are needed in order to match the observed peak luminosities of SLSNe. This is a factor 2 higher than our most optimistic mass estimates. There is also a larger problem, which is independent of our mass estimates. The 120\,\M~model of \citet{kas2011} has \tr$\,=72\,$d and \td$\,=99\,$d, defined in the same way as for our objects (in these PISN models, the decline timescale is mainly set by \Co~decay, which has a timescale of 111 d). A few of our objects have \td$\,\ga80\,$d, but all have \tr~shorter than the PISN model by at least a factor of 2. This is the same problem found by \citet{nic2013} and \citet{mcc2014} when investigating PISN models for PTF12dam and PS1-11ap. Our measurements here support their conclusion: that even the 2007bi-like SLSNe do not quantitatively match the pair-instability explosion models.

\section{Summary of results}\label{sum}

For convenience, we here summarise our main findings:
\begin{itemize} 
\item{SLSNe Ic typically have $griz$ pseudobolometric magnitudes in the range $-20\ga M_{griz} \ga-21.5$, with a mean of -20.72 and a standard deviation of 0.55.}
\item{Their light curves obey a fairly tight relationship between rise and decline rates around maximum light, with \td$\,\approx2\,$\tr. This is naturally produced by simple diffusion models, but the lack of scatter may be difficult to explain with models requiring circumstellar interaction. For CSM interaction models to be the sole explanation, they must have CSM mass comparable to the ejected mass (at a large radius). In addition they require the CSM to have  a surprisingly narrow range of densities across all the objects.}
\item{The shape of the light curve around peak is intrinsically very similar to that of normal SNe Ibc, except SLSNe are broader and brighter. Broad-lined SNe Ic, including GRB-SNe, may span the magnitude gap between spectroscopically normal SNe Ic and SLSNe.}
\item{SLSNe Ic span a wide range of light curve timescales, but there is not yet statistically significant evidence for separate populations of fast- and slowly-evolving objects. It may be one continuous distribution.}
\item{The spectroscopic evolution requires sustained heating around peak, to maintain a temperature \Tcol$\,\sim15000\,$K despite weeks of expansion. At around one month after maximum light, SLSNe have cooled to temperatures comparable to normal SNe Ic at maximum, at which point the spectra show very similar photospheric-phase lines. This implies similar ejecta composition.}
\item{The temperature evolution is consistent with both central-engine and circumstellar-interaction light curve models.}
\item{The broad lines at all phases, lack of narrow lines in any object, similarity to SNe Ic, and the overall homogeneity of SLSNe spectra all argue against significant modification of the spectra by circumstellar material.}
\item{SLSNe and normal SNe Ic have similar photospheric velocities around maximum light. After 0-10 days from peak, this velocity stays remarkably constant in time for many SLSNe. This could be explained by a re-shock from their central power source, sweeping the ejecta up into a uniform shell \citep{kas2010}.}
\item{For a given opacity, the ejected mass, derived from measured Fe II $\lambda$5169 velocities and simple estimates of the diffusion timescale, are on average $\ga2$ times higher than the masses in normal and energetic SNe Ibc (although there is overlap). However, an alternative explanation is that the opacity in SLSNe is twice as high as in other SNe Ibc. Such an effect could possibly arise from a magnetar ionisation wind \citep{met2014}. There is no statistical evidence for a bimodal mass distribution in SLSNe.}
\item{SLSNe may eject as much as several tens of solar masses of material. Our estimates are consistent with masses derived from magnetar light curve fits. The masses seem to be too low (and the light curve evolution too fast) to be consistent with pair-instability models.}
\end{itemize}

\section{Conclusions}\label{conc}

We have investigated the physical properties of the largest sample of super-luminous supernovae constructed to date, and have found that they appear to be closely linked with normal SNe Ic. All of the properties of the class can be explained by taking a normal SN Ic, increasing the ejected mass, and re-shocking the ejecta using a powerful central energy input, such as the emission from magnetar spin-down. The key observables are a peak luminosity boosted by $\sim3$ magnitudes, a broader light curve from the large diffusion mass, and a flat velocity evolution caused by the hydrodynamical impact of the additional energy source. At late times, the light curve shape deviates from that of a normal SN Ic, which follows the radioactive decay of \Co~after a few diffusion times from peak. In super-luminous objects, the \Co~decay is masked by some other dominant heat source (e.g.~magnetar).

Other central engine models have been proposed, for example fallback accretion onto a newly-formed black hole \citep{dex2013}, but it is magnetar models that have been explored most in the literature. The fundamental lower limit on a neutron star spin period is $\sim1\,$ms, and Galactic magnetars have magnetic fields $<10^{15}\,$G. None of the SLSNe yet discovered---neither the slowly declining objects, such as PTF12dam \citep{nic2013}, nor the brightest SLSNe, like CSS121015 \citep{ben2014}---have emitted an integrated luminosity higher than that expected for magnetars with spin periods in the range 1-10$\,$ms and magnetic fields of a few times $10^{14}\,$G. It would be a problem for the magnetar model of SLSNe if an object was discovered that could only be fit with a sub-ms spin period or magnetic field $>10^{15}\,$G, but otherwise looked like a typical SLSN Ic. This could be one way to discriminate between competing models with different central engines.

Another important test will be detailed calculations of synthetic spectra. So far, only \citet{des2012} have presented calculations of model SLSN spectra based on magnetar radiation. \citet{how2013} presented parameterised models, putting a bright central energy source inside carbon- and oxygen-rich ejecta, such as might be expected from a stripped-envelope SN. In both cases, good matches were found to observational data. The main competing theory, that SLSNe are powered by hydrogen-poor circumstellar interaction, will also have to pass this test, but calculating such a spectrum is at the limit of current modelling capabilities. Another issue is that the interaction model has so many tuneable parameters that a wide range of models will need to be produced to compare with observations. However, light curve fitting shows that spectral modellers should focus on the regime with several solar masses of ejecta and CSM, with $\rho_{\rm CSM} \sim 10^{-12}\,$g$\,$cm$^{-3}$. \citep[We also note that interaction models provide a good description of the photometric and spectroscopic evolution of SLSNe II, which in some cases share characteristics with SLSNe Ic;][]{ben2014}. Based on our analysis here, we prefer magnetar-like models for SLSNe Ic, at least until such time as synthetic spectra exist for the CSM model.

On the observational side, future work should focus on finding more low redshift SLSNe, in order to improve on the statistics presented here. This should reveal more clearly whether, for example, all SLSNe Ic are drawn from a continuous population, if there is an excess of very massive objects, etc. Understanding the properties of these remarkable events is essential, particularly in preparation for the coming era of the James Webb Space Telescope and 30m-class ground-based facilities, which will allow us to find SLSNe at redshifts up to $z\sim10$. By constraining their physics at low redshift, we may be able to use SLSNe to probe early cosmological expansion, high-redshift dwarf galaxies, and the first generation of star formation in the Universe.

%I would add here that this is somehow complementary to GRBs. We have been using them to trace star-formation up to 8.2 (e.g. Tanvir et al. 2012; Hjorth et al. 2012).

%It is complementary because GRBs and SLSNe require different progenitors and hence are possibly formed during a different phase of the starburst (Leloudas et al. 2014). In Leloudas et al. (2014) we proposed that the progenitors of SLSNe are formed before GRB progenitors at least at low redshift. If this trend holds at higher redshift, then SLSNe could be detected at higher redshifts than GRBs. But even if not GRBs + SLSNe will give us a more complete view on massive star formation at very low and high redshift. 

%There is one issue with SLSN as tracers of SF. They are selected according to the observed luminosity + the survey selection function. Because of that, we will only be able to pick the barely extinguished ones.
%\\
%\\

\bigskip
\noindent
{\bf ACKNOWLEDGMENTS}
We thank B.~Metzger for helpful comments that improved the manuscript. The research leading to these results has received funding from the European Research Council under the European Union's Seventh Framework Programme (FP7/2007-2013)/ERC Grant agreement n$^{\rm o}$ [291222]  (PI : S. J. Smartt) and  STFC grants ST/I001123/1 and ST/L000709/1. M.N. acknowledges a studentship from DEL. This work is based (in part) on observations collected at the European Organisation for Astronomical Research in the Southern Hemisphere, Chile as part of PESSTO, (the Public ESO Spectroscopic Survey for Transient Objects Survey) ESO program 188.D-3003, 191.D-0935. M.F. is supported by the European Union FP7 programme through ERC grant number 320360. S.B. is partially supported by the PRIN-INAF 2014 with the project ÒTransient Universe: unveiling new types of stellar explosions with PESSTOÓ. N.E.R. acknowledges the support from the European Union Seventh Framework Programme (FP7/2007-2013) under grant agreement n. 267251 "Astronomy Fellowships in Italy" (AstroFIt). Research with SkyMapper was conducted in part by the Australian Research Council Centre of Excellence for All-sky Astrophysics (CAASTRO), through project number CE110001020. B.P.S. acknowledges support from the Australian Research Council Laureate Fellowship Grant LF0992131. DARK is funded by DNRF. M.S. acknowledges support from the Royal Society and EU/FP7-ERC grant no [615929]. F.E.B. acknowledges support by CONICYT through FONDECYT grant 1141218, and "EMBIGGEN" Anillo ACT1101, L.G. support through FONDECYT grant 3140566, and S.S. support through FONDECYT 3140534. L.G. and S.S. also acknowledge Basal-CATA PFB-06/2007. F.E.B., L.G. and S.S. acknowledge Project IC120009 ``Millennium Institute of Astrophysics (MAS)«« of Iniciativa Cient\'{\i}fica Milenio del Ministerio de Econom\'{\i}a, Fomento y Turismo. The Yale group thanks the Office of Science of the US Department of Energy, Grant no DE-FG02-92ER40704 and the Provosts Office at Yale for their support. A. G.-Y. acknowledges support by the EU/FP7 via ERC grant 307260; ISF, Minerva, and Weizmann-UK grants; as well as the ÒQuantum UniverseÓ I-Core Program of the Planning and Budgeting Committee and the Israel Science Foundation and the Kimmel Award. K.M. acknowledges support from a Marie Curie Intra-European Fellowship, within the 7th European Community Framework Programme (FP7).

\bibliographystyle{mn2e}

\bibliography{/Users/matt/Documents/Papers/bib_lib}

\begin{thebibliography}{103}
\expandafter\ifx\csname natexlab\endcsname\relax\def\natexlab#1{#1}\fi

\bibitem[{Arnett(1980)}]{arn1980}
Arnett W.~D., 1980, The Astrophysical Journal, 237, 541

\bibitem[{Arnett(1982)}]{arn1982}
Arnett W.~D., 1982, The Astrophysical Journal, 253, 785

\bibitem[{Baltay {et~al}\mbox{.}(2013)Baltay, Rabinowitz, Hadjiyska, Walker,
  Nugent, Coppi, Ellman, Feindt, McKinnon, Horowitz, {et~al.}}]{balt2013}
Baltay C. {et~al.}, 2013, Publications of the Astronomical Society of the
  Pacific, 125, 683

\bibitem[{Barbary {et~al}\mbox{.}(2009)Barbary, Dawson, Tokita, Aldering,
  Amanullah, Connolly, Doi, Faccioli, Fadeyev, Fruchter, {et~al.}}]{bar2009}
Barbary K. {et~al.}, 2009, The Astrophysical Journal, 690, 1358

\bibitem[{Barkat, Rakavy \& Sack(1967)Barkat, Rakavy, \& Sack}]{bar1967}
Barkat Z., Rakavy G., Sack N., 1967, Physical Review Letters, 18, 379

\bibitem[{Ben-Ami {et~al}\mbox{.}(2014)Ben-Ami, Gal-Yam, Mazzali, Gnat, Modjaz,
  Rabinak, Sullivan, Bildsten, Poznanski, Yaron, Arcavi, Bloom, Horesh,
  Kasliwal, Kulkarni, Nugent, Ofek, Perley, Quimby, \& Xu}]{ami2014}
Ben-Ami S. {et~al.}, 2014, The Astrophysical Journal, 785, 37

\bibitem[{Benetti {et~al}\mbox{.}(2014)Benetti, Nicholl, Cappellaro,
  Pastorello, Smartt, Elias-Rosa, Drake, Tomasella, Turatto, Harutyunyan,
  {et~al.}}]{ben2014}
Benetti S. {et~al.}, 2014, Monthly Notices of the Royal Astronomical Society,
  441, 289

\bibitem[{Berger {et~al}\mbox{.}(2012)Berger, Chornock, Lunnan, Foley, Czekala,
  Rest, Leibler, Soderberg, Roth, \& Narayan}]{ber2012}
Berger E. {et~al.}, 2012, The Astrophysical Journal Letters, 755, L29

\bibitem[{Bucciantini {et~al}\mbox{.}(2007)Bucciantini, Quataert, Arons,
  Metzger, \& Thompson}]{buc2007}
Bucciantini N., Quataert E., Arons J., Metzger B., Thompson T.~A., 2007,
  Monthly Notices of the Royal Astronomical Society, 380, 1541

\bibitem[{Bucciantini {et~al}\mbox{.}(2009)Bucciantini, Quataert, Metzger,
  Thompson, Arons, \& Del~Zanna}]{buc2009}
Bucciantini N., Quataert E., Metzger B., Thompson T., Arons J., Del~Zanna L.,
  2009, Monthly Notices of the Royal Astronomical Society, 396, 2038

\bibitem[{Cano {et~al}\mbox{.}(2011)Cano, Bersier, Guidorzi, Kobayashi, Levan,
  Tanvir, Wiersema, D'Avanzo, Fruchter, Garnavich, {et~al.}}]{cano2011}
Cano Z. {et~al.}, 2011, The Astrophysical Journal, 740, 41

\bibitem[{Chatzopoulos, Wheeler \& Vinko(2012)Chatzopoulos, Wheeler, \&
  Vinko}]{cha2012}
Chatzopoulos E., Wheeler J.~C., Vinko J., 2012, The Astrophysical Journal, 746,
  121

\bibitem[{Chatzopoulos {et~al}\mbox{.}(2013)Chatzopoulos, Wheeler, Vinko,
  Horvath, \& Nagy}]{cha2013}
Chatzopoulos E., Wheeler J.~C., Vinko J., Horvath Z., Nagy A., 2013, The
  Astrophysical Journal, 773, 76

\bibitem[{Chatzopoulos {et~al}\mbox{.}(2011)Chatzopoulos, Wheeler, Vinko,
  Quimby, Robinson, Miller, Foley, Perley, Yuan, Akerlof, {et~al.}}]{cha2011}
Chatzopoulos E. {et~al.}, 2011, The Astrophysical Journal, 729, 143

\bibitem[{{Chen} {et~al}\mbox{.}(2013){Chen}, {Smartt}, {Bresolin},
  {Pastorello}, {Kudritzki}, {Kotak}, {McCrum}, {Fraser}, \&
  {Valenti}}]{chen2013}
{Chen} T.-W. {et~al.}, 2013, The Astrophysical Journal, 763, L28

\bibitem[{Chen {et~al}\mbox{.}(2014)Chen, Smartt, Jerkstrand, Nicholl,
  Bresolin, Kotak, Polshaw, Rest, Kudritzki, Zheng, {et~al.}}]{chen2014}
Chen T.-W. {et~al.}, 2014, arXiv preprint arXiv:1409.7728

\bibitem[{Chevalier \& Fransson(1994)}]{che1994}
Chevalier R.~A., Fransson C., 1994, The Astrophysical Journal, 420, 268

\bibitem[{Chevalier \& Irwin(2011)}]{che2011}
Chevalier R.~A., Irwin C.~M., 2011, The Astrophysical Journal Letters, 729, L6

\bibitem[{Chomiuk {et~al}\mbox{.}(2011)Chomiuk, Chornock, Soderberg, Berger,
  Chevalier, Foley, Huber, Narayan, Rest, Gezari, {et~al.}}]{chom2011}
Chomiuk L. {et~al.}, 2011, The Astrophysical Journal, 743, 114

\bibitem[{Clausen, Piro \& Ott(2015)Clausen, Piro, \& Ott}]{cla2015}
Clausen D., Piro A.~L., Ott C.~D., 2015, The Astrophysical Journal, 799, 190

\bibitem[{Cooke {et~al}\mbox{.}(2012)Cooke, Sullivan, Gal-Yam, Barton,
  Carlberg, Ryan-Weber, Horst, Omori, \& D{\'\i}az}]{coo2012}
Cooke J. {et~al.}, 2012, Nature, 491, 228

\bibitem[{Davies {et~al}\mbox{.}(2009)Davies, Figer, Kudritzki, Trombley,
  Kouveliotou, \& Wachter}]{dav2009}
Davies B., Figer D.~F., Kudritzki R.-P., Trombley C., Kouveliotou C., Wachter
  S., 2009, ApJ, 707, 844

\bibitem[{Dessart {et~al}\mbox{.}(2012)Dessart, Hillier, Waldman, Livne, \&
  Blondin}]{des2012}
Dessart L., Hillier D.~J., Waldman R., Livne E., Blondin S., 2012, Monthly
  Notices of the Royal Astronomical Society: Letters, 426, L76

\bibitem[{Dessart, O'Connor \& Ott(2012)Dessart, O'Connor, \& Ott}]{des2012c}
Dessart L., O'Connor E., Ott C.~D., 2012, The Astrophysical Journal, 754, 76

\bibitem[{Dexter \& Kasen(2013)}]{dex2013}
Dexter J., Kasen D., 2013, The Astrophysical Journal, 772, 30

\bibitem[{Drake {et~al}\mbox{.}(2009)Drake, Djorgovski, Mahabal, Beshore,
  Larson, Graham, Williams, Christensen, Catelan, Boattini, {et~al.}}]{dra2009}
Drake A. {et~al.}, 2009, The Astrophysical Journal, 696, 870

\bibitem[{Drake {et~al}\mbox{.}(2010)Drake, Djorgovski, Prieto, Mahabal, Balam,
  Williams, Graham, Catelan, Beshore, \& Larson}]{dra2010}
Drake A. {et~al.}, 2010, The Astrophysical Journal Letters, 718, L127

\bibitem[{Drout {et~al}\mbox{.}(2011)Drout, Soderberg, Gal-Yam, Cenko, Fox,
  Leonard, Sand, Moon, Arcavi, \& Green}]{dro2011}
Drout M.~R. {et~al.}, 2011, The Astrophysical Journal, 741, 97

\bibitem[{Duncan \& Thompson(1992)}]{dunc1992}
Duncan R.~C., Thompson C., 1992, The Astrophysical Journal, 392, L9

\bibitem[{Filippenko(1997)}]{fil1997}
Filippenko A., 1997, Annu. Rev. Astro. Astrophys., 35, 309

\bibitem[{Filippenko {et~al}\mbox{.}(1995)Filippenko, Barth, Matheson, Armus,
  Brown, Espey, Fan, Goodrich, Ho, Junkkarinen, {et~al.}}]{fil1995}
Filippenko A.~V. {et~al.}, 1995, The Astrophysical Journal Letters, 450, L11

\bibitem[{Gaensler {et~al}\mbox{.}(2005)Gaensler, McClure-Griffiths, Oey,
  Haverkorn, Dickey, \& Green}]{gae2005}
Gaensler B., McClure-Griffiths N., Oey M., Haverkorn M., Dickey J., Green A.,
  2005, The Astrophysical Journal Letters, 620, L95

\bibitem[{Gal-Yam(2012)}]{gal2012}
Gal-Yam A., 2012, Science, 337, 927

\bibitem[{Gal-Yam {et~al}\mbox{.}(2009)Gal-Yam, Mazzali, Ofek, Nugent,
  Kulkarni, Kasliwal, Quimby, Filippenko, Cenko, Chornock, {et~al.}}]{gal2009}
Gal-Yam A. {et~al.}, 2009, Nature, 462, 624

\bibitem[{Gal-Yam, Ofek \& Shemmer(2002)Gal-Yam, Ofek, \& Shemmer}]{gal2002}
Gal-Yam A., Ofek E.~O., Shemmer O., 2002, Monthly Notices of the Royal
  Astronomical Society, 332, L73

\bibitem[{Gezari {et~al}\mbox{.}(2009)Gezari, Halpern, Grupe, Yuan, Quimby,
  McKay, Chamarro, Sisson, Akerlof, \& Wheeler}]{gez2009}
Gezari S. {et~al.}, 2009, ApJ, 690, 1313

\bibitem[{Ginzburg \& Balberg(2012)}]{gin2012}
Ginzburg S., Balberg S., 2012, The Astrophysical Journal, 757, 178

\bibitem[{Hartigan \& Hartigan(1985)}]{har1985}
Hartigan J.~A., Hartigan P., 1985, The Annals of Statistics, 70

\bibitem[{Heger {et~al}\mbox{.}(2003)Heger, Fryer, Woosley, Langer, \&
  Hartmann}]{heg2003}
Heger A., Fryer C., Woosley S., Langer N., Hartmann D., 2003, ApJ, 591, 288

\bibitem[{Howell {et~al}\mbox{.}(2013)Howell, Kasen, Lidman, Sullivan, Conley,
  Astier, Balland, Carlberg, Fouchez, Guy, {et~al.}}]{how2013}
Howell D. {et~al.}, 2013, The Astrophysical Journal, 779, 98

\bibitem[{Inserra {et~al}\mbox{.}(2013)Inserra, Smartt, Jerkstrand, Valenti,
  Fraser, Wright, Smith, Chen, Kotak, Pastorello, {et~al.}}]{ins2013}
Inserra C. {et~al.}, 2013, The Astrophysical Journal, 770, 128

\bibitem[{Inserra \& Smartt(2014)}]{ins2014}
Inserra C., Smartt S.~J., 2014, The Astrophysical Journal, 796, 87

\bibitem[{Kaiser {et~al}\mbox{.}(2010)Kaiser, Burgett, Chambers, Denneau,
  Heasley, Jedicke, Magnier, Morgan, Onaka, \& Tonry}]{kai2010}
Kaiser N. {et~al.}, 2010, in SPIE Astronomical Telescopes+ Instrumentation,
  International Society for Optics and Photonics, pp. 77330E--77330E

\bibitem[{Kasen \& Bildsten(2010)}]{kas2010}
Kasen D., Bildsten L., 2010, The Astrophysical Journal, 717, 245

\bibitem[{Kasen, Woosley \& Heger(2011)Kasen, Woosley, \& Heger}]{kas2011}
Kasen D., Woosley S., Heger A., 2011, ApJ, 734, 102

\bibitem[{Keller {et~al}\mbox{.}(2007)Keller, Schmidt, Bessell, Conroy,
  Francis, Granlund, Kowald, Oates, Martin-Jones, Preston, {et~al.}}]{kel2007}
Keller S.~C. {et~al.}, 2007, Publications of the Astronomical Society of
  Australia, 24, 1

\bibitem[{{Leloudas} {et~al}\mbox{.}(2014){Leloudas}, {Ergon}, {Taddia},
  {Nyholm}, {Sollerman}, {Inserra}, {Scalzo}, {Benetti}, {Pastorello},
  {Smartt}, {Smith}, {Young}, {Sullivan}, {Taubenberger}, {Valenti}, {Fraser},
  {Yaron}, {Gal-Yam}, {Manulis}, {Knapic}, {Smareglia}, {Molinaro}, {Baltay},
  {Ellman}, {Hadjiyska}, {McKinnon}, {Rabinowitz}, {Walker}, {Feindt},
  {Kowalski}, \& {Nugent}}]{14mo_atel}
{Leloudas} G. {et~al.}, 2014, The Astronomer's Telegram, 5839, 1

\bibitem[{{Leloudas} {et~al}\mbox{.}(2015){Leloudas}, {Schulze}, {Kr{\"u}hler},
  {Gorosabel}, {Christensen}, {Mehner}, {de Ugarte Postigo}, {Amor{\'{\i}}n},
  {Th{\"o}ne}, {Anderson}, {Bauer}, {Gallazzi}, {He{\l}miniak}, {Hjorth},
  {Ibar}, {Malesani}, {Morell}, {Vinko}, \& {Wheeler}}]{lel2015}
{Leloudas} G. {et~al.}, 2015, Monthly Notices of the Royal Astronomical
  Society,, 449, 917

\bibitem[{{Li} {et~al}\mbox{.}(2011){Li}, {Leaman}, {Chornock}, {Filippenko},
  {Poznanski}, {Ganeshalingam}, {Wang}, {Modjaz}, {Jha}, {Foley}, \&
  {Smith}}]{li2011}
{Li} W. {et~al.}, 2011, Monthly Notices of the Royal Astronomical Society, 412,
  1441

\bibitem[{Lunnan {et~al}\mbox{.}(2014)Lunnan, Chornock, Berger, Laskar, Fong,
  Rest, Sanders, Challis, Drout, Foley, {et~al.}}]{lun2014}
Lunnan R. {et~al.}, 2014, The Astrophysical Journal, 787, 138

\bibitem[{Lunnan {et~al}\mbox{.}(2013)Lunnan, Chornock, Berger, Milisavljevic,
  Drout, Sanders, Challis, Czekala, Foley, Fong, {et~al.}}]{lun2013}
Lunnan R. {et~al.}, 2013, The Astrophysical Journal, 771, 97

\bibitem[{Lunnan {et~al}\mbox{.}(2015)Lunnan, Chornock, Berger, Rest, Fong,
  Scolnic, Jones, Soderberg, Challis, Drout, {et~al.}}]{lun2015}
Lunnan R. {et~al.}, 2015, The Astrophysical Journal, 804, 90

\bibitem[{Mazzali {et~al}\mbox{.}(2002)Mazzali, Deng, Maeda, Nomoto, Umeda,
  Hatano, Iwamoto, Yoshii, Kobayashi, Minezaki, {et~al.}}]{maz2002}
Mazzali P. {et~al.}, 2002, The Astrophysical Journal Letters, 572, L61

\bibitem[{McCrum {et~al}\mbox{.}(2014)McCrum, Smartt, Kotak, Rest, Jerkstrand,
  Inserra, Rodney, Chen, Howell, Huber, {et~al.}}]{mcc2014}
McCrum M. {et~al.}, 2014, Monthly Notices of the Royal Astronomical Society,
  437, 656

\bibitem[{McCrum {et~al}\mbox{.}(2015)McCrum, Smartt, Rest, Smith, Kotak,
  Rodney, Young, Chornock, Berger, Foley, Fraser, Wright, Scolnic, Tonry,
  Urata, Huang, Pastorello, Botticella, Valenti, Mattila, Kankare, Farrow,
  Huber, Stubbs, Kirshner, Bresolin, Burgett, Chambers, Draper, Flewelling,
  Jedicke, Kaiser, Magnier, Metcalfe, Morgan, Price, Sweeney, Wainscoat, \&
  Waters}]{mcc2015}
McCrum M. {et~al.}, 2015, Monthly Notices of the Royal Astronomical Society,
  448, 1206

\bibitem[{McKee \& Draine(1991)}]{mck1991}
McKee C.~F., Draine B.~T., 1991, Science, 252, 397

\bibitem[{{Metzger} {et~al}\mbox{.}(2011){Metzger}, {Giannios}, {Thompson},
  {Bucciantini}, \& {Quataert}}]{met2011}
{Metzger} B.~D., {Giannios} D., {Thompson} T.~A., {Bucciantini} N., {Quataert}
  E., 2011, Monthly Notices of the Royal Astronomical Society, 413, 2031

\bibitem[{{Metzger} \& {Piro}(2014)}]{met2014}
{Metzger} B.~D., {Piro} A.~L., 2014, Monthly Notices of the Royal Astronomical
  Society, 439, 3916

\bibitem[{Miller {et~al}\mbox{.}(2009)Miller, Chornock, Perley, Ganeshalingam,
  Li, Butler, Bloom, Smith, Modjaz, Poznanski, {et~al.}}]{mil2009}
Miller A. {et~al.}, 2009, The Astrophysical Journal, 690, 1303

\bibitem[{Modjaz {et~al}\mbox{.}(2009)Modjaz, Li, Butler, Chornock, Perley,
  Blondin, Bloom, Filippenko, Kirshner, Kocevski, {et~al.}}]{mod2009}
Modjaz M. {et~al.}, 2009, The Astrophysical Journal, 702, 226

\bibitem[{{Moriya} \& {Maeda}(2012)}]{mor2012}
{Moriya} T.~J., {Maeda} K., 2012, The Astrophysical Journal Letters, 756, L22

\bibitem[{Neill {et~al}\mbox{.}(2011)Neill, Sullivan, Gal-Yam, Quimby, Ofek,
  Wyder, Howell, Nugent, Seibert, Martin, {et~al.}}]{nei2011}
Neill J.~D. {et~al.}, 2011, The Astrophysical Journal, 727, 15

\bibitem[{Nicholl {et~al}\mbox{.}(2013)Nicholl, Smartt, Jerkstrand, Inserra,
  McCrum, Kotak, Fraser, Wright, Chen, Smith, {et~al.}}]{nic2013}
Nicholl M. {et~al.}, 2013, Nature, 502, 346

\bibitem[{Nicholl {et~al}\mbox{.}(2015)Nicholl, Smartt, Jerkstrand, Sim,
  Inserra, Anderson, Baltay, Benetti, Chambers, Chen, {et~al.}}]{nic2015b}
Nicholl M. {et~al.}, 2015, arXiv preprint arXiv:1505.01078

\bibitem[{Nicholl {et~al}\mbox{.}(2014)Nicholl, Smartt, Jerkstrand, Inserra,
  Anderson, Baltay, Benetti, Chen, Elias-Rosa, Feindt, Fraser, Gal-Yam,
  Hadjiyska, Howell, Kotak, Lawrence, Leloudas, Margheim, Mattila, McCrum,
  McKinnon, Mead, Nugent, Rabinowitz, Rest, Smith, Sollerman, Sullivan, Taddia,
  Valenti, Walker, \& Young}]{nic2014}
Nicholl M. {et~al.}, 2014, Monthly Notices of the Royal Astronomical Society,
  444, 2096

\bibitem[{O'Connor \& Ott(2011)}]{oco2011}
O'Connor E., Ott C.~D., 2011, The Astrophysical Journal, 730, 70

\bibitem[{Ofek {et~al}\mbox{.}(2007)Ofek, Cameron, Kasliwal, Gal-Yam, Rau,
  Kulkarni, Frail, Chandra, Cenko, Soderberg, {et~al.}}]{ofek2007}
Ofek E. {et~al.}, 2007, The Astrophysical Journal Letters, 659, L13

\bibitem[{Ofek {et~al}\mbox{.}(2010)Ofek, Rabinak, Neill, Arcavi, Cenko,
  Waxman, Kulkarni, Gal-Yam, Nugent, Bildsten, {et~al.}}]{ofek2010}
Ofek E. {et~al.}, 2010, The Astrophysical Journal, 724, 1396

\bibitem[{Pastorello {et~al}\mbox{.}(2010)Pastorello, Smartt, Botticella,
  Maguire, Fraser, Smith, Kotak, Magill, Valenti, Young, {et~al.}}]{pas2010}
Pastorello A. {et~al.}, 2010, The Astrophysical Journal Letters, 724, L16

\bibitem[{Patat {et~al}\mbox{.}(2001)Patat, Cappellaro, Danziger, Mazzali,
  Sollerman, Augusteijn, Brewer, Doublier, Gonzalez, Hainaut,
  {et~al.}}]{pat2001}
Patat F. {et~al.}, 2001, The Astrophysical Journal, 555, 900

\bibitem[{Piro(2015)}]{piro2015}
Piro A.~L., 2015, arXiv preprint arXiv:1505.07103

\bibitem[{Quimby {et~al}\mbox{.}(2007)Quimby, Aldering, Wheeler, H{\"o}flich,
  Akerlof, \& Rykoff}]{qui2007}
Quimby R.~M., Aldering G., Wheeler J.~C., H{\"o}flich P., Akerlof C.~W., Rykoff
  E.~S., 2007, The Astrophysical Journal Letters, 668, L99

\bibitem[{Quimby {et~al}\mbox{.}(2011)Quimby, Kulkarni, Kasliwal, Gal-Yam,
  Arcavi, Sullivan, Nugent, Thomas, Howell, Nakar, {et~al.}}]{qui2011}
Quimby R.~M. {et~al.}, 2011, Nature, 474, 487

\bibitem[{Quimby {et~al}\mbox{.}(2013)Quimby, Yuan, Akerlof, \&
  Wheeler}]{qui2013}
Quimby R.~M., Yuan F., Akerlof C., Wheeler J.~C., 2013, Monthly Notices of the
  Royal Astronomical Society, 431, 912

\bibitem[{Rakavy \& Shaviv(1967)}]{rak1967}
Rakavy G., Shaviv G., 1967, The Astrophysical Journal, 148, 803

\bibitem[{Rau {et~al}\mbox{.}(2009)Rau, Kulkarni, Law, Bloom, Ciardi,
  Djorgovski, Fox, Gal-Yam, Grillmair, Kasliwal, {et~al.}}]{rau2009}
Rau A. {et~al.}, 2009, Publications of the Astronomical Society of the Pacific,
  121, 1334

\bibitem[{Rest {et~al}\mbox{.}(2011)Rest, Foley, Gezari, Narayan, Draine,
  Olsen, Huber, Matheson, Garg, Welch, {et~al.}}]{rest2011}
Rest A. {et~al.}, 2011, The Astrophysical Journal, 729, 88

\bibitem[{Richardson {et~al}\mbox{.}(2002)Richardson, Branch, Casebeer,
  Millard, Thomas, \& Baron}]{ric2002}
Richardson D., Branch D., Casebeer D., Millard J., Thomas R., Baron E., 2002,
  The Astronomical Journal, 123, 745

\bibitem[{{Richardson} {et~al}\mbox{.}(2014){Richardson}, {Jenkins}, {Wright},
  \& {Maddox}}]{rich2014}
{Richardson} D., {Jenkins}, III R.~L., {Wright} J., {Maddox} L., 2014, The
  Astrophysical Journal, 147, 118

\bibitem[{Richmond {et~al}\mbox{.}(1996)Richmond, Van~Dyk, Ho, Peng, Paik,
  Treffers, Filippenko, Bustamante-Donas, Moeller, Pawellek,
  {et~al.}}]{ric1996}
Richmond M.~W. {et~al.}, 1996, The Astronomical Journal, 111, 327

\bibitem[{Sanders {et~al}\mbox{.}(2012)Sanders, Soderberg, Valenti, Foley,
  Chornock, Chomiuk, Berger, Smartt, Hurley, Barthelmy, {et~al.}}]{san2012}
Sanders N.~E. {et~al.}, 2012, The Astrophysical Journal, 756, 184

\bibitem[{Schlafly \& Finkbeiner(2011)}]{schlaf2011}
Schlafly E.~F., Finkbeiner D.~P., 2011, The Astrophysical Journal, 737, 103

\bibitem[{Schulze {et~al}\mbox{.}(2014)Schulze, Malesani, Cucchiara, Tanvir,
  Kr{\"u}hler, de~Ugarte~Postigo, Leloudas, Lyman, Bersier, Wiersema,
  {et~al.}}]{sch2014}
Schulze S. {et~al.}, 2014, Astronomy \& Astrophysics, 566, A102

\bibitem[{{Smartt} {et~al}\mbox{.}(2014){Smartt}, {Valenti}, {Fraser},
  {Inserra}, {Young}, {Sullivan}, {Pastorello}, {Benetti}, {Gal-Yam}, {Knapic},
  {Molinaro}, {Smareglia}, {Smith}, {Taubenberger}, {Yaron}, {Anderson},
  {Ashall}, {Balland}, {Baltay}, {Barbarino}, {Bauer}, {Baumont}, {Bersier},
  {Blagorodnova}, {Bongard}, {Botticella}, {Bufano}, {Bulla}, {Cappellaro},
  {Campbell}, {Cellier-Holzem}, {Chen}, {Childress}, {Clocchiatti},
  {Contreras}, {Dall Ora}, {Danziger}, {de Jaeger}, {Della Valle}, {Dennefeld},
  {Elias-Rosa}, {Elman}, {Feindt}, {Fleury}, {Gall}, {Gonzalez-Gaitan},
  {Galbany}, {Greggio}, {Guillou}, {Hachinger}, {Hadjiyska}, {Hage},
  {Hillebrandt}, {Hodgkin}, {Hsiao}, {James}, {Jerkstrand}, {Kangas},
  {Kankare}, {Kotak}, {Kromer}, {Kuncarayakti}, {Leloudas}, {Lundqvist},
  {Hook}, {Maguire}, {Manulis}, {Margheim}, {Mattila}, {Maund}, {Mazzali},
  {McCrum}, {McKinnon}, {Moreno-Raya}, {Nicholl}, {Nugent}, {Pain}, {Phillips},
  {Pignata}, {Polshaw}, {Pumo}, {Rabinowitz}, {Reilly}, {Romero-Canizales},
  {Scalzo}, {Schmidt}, {Schulze}, {Sim}, {Sollerman}, {Taddia}, {Tartaglia},
  {Terreran}, {Tomasella}, {Turatto}, {Walker}, {Walton}, {Wyrzykowski},
  {Yuan}, \& {Zampieri}}]{sma2014}
{Smartt} S.~J. {et~al.}, 2014, ArXiv e-prints

\bibitem[{Smith {et~al}\mbox{.}(2008)Smith, Chornock, Li, Ganeshalingam,
  Silverman, Foley, Filippenko, \& Barth}]{smi2008}
Smith N., Chornock R., Li W., Ganeshalingam M., Silverman J.~M., Foley R.~J.,
  Filippenko A.~V., Barth A.~J., 2008, The Astrophysical Journal, 686, 467

\bibitem[{Smith {et~al}\mbox{.}(2007)Smith, Li, Foley, Wheeler, Pooley,
  Chornock, Filippenko, Silverman, Quimby, Bloom, {et~al.}}]{smi2007b}
Smith N. {et~al.}, 2007, The Astrophysical Journal, 666, 1116

\bibitem[{Soderberg {et~al}\mbox{.}(2008)Soderberg, Berger, Page, Schady,
  Parrent, Pooley, Wang, Ofek, Cucchiara, Rau, {et~al.}}]{sod2008}
Soderberg A. {et~al.}, 2008, Nature, 453, 469

\bibitem[{Stritzinger {et~al}\mbox{.}(2002)Stritzinger, Hamuy, Suntzeff, Smith,
  Phillips, Maza, Strolger, Antezana, Gonz{\'a}lez, Wischnjewsky,
  {et~al.}}]{str2002}
Stritzinger M. {et~al.}, 2002, The Astronomical Journal, 124, 2100

\bibitem[{{Taddia} {et~al}\mbox{.}(2015){Taddia}, {Sollerman}, {Leloudas},
  {Stritzinger}, {Valenti}, {Galbany}, {Kessler}, {Schneider}, \&
  {Wheeler}}]{tad2014}
{Taddia} F. {et~al.}, 2015, Astronomy \& Astrophysics, 574, A60

\bibitem[{Taubenberger {et~al}\mbox{.}(2006)Taubenberger, Pastorello, Mazzali,
  Valenti, Pignata, Sauer, Arbey, B{\"a}rnbantner, Benetti, Della~Valle,
  {et~al.}}]{tau2006}
Taubenberger S. {et~al.}, 2006, Monthly Notices of the Royal Astronomical
  Society, 371, 1459

\bibitem[{Thompson, Chang \& Quataert(2004)Thompson, Chang, \&
  Quataert}]{tho2004}
Thompson T.~A., Chang P., Quataert E., 2004, The Astrophysical Journal, 611,
  380

\bibitem[{Valenti {et~al}\mbox{.}(2008{\natexlab{a}})Valenti, Benetti,
  Cappellaro, Patat, Mazzali, Turatto, Hurley, Maeda, Gal-Yam, Foley,
  {et~al.}}]{val2008b}
Valenti S. {et~al.}, 2008{\natexlab{a}}, Monthly Notices of the Royal
  Astronomical Society, 383, 1485

\bibitem[{Valenti {et~al}\mbox{.}(2008{\natexlab{b}})Valenti, Elias-Rosa,
  Taubenberger, Stanishev, Agnoletto, Sauer, Cappellaro, Pastorello, Benetti,
  Riffeser, {et~al.}}]{val2008}
Valenti S. {et~al.}, 2008{\natexlab{b}}, The Astrophysical Journal Letters,
  673, L155

\bibitem[{Valenti {et~al}\mbox{.}(2011)Valenti, Fraser, Benetti, Pignata,
  Sollerman, Inserra, Cappellaro, Pastorello, Smartt, Ergon,
  {et~al.}}]{val2011}
Valenti S. {et~al.}, 2011, Monthly Notices of the Royal Astronomical Society,
  416, 3138

\bibitem[{Valenti {et~al}\mbox{.}(2012)Valenti, Taubenberger, Pastorello,
  Aramyan, Botticella, Fraser, Benetti, Smartt, Cappellaro, Elias-Rosa,
  {et~al.}}]{val2012}
Valenti S. {et~al.}, 2012, The Astrophysical Journal Letters, 749, L28

\bibitem[{Vreeswijk {et~al}\mbox{.}(2014)Vreeswijk, Savaglio, Gal-Yam, De~Cia,
  Quimby, Sullivan, Cenko, Perley, Filippenko, Clubb, {et~al.}}]{vre2014}
Vreeswijk P.~M. {et~al.}, 2014, The Astrophysical Journal, 797, 24

\bibitem[{{Wheeler}, {Johnson} \& {Clocchiatti}(2014){Wheeler}, {Johnson}, \&
  {Clocchiatti}}]{whe2014}
{Wheeler} J.~C., {Johnson} V., {Clocchiatti} A., 2014, ArXiv e-prints

\bibitem[{Woosley(2010)}]{woo2010}
Woosley S., 2010, The Astrophysical Journal Letters, 719, L204

\bibitem[{Woosley, Blinnikov \& Heger(2007)Woosley, Blinnikov, \&
  Heger}]{woo2007}
Woosley S., Blinnikov S., Heger A., 2007, Nature, 450, 390

\bibitem[{Woosley \& Bloom(2006)}]{woo2006}
Woosley S., Bloom J., 2006, Annu. Rev. Astron. Astrophys., 44, 507

\bibitem[{Wyrzykowski {et~al}\mbox{.}(2014)Wyrzykowski, Kostrzewa-Rutkowska,
  Kozlowski, Udalski, Poleski, Skowron, Blagorodnova, Kubiak, Szymanski,
  Pietrzynski, {et~al.}}]{wyr2014}
Wyrzykowski L. {et~al.}, 2014, Acta Astronomica, 64, 197

\bibitem[{Yaron \& Gal-Yam(2012)}]{yar2012}
Yaron O., Gal-Yam A., 2012, Wiserep — an interactive supernova data
  repository

\bibitem[{Young {et~al}\mbox{.}(2010)Young, Smartt, Valenti, Pastorello,
  Benetti, Benn, Bersier, Botticella, Corradi, Harutyunyan, {et~al.}}]{you2010}
Young D. {et~al.}, 2010, Astronomy \& Astrophysics, 512, A70

\end{thebibliography}

\end{document}